\documentclass[aps,prd,reprint,showpacs,floatfix,longbibliography,nofootinbib,superscriptaddress]{revtex4-1}
\usepackage{tikz,graphicx,booktabs,multirow,array,microtype,mathtools,float}
\graphicspath{{figures/}}
\usepackage{enumerate}  
\extrafloats{200}
\usepackage[colorlinks=true,backref=false, linktocpage=true,
citecolor=blue,
urlcolor=blue,linkcolor=blue,
pdfpagemode=UseOutlines]{hyperref}
\newcolumntype{C}{>{$}c<{$}}
\usepackage{dcolumn}

\usepackage{anyfontsize}
\usepackage[shortlabels]{enumitem}

\usepackage[noabbrev,capitalize]{cleveref} 

\usepackage{longtable}
\newcolumntype{L}{>{$}l<{$}}
\newcolumntype{C}{>{$}c<{$}}
\newcolumntype{R}{>{$}r<{$}}
\renewcommand{\thetable}{\arabic{table}}  

\usepackage{comment} 
\excludecomment{supplement}

\usepackage{amsfonts}
\AtBeginDocument{
\heavyrulewidth=.08em
\lightrulewidth=.05em
\cmidrulewidth=.03em
\belowrulesep=.65ex
\belowbottomsep=0pt
\aboverulesep=.4ex
\abovetopsep=0pt
\cmidrulesep=\doublerulesep
\cmidrulekern=.5em
\defaultaddspace=.5em
}

\usepackage{amsmath} 
\usepackage{dsfont}  
\usepackage{bm}      

\def\reals{\mathds{R}}
\def\beq{\begin{equation}}
\def\eeq{\end{equation}}
\def\beqs#1\eeqs{\beq\begin{split} #1 \end{split}\eeq}

\long\def\comment#1{}

\usepackage{mleftright,xparse}
\NewDocumentCommand\xDeclarePairedDelimiter{mmm}
{%
	\NewDocumentCommand#1{som}{%
		\IfNoValueTF{##2}
		{\IfBooleanTF{##1}{#2##3#3}{\mleft#2##3\mright#3}}
		{\mathopen{\csname##2\endcsname#2}##3\mathclose{\csname##2\endcsname#3}}%
	}%
}
\xDeclarePairedDelimiter{\av}{\langle}{\rangle}
\xDeclarePairedDelimiter{\ket}{|}{\rangle}
\xDeclarePairedDelimiter{\bra}{\langle}{|}
\NewDocumentCommand\braket{somm}{%
	\IfNoValueTF{#2}{\mleft\langle #3\,|#4\mright\rangle}{NOTIMPLEMENTED}
}
\NewDocumentCommand\opbraket{sommm}{%
	\IfNoValueTF{#2}
	{\IfBooleanTF{#1}{\langle#3|#4|#5\rangle}{\mleft\langle #3 \left| #4 \right| #5 \mright\rangle}}
	{\mathopen{\csname#2\endcsname\langle}#3\mathopen{\csname#2\endcsname|} #4 \mathclose{\csname#2\endcsname|} #5\mathclose{\csname#2\endcsname\rangle}}
}

\begin{document}
\title{
Higher order finite volume quantization conditions for two spinless particles}
\author{Frank X. Lee}
\email{fxlee@gwu.edu}
\affiliation{Physics Department, The George Washington University, Washington, DC 20052, USA}
\author{Andrei Alexandru}
\email{aalexan@gwu.edu}
\affiliation{Physics Department, The George Washington University, Washington, DC 20052, USA}
\author{Ruair\'i Brett}
\email{rbrett@gwu.edu}
\affiliation{Physics Department, The George Washington University, Washington, DC 20052, USA}
\date{\today}

\begin{abstract}
Lattice QCD calculations of scattering phase shifts and resonance parameters in the two-body sector are becoming precision studies. 
Early calculations employed L\"uscher's formula for extracting these quantities at lowest order.
As the calculations become more ambitious, higher-order relations are required.
In this study we derive higher-order quantization conditions and introduce a method to transparently cross-check our results. 
This is an important step given the involved derivations of these formulas.
We derive quantization conditions up to $\ell=5$ partial waves in both cubic and elongated geometries, and for states with zero and nonzero total momentum.
All 45 quantization conditions we include here (22 in cubic box, 23 in elongated box) pass our cross-check test.
\end{abstract} 


\maketitle

\section{Introduction} \label{sec:intro}

Hadron structure and interactions are controlled by the quark and gluon
dynamics as described by quantum chromodynamics (QCD). Inside the hadrons the quarks
and gluons interact strongly and nonperturbative methods are required to
describe the interactions. Lattice QCD is used to study this dynamics
in the Euclidean time framework. In this framework QCD spectrum for one
or multiparticle states can be accessed directly. On the other hand,
information about hadron interactions, in particular scattering properties,
are accessed indirectly by calculating the energy of two-hadron states
in finite volume. An intuition about this connection comes from understanding
that the finite-volume energy shift for these states compared to the infinite volume 
setup is due to the interactions between the particles which in finite volume have
a nonvanishing probability of being separated by distances comparable to
the interaction range. The full details for this relation were worked out by L\"uscher~\cite{Luscher:1990ux}: he showed that the energy shifts for
two-particle states in finite volume (with periodic boundary conditions)
are directly controlled by the scattering phase shifts and the relations
are model independent. These relations are exact, up to exponentially small
finite-volume corrections, as long as the energy of the two particle states
is below the inelastic threshold.

In the field of nuclear and particle physics, the method has proven especially successful.
Various extensions to the method have since been made to enhance its applicability, 
including moving frames~\cite{Rummukainen:1995vs,Kim_2005,Fu:2011xz,Leskovec:2012gb,Gockeler:2012yj,Li:2021mob,Doring:2012eu},
asymmetric boxes~\cite{Feng:2004ua,Lee:2017igf,Li:2021mob}, multiple partial waves and coupled channel scattering~\cite{LIU_2006,LAGE2009439,Bernard:2008ax,Doring:2011wz,Hansen_2012,Luu:2011ep,Li:2012bi,Briceno:2012yi,Briceno:2014oea,Morningstar_2017,Li:2019qvh}.
The use of asymmetric lattices has proven to be computationally efficient~\cite{Pelissier:2012pi,Guo:2016zos,Guo:2018zss,Culver:2019qtx,Li:2021mob} 
so we will include it in our discussions. 
The method has been widely applied to a multitude of meson-meson scattering processes, along with some meson-baryon and baryon-baryon systems over the past decade~\cite{Pelissier:2012pi, Guo:2016zos,Guo:2018zss,Culver:2019qtx,Mai:2019pqr,Bulava:2016mks,Brett:2018jqw,Andersen:2018mau,Alexandrou:2017mpi,Bali:2015gji,Feng:2014gba,Feng:2010es,Orginos:2015aya,Beane:2011sc,Aoki:2011yj,Dudek:2012xn,Dudek:2012gj,Mohler:2013rwa,Prelovsek:2013ela,Mohler:2012na,Lang:2012sv,Lang:2011mn,Guo:2016zos,Helmes:2017smr,Liu:2016cba,Helmes:2015gla,Wilson:2014cna,Wilson:2015dqa,Moir:2016srx}. 
Significant progress towards a complete three-body scattering quantization condition has also been made in recent years, though we do not discuss it here. See Refs.~\cite{mai2021threebody,Mai:2021lwb,Fischer:2020jzp,Hansen:2019nir,Rusetsky:2019gyk,Culver:2019vvu,Blanton_2020,Alexandru:2020xqf,Brett:2021wyd,Hansen:2020otl,blanton2021interactions} for reviews of theoretical developments, and some first applications to three-pion and kaon scattering.


The {\em quantization conditions} (QCs), that is the equations connecting
the phase shifts to finite volume spectrum, are rather complex.
Early lattice QCD scattering studies have employed only the lowest-order
quantization conditions, which can be used to directly extract the
phase shifts from finite-volume energies. As these calculations become more ambitious higher-order
conditions, where several partial waves are involved, are required~\cite{Dudek2012,Dudek2013,Wilson2015,Morningstar_2017}.
L\"uscher worked out these formulas for zero-momentum states of
two equal-mass, spinless particles in a cubic box. Further work
extended these relations for different 
kinematics~\cite{Rummukainen:1995vs,Fu:2011xz,Leskovec:2012gb,Gockeler:2012yj,Feng:2004ua,Lee:2017igf,Morningstar_2017,Luu:2011ep,Li:2012bi,Briceno:2012yi,Briceno:2014oea}.
These relations need to take into account the geometry of the box,
the mass and spin of the hadrons, scattering channels, etc.
In this work we compute the quantization conditions for two spinless
particles in different kinematic situation for partial waves as high as $\ell=5$.

We note that the derivation of these formulas is fairly involved and the quantization
conditions are expressed in terms of special
functions that are nontrivial to calculate. Furthermore the coefficients 
appearing in these formulas depend on the symmetry group and the relevant 
irreducible representation. Some of the lower-order quantization conditions
have been checked thoroughly, but the ones we derive in this work, for special cases and higher order,
need to be cross-checked before they are used to extract scattering information
from noisy lattice QCD data. To this end, we propose a method to check our derivations
that is relatively transparent, especially for lattice
QCD practitioners, and we use it to test our results with high accuracy 
in all possible channels for two spinless particles.

The paper is organized as follows.  
In Sec.~\ref{sec:QC} we review how the QC are derived and how they are connected to the irreducible representations of the symmetry group. We also discuss moving frames in  nonrelativistic kinematics, and derive all the QCs we will investigate in this work.  
Then in Sec.~\ref{sec:scheq} we discuss the method we used to check our results, that is our approach to computing the two-particle spectrum
in a finite-volume box by solving the associated Schr\"odinger equation. 
The role of symmetries and their influence of the spectrum is first 
discussed here.
In Sec.~\ref{sec:validation}, we detail our numerical checks and compare  the QC results with the spectrum derived in Sec.~\ref{sec:scheq}. 
Some examples will be given. The rest will be available as Supplemental Material~\cite{supp}.
In Sec.~\ref{sec:con}, we summarize our findings and give future outlook. All group theory details are collected in Appendix~\ref{sec:group} and all matrix elements in Appendix~\ref{sec:ME}.

\section{Quantization condition}
\label{sec:QC}

Scattering is omnipresent in understanding the nature of interactions between particles.
In infinite volume, nonrelativistic two-particle scattering can be captured by solving the Schr\"odinger equation in the center of mass (CM) frame,
\beq
\left[ -\frac{\hbar^2}{2\tilde m} \nabla^2  + V(r)\right] \psi(\bm r) = E \psi(\bm r),
\label{eq:qm}
\eeq
where $\tilde m=m_1m_2/(m_1+m_2)$ is the reduced mass of the system.
Elastic scattering phase shift is defined as the change in phase 
in the scattered wave relative to the incident wave in the asymptotic region where the interaction can be neglected.
In the partial-wave expansion, the wave function satisfies the asymptotic condition,
\beq 
\psi(\bm r) = e^{ikz} + f(\theta) \frac{e^{ikr}}{r}
\eeq 
where $k$ is the relative CM momentum related to the energy $E=\hbar^2k^2/(2\tilde m)$ and 
\beq
f(\theta)=\sum_{l=0}^\infty (2l+1)f_lP_l(\cos\theta)
\eeq 
is the scattering amplitude. Phase shift $\delta_l$ enters via the partial-wave amplitudes
\beq
f_l\equiv \frac{e^{2i\delta_l}-1}{2 i k} 
= \frac{S_l-1}{2 i k}
= \frac{T_l}{k}
= \frac{K_l}{k(1-iK_l)},
\eeq 
where alternative definitions via S matrix, T matrix and K matrix are also indicated.
The phase shift is a real valued function of the interaction energy and carries information about the nature of the interaction, 
such as whether the force is 
attractive ($\delta<0$) or repulsive ($\delta>0$), whether a resonance is formed in the scattering, etc. 
In the exterior region  ($r>R$) where the interaction is vanishing, the
Schr\"{o}dinger equation has the form of a Helmholtz equation,
\beq
(\nabla^2 + k^2) \psi(\bm r)=0.
\label{eq:hemholtz}
\eeq
Its radial wave function can be expressed as a linear combination of spherical Bessel functions
$ u_l(r) \propto   a_l j_l(kr) + b_l n_l(kr)$, where the coefficients can be found by 
matching up with the wave function in the interior ($r<R$).
The phase shift can then be computed from the coefficients by
\beq
e^{2i\delta_l(k)} = \frac{ a_l(k) + i b_l(k) }{ a_l(k) - i b_l(k)  }\,.
\label{eq:phaseAB}
\eeq

\begin{figure}[b]
\includegraphics[width=0.375\columnwidth]{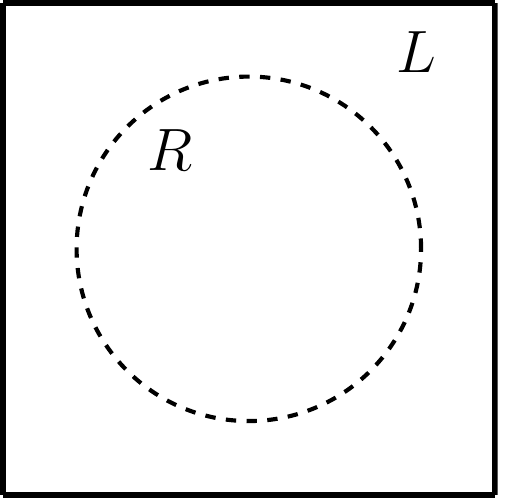}
\caption{Schematic sketch of the L\"uscher method. The system is enclosed in a box of size $L$. The range of the interaction is $R$. Periodic boundary conditions are imposed across the box surfaces.}
\label{fig:box}
\end{figure}
In finite volume, a similar procedure can be realized as detailed 
by the pioneering work of  L\"{u}scher~\cite{Luscher:1990ux}. The system is 
now confined in a box of size $L$  where we assume its size is big enough so that  the interaction range $R<L/2$, as shown in Fig.~\ref{fig:box}. 
Periodic boundary conditions are imposed on the wave function across the box surface,  
\beq 
\psi(\bm r + \bm n L) =\psi(\bm r ).
 \label{eq:bccm}\eeq
As we will see below, one basically ends up with a new relation that connects the same infinite-volume phase shifts with the
discrete energies of two-body states in the box, 
in the form of a quantization condition
\beq
\det \left [e^{2i\delta(k)} - \frac{ M(k,L) +i }{ M(k,L) -i} \right ] =0.
\label{eq:phaselat}
\eeq
Here $e^{2i\delta(k)}\equiv \text{diag}\{e^{2i\delta_{l}(k)}\}$ is a shorthand for diagonal matrix of all partial waves. The $M(k,L)$ is a Hermitian matrix function of CM momentum and box size. It is at the heart of the entire approach.

The L\"{u}scher method is very general, not just limited to the simple illustration above.  
It does not matter how the energy levels are obtained, be it in quantum mechanics, effective field theories, lattice QCD, or any other method. 
The same quantization condition applies and the results are the same up to exponentially suppressed finite-volume corrections. 
For this reason, it has become the method of choice for studying strongly interacting systems where traditional methods 
like perturbation theory do not apply.


The derivation of the QC in Eq.\eqref{eq:phaselat} is fairly involved~\cite{Luscher:1990ux}. 
The basic idea mimics the matching of interior and exterior wave functions  in standard scattering theory.
The complication comes from enclosing the system in a periodic box.
To make the presentation reasonably self-contained, we outline the essential steps here.
A solution to  Eq.\eqref{eq:hemholtz} in the region $r>R$ that satisfies the periodic boundary conditions in Eq.\eqref{eq:bccm}  
is given by the Green's function,
\beq
G(\bm r, k^2) = \frac{1}{L^3}  \sum_{\bm p} \frac{e^{i\bm p\cdot\bm r}}{\bm{p}^2-k^2},
\eeq
where the sum is over quantized momenta in the box.
A complete basis can be generated by taking its derivatives  
\beq
G_{lm}(\bm r, k^2)   =\mathcal{Y}_{lm}(\nabla) G(\bm r, k^2),
\label{eq:nabla}
\eeq
where  $\mathcal{Y}_{lm}(\bm r)\equiv r^l Y_{lm}(\theta,\phi)$ are the homogenous harmonic polynomials. 
The expansion of $G_{lm}$ in terms of $n_l$, $j_l$, and $Y_{lm}$ is needed for the matching.
The action of the differential operator on the singular and regular  terms produces the following identities,
\beq
\mathcal{Y}_{lm}(\nabla) \,n_0(kr)  =(-k)^l n_l(kr) Y_{lm}(\theta,\phi),
\eeq
and
\beqs
& \mathcal{Y}_{lm}(\nabla)\, j_l(kr) Y_{js}(\theta,\phi) \\
&= \frac{k^l}{\sqrt{4\pi}} \sum_{l'=|j-l|}^{j+l} \sum_{m'=-l'}^{l'} C_{ lm,js,|l'm'} j_{l'}(kr) Y_{l'm'}(\theta,\phi),
\eeqs
where the tensor coefficient is given in Wigner $3j$ symbols,
\beqs
C_{ lm,js,|l'm'}&= (-1)^{m'} i^{l-j+l'} \sqrt{(2l+1)(2j+1)(2l'+1)}  \\
 & \quad \times
\begin{pmatrix} l&j&l' \\ 0&0&0 \end{pmatrix}
\begin{pmatrix} l&j&l' \\ m&s&-m' \end{pmatrix}.
 \eeqs
Applying the identities, the basis functions can be expanded as
\beqs
&G_{lm}(\bm r, k^2) = \frac{(-1)^l k^{l+1}}{4\pi} \Big [ n_l(kr) Y_{lm}(\theta,\phi)   \\ 
&  +\sum_{l'=0}^{\infty} \sum_{m'=-l'}^{l'}  M_{l m,l' m' } j_{l'}(kr) Y_{l'm'}(\theta,\phi) \Big],
\eeqs
where matrix $M$ is introduced as a conduit to connect with the phase shifts.
Expanding the wave function in this basis, and matching it with the interior one in the region between the sphere and the box ($R<r<L/2$), one has
\beq
\psi=\sum_{lm} d_{lm} G_{lm}(\bm r, k^2)
= \sum_{lm} c_{lm} [a_l j_l +b_l n_l] Y_{lm}. 
\eeq
By equating the coefficients of $n_l$ and $j_l$, the following condition emerges (eliminate $d_{lm}$ in favor of $c_{lm}$)
\beq
\sum_{l'm'} c_{l'm'} \left[b_{l'} M_{l' m',l m }  - a_{l'} \delta_{ll'}\delta_{mm'} \right] =0. 
\eeq
By requiring nontrial solution of the linear system we get a determinant condition,
\beq
\det [B M -A] =0,
\eeq
where $A$ and $B$ are diagonal matrices from $a_l$ and $b_l$, respectively.
Finally, using the matrix version of Eq.\eqref{eq:phaseAB} to connect with the phase shifts,  
\beq
e^{2i\delta} = \frac{A+ i B}{A - i B },
\eeq
one arrives at the QC introduced in Eq.\eqref{eq:phaselat}.
Note that the QC is a single condition that connects all partial waves with all energy levels in the box.
At face value, it has very limited predictive power.
Later we will see how the QC can be reduced into pieces and used to make approximate predictions.

The explicit form of the matrix is given by 
\beq
M_{l m,l' m' } = \frac{ (-1)^l }{ \eta \pi^{3/2} } \sum_{j=[l-l'|}^{l+l'} \sum_{s=-j}^{j} 
\frac{ i^j }{ q^{j+1} } Z_{js}(q^2,\eta) C_{ lm,js,l'm'},
\label{eq:mmat}
\eeq
where we have adapted it to include $z$ elongated box geometry via $\eta$\footnote{Although we treat the cubic and elongated geometries jointly via the elongation factor $\eta$. (setting $\eta=1$ for cubic),
they are handled differently by their group symmetries.}
The zeta function is defined by
\beq
\mathcal{Z}_{lm} (q^2,\eta) = \sum_{\widetilde{\bm n}} \frac{\mathcal{Y}_{lm}(\widetilde{\bm n})}{\widetilde{\bm n}^2-q^2},
\label{eq:zfun}
\eeq
where the summation index $\widetilde{\bm n}$ and the dimensionless $q$ are defined as 
\beq
\widetilde{\bm n}= (n_x,n_y,n_z/\eta), \quad q=\frac{kL}{2\pi}.
\label{eq:psum}
\eeq
We see that the zeta function is a pure, dimensionless mathematical function with dimensionless variables.
The same is true for the matrix $M$ that appears in the quantization conditions.
The poles of the zeta function $\tilde{n}^2=q^2$ correspond to free-particle energies in the box. Deviations from the poles due to interactions are connected to phase shifts.
It is customary to introduce the shorthand notation,\footnote{Another convention in the literature has the factor $\sqrt{2l+1}$ dividing this expression.}
\beq
w_{lm}(q^2,\eta) \equiv  \frac{\mathcal{Z}_{lm}(q^2,\eta)}{\eta \pi^{3/2} q^{l+1}},
\label{eq:wfun}
\eeq
so $M$ can be expressed as a linear combination of $w_{lm}$ with purely numerical coefficients, and the simplest phase shift formula from the QC reads $\cot\delta=w_{00}$. 
The $M$ matrix plays a central role in the methodology and will be discussed extensively below.

\subsection{Symmetry-adapted quantization conditions}
\label{sec:block}
The quantization condition in Eq.\eqref{eq:phaselat} must be adapted to the symmetry under consideration.  The issue arises because symmetries in the infinite volume are reduced to the symmetries in the box.  For example, rotational symmetry
is no longer continuous, but is reduced to a limited number of possibilities. 
We start by writing Eq.\eqref{eq:phaselat} as
\beq
\det \left [e^{2i\delta} (M-i) - (M+i) \right ] =0\,,
\eeq
after dropping the nonzero $\det[M-i]$. 
This is another form of the QC commonly in use.
It can be rearranged into
\beq
\det \left [(e^{2i\delta}-1) \left( M- i\, \frac{e^{2i\delta}+1 }{ e^{2i\delta}-1}\right )  \right ] =0.
\eeq
Dropping the constant factor, we can write the QC in the form using matrix elements,
\beq
\det  [ M_{lm,l'm'}- \delta_{ll'}  \delta_{mm'} \cot \delta_{l} ] =0.
\label{eq:phaselat2}
\eeq
The goal is to further reduce the matrix according to the irreducible representations (irreps) of the  symmetry group.
Operationally, it is equivalent to reducing the matrix in the QC into its block diagonal form with each block representing an irrep. Then the QC is a product of the determinant of the blocks. This is achieved by a change of basis, using the basis vectors of the symmetry group,
expressed as
\beq
|\Gamma \alpha l n\rangle = \sum_m C^{\Gamma \alpha n}_{lm} | lm  \rangle,
\eeq
where $\Gamma$ stands for a given irrep of the group and $\alpha$ runs from 1 to the dimension of the irrep, 
$n$ runs from 1 to $n(\Gamma,l)$, the multiplicity of $l$ in irrep $\Gamma$. The coefficients are discussed in Appendix~\ref{sec:basis}.
In the new basis,  $M$ is block-diagonalized by irreps
\beqs
&\langle \Gamma \alpha l n |M| \Gamma' \alpha' l' n'  \rangle   \\
&=
\sum_{mm'}  \left (C^{\Gamma \alpha n}_{lm}\right )^*  C^{\Gamma' \alpha' n'}_{l' m'}  M_{lm,l'm' } \\
& = \delta_{\Gamma \Gamma'} \delta_{\alpha \alpha'}  M^\Gamma_{l n,l' n' },
\label{eq:ME}
 \eeqs
where the orthogonality relation for irreducible representations (Schur's lemma)  is used in the last step. For multidimensional irreps, 
we average over the components $\alpha$ since they are not observables.
The final form for the symmetry-adapted QC is
\beq
\prod_\Gamma  \det  \left[ M^\Gamma_{ln,l' n' }- \delta_{ll'}  \delta_{nn'} \cot \delta_{l} \right ]=0.
\quad (\text{\bf QC1}) 
\label{eq:QC1}
\eeq
The QC can now be investigated irrep by irrep. Since total  angular momentum is the same as orbital  angular momentum  ($J=l$) for spinless particles, we keep the notation simple by using only $l$.  For particles with spin, one needs to keep $J$ and $l$ separate for basis vectors $|\Gamma \alpha J l n\rangle$, matrix elements $M^\Gamma_{Jln,Jl' n' }$, and phase shifts $\delta_{Jl}$.
The corresponding symmetry-adapted version of the original QC in Eq.\eqref{eq:phaselat} can be written as a matrix equation  for each irrep,
\beqs
 &\det \left [S - U \right ]=0,  \quad(\text{\bf QC2}) \\
&S  =\text{diag}\{e^{2i\delta_{l}(k)}\}, \,
U = \frac{M^\Gamma(k,L) +i}{ M^\Gamma(k,L) -i }.
\label{eq:QC2}
\eeqs
We will refer to Eq.\eqref{eq:QC1} as QC1 and  Eq.\eqref{eq:QC2} as QC2 as already indicated.
They have the same solutions, but different features.
The determinant in QC1 is real valued and is unbounded due to singularities (free-particle poles). The one in QC2 is complex valued and bounded.
The $M-i$ in the denominator in QC2 removes the noninteracting divergences while leaving the zeroes of the determinant unchanged. 
Both QCs will be employed in this work.

In the following, we present the matrix elements defined in Eq.\eqref{eq:ME} 
for four different total momenta:  rest frame and three moving frames, in both cubic and elongated boxes.
Some already exist in the literature, but we find it necessary to extend to higher partial waves.
We need up to five partial waves in each irrep, depending on its angular momentum content.
So we decide to take a fresh look and set out to derive all the QCs studied in this work. Some are rederived, some are new.

\subsection{Rest and moving frames} 
In group theory language, the symmetry group for states at rest is $O_h$ in cubic box,  $D_{4h}$ in $z$ elongated box.
For moving states, the symmetry is described by the so-called little groups, depending in which direction the system is moving in the fixed box. 
Table~\ref{tab:boostAll} summarizes all the possibilities. Only the lowest distinct momenta are given, which should be sufficient in most applications. The momenta is given in units of lowest nonzero momentum allowed on periodic lattices $\bm P = (2\pi/L)\bm d $. Note that this is the lowest momentum in the traverse directions, when considering elongated boxes. If needed, then one can go higher by following the rules in the table. 
We will consider four distinct moving frames,  $d=(0,0,1)$,  $d=(1,1,0)$, $d=(1,1,1)$, and  $d=(0,1,2)$. They correspond to the lowest momentum square norms  $|d|^2=1,2,3,5$ [$|d|^2=4$ is a multiple of $d=(0,0,1)$].
In both cubic and $z$ elongated boxes,  $d=(0,0,1)$ has $C_{4v}$ as the little group, $d=(1,1,0)$ corresponds to $C_{2v}$, and $d=(0,1,2)$ corresponds to $C_{1v}$.
However, for $d=(1,1,1)$, the little group is $C_{3v}$ in cubic box and $C_{1v}$ in $z$ elongated box.

\begin{table}[b]
\caption{The little groups for moving states in cubic and elongated boxes. The lowest distinct patterns for the boost vector $\bm d=(n_x,n_y,n_z)$ are shown, but integer multiples $n\bm d$ with $n \in Z^3$ have the same little groups. Furthermore, the momenta related via a lattice symmetry with the ones below have the same little groups (this means all permutations for cubic, and permutations in $n_x$ and $n_y$ for $z$ elongated). 
Additionally, all momenta for $C_{1v}$ in the $z$ elongated case can have $n_z$ increase independently.
}
\label{tab:boostAll}          
              
$                                
\renewcommand{\arraystretch}{1.3}
\begin{array}{c | c | c }
\toprule                         
 & \text{cubic  } (O_h) & \text{$z$ elongated  }  (D_{4h})      \\
\hline                           
 C_{4v} & (0,0,1)  & (0,0,1)  \\
 \hline      
 C_{3v} & (1,1,1)  &    \text{none}  \\
 \hline      
 C_{2v} & (1,1,0)  &  (1,0,0), (1,1,0)     \\
 \hline      
 C_{1v} & (0,1,2), (1,1,2), (1,2,2)  &  (0,1,1), (1,1,1), \\ 
 && (1,2,0), (0,1,2), (0,2,1),\\
 &&  (1,2,1), (2,2,1) \\
\bottomrule                      
\end{array}                      
$                                
                   
\end{table}                     

The derivations for the matrix elements $M^\Gamma_{l n,l' n' }$ in the QCs involve extensive use of group theory. 
To improve readability, we highlight some important consequences from  symmetries here and 
relegate the details to Appendix~\ref{sec:group}. 
All the tables for $M^\Gamma_{l n,l' n' }$ are placed in Appendix~\ref{sec:ME}.
To eliminate typos, we construct the tables in {\it Mathematica} and copy and paste in LaTeX format.
The same expressions in the tables are also used directly in the numerical tests to be discussed later.

In Table~\ref{tab:all}, we give an overview of the total angular momentum content in each irrep 
(or QC), as part of a larger summary. 
It is important to realize that each QC is a single condition that couples to an infinite tower of $l$ values; 
only the lowest few are shown.
The lowest partial wave in each irrep can be computed using the energy levels in the box and if the higher partial waves can be neglected.~\footnote{We will refer to the lowest-order QC as the ``L\"uscher formula", and the general QC as the ``L\"uscher method".} In this sense, the decomposition of angular momentum  can be regarded as a blessing in disguise: it provides means to predict individual partial waves via the L\"uscher formula by picking irreps and dialing the box geometry.
In the table for each geometry going from top to bottom the symmetry is reduced which leads to more and more mixing of partial waves.
For example, the gap between the two lowest $l$ in $A_{1g}$ or $A_1$ is 4 in $O_h$, 2 in $D_{4h}$, 1 in $C_{nv}$.
There are additional indicators of mixing: appearance of multiplicities, loss of parity, and lower starting values of $l$.

In evaluating the matrix elements, a lot of the $w_{lm}$ functions vanish or satisfy certain constraints due to symmetry present in the system. 
This can be traced back to how the spherical harmonics
behave under the group operations.
The following properties apply to both cubic and $z$ elongated boxes in the rest frame.
\begin{enumerate}[(i)]
\item 
The standard property  $Y_{l-m}=(-1)^m Y^*_{lm}$ translates directly to 
$w_{l-m}=(-1)^m w^*_{lm}$. This holds in general.
\item 
The system is invariant under a mirror reflection about the $xy$ plane.
It leads to $Y_{lm}(\theta,\phi)=Y_{lm}(\pi-\theta,\phi)=(-1)^{(l-m)} Y_{lm}(\theta,\phi)$, which means 
$w_{lm}=0 \text{  for  } l-m = \text{odd.  In particular  } w_{l0}=0  \text{  for  } l = 1,3,5,\cdots.  $
This is valid for all systems with inversion symmetry, which leads 
to a separation into sectors by parity.
\item 
The system is invariant under a $\pi/2$ rotation about the $z$ axis which leads to the constraint $e^{im\pi/2}=1$ due to the $e^{im\phi}$ dependence in $Y_{lm}$.
This means 
$ w_{lm}=0 \text{  for  }  m \neq  0, 4, 8, \cdots, \text{ regardless of } l$.
\item 
The system is invariant under a mirror reflection about the $xz$ plane,
which leads to 
$Y_{lm}(\theta,\phi)=Y_{lm}(\theta,2\pi-\phi)= Y^*_{lm}(\theta,\phi)$.
This means all the $w_{lm}$ functions are real valued. However, the matrix elements $M$ can have complex valued coefficients depending on basis vectors.
\end{enumerate}
 We take advantage of these properties to simplify the presentation of the QCs.
 We use the minimum number of nonzero elements.
 Moreover, the $M$ matrix is Hermitian so we only list the upper triangular part of the matrix.

In comparing with literature one needs to pay attention to different notations and conventions.
A feature of a generic QC is that it is invariant under a change of basis (similarity transform),
\beqs
& | BM_{ll'}B^{-1}-\delta_{ll'} \cot\delta | \\& = |B M_{ll'}B^{-1} -   B \delta_{ll'}  B^{-1} \cot\delta| \\
& =|B(M_{ll'} - \delta_{ll'} \cot\delta)  B^{-1}| \\& =|B| \,|M_{ll'}-\delta_{ll'} \cot\delta| \,|B^{-1}|
 = |M_{ll'}-\delta_{ll'} \cot\delta|.
\eeqs
So the same QC can take different analytical forms depending on  the basis vectors and coordinate systems used, but the physics content is the same. Numerically they should produce the same determinant.


So far the discussion is limited to systems that are at rest; 
the two particles have back-to-back nonzero momentum, but the total momentum $\bm P=0$. 
The total energy is the same in both the lab and CM frames.  
Now we consider moving frames, that is, systems with nonzero $\bm P$ in the lab frame,
\beq 
\bm P=\bm p_1 + \bm p_2. 
\label{eq:kinP}
\eeq
In relativistic kinematics, moving frames are also known as Lorentz boosting. In nonrelativistic kinematics we only have ``Galilean boosting", thus no length contraction in the direction of motion and no mixing of energy and momentum in the transformation. For this reason, we will remove all references to the relativistic factor $\gamma$ (or set $\gamma=1$  in practice). Although the current formalism uses $\gamma=1$ for our purposes, we note that $\gamma$ is significant larger than 1 in the majority of lattice QCD calculations that employ moving frames.  We should point out that aside from kinematics, all other ingredients of the L\"uscher method remain the same.   
The momenta are quantized in the box.
We use the notation for lab momenta, 
\beqs 
\bm P&= \frac{2\pi }{ L} \widetilde{\bm d}= \frac{2\pi }{ L} (d_x, d_y, d_z/\eta)\\
\bm p_{1}&= \frac{2\pi }{ L}\widetilde{\bm n}_1= \frac{2\pi }{ L} (n_{1x}, n_{1y}, n_{1z}/\eta),  \\
\bm p_{2}&= \frac{2\pi }{ L}\widetilde{\bm n}_2= \frac{2\pi }{ L} (n_{2x}, n_{2y}, n_{2z}/\eta), \\
\bm d= (d_x, d_y, d_z)&= (n_{1x}, n_{1y}, n_{1z})+(n_{2x}, n_{2y}, n_{2z})
\label{eq:momenta}
\eeqs
where we distinguish the input vector $ \widetilde{\bm d}$ from the summation vectors $\widetilde{\bm n}$.
The energy of the system in the lab frame is given by
\beq
 E_{\rm lab}=\frac{p^2_1}{ 2 m_1}+\frac{p^2_2}{ 2 m_2}.
 \label{eq:elab}
 \eeq
In the CM frame, the energy is given by 
\beq
 E_{\rm cm}=\frac{k^2}{ 2 \tilde m},
  \label{eq:ecm} 
 \eeq
 where $\tilde m$ is the reduced mass and $\bm k$ is the relative CM momentum.
The advantage of moving frames is that it can lower the center-of-mass energy,
\beq
E_{\rm cm}=E_{\rm lab}-\frac{P^2}{2(m_1+m2)},
\label{eq:ecm2}
\eeq
thus providing a dial for wider energy coverage.
The procedure to extract infinite-volume phase shift is to first measure the interaction energy $E_{lab}$ in the box, then determine $k$ via kinematics [Eq.\eqref{eq:ecm} and Eq.\eqref{eq:elab}], then $\delta(k)$ via the QC.

To find out how  $\bm k$ is quantized in terms of lab momenta, we perform Galilean transformations
\beq
\bm p_1 = \bm k + \frac{m_1}{m_1+ m_2}\bm P, \quad \bm p_2 = - \bm k + \frac{m_2}{m_1+ m_2}\bm P, 
\label{eq:kin}
\eeq
where we assume particle 1 has $\bm k$ with the same sign as $\bm P$. 
Solving for $\bm k$, we have
\beqs
\bm k  & = \frac{1}{2} (\bm p_1 -\bm p_2) + \frac{m_1-m_2}{ 2(m_1+ m_2)}\bm P \\
&  =\bm p_1 - \frac{1}{2} \bm P +  \frac{m_1-m_2}{ 2(m_1+ m_2)}\bm P \\
& =  \frac{2\pi}{ L} \left (\widetilde{\bm n} - \frac{1}{2} A \widetilde{\bm d} \right )
\eeqs
where in the last step we have inserted the box momenta and defined the factor 
\beq
A\equiv 1+ \frac{m_2 - m_1 }{m_2 + m_1}.
\eeq
This is to be contrasted with the relativistic version 
$A=1+ (m_1^2-m_2^2) / W^2$ where $W=\sqrt{m_1^2 + k^2} + \sqrt{m_2^2 + k^2}$ is the invariant energy of the system.
Note that if we assume the other possibility (particle 2 has the same sign as $\bm P$)  in Eq.\eqref{eq:kin},  the order of $m_1$ and $m_2$ in $A$ is switched, but it does not affect the quantization condition as we will see below. The system is symmetric about $A=1$.

Another effect of moving frames is that the periodic boundary condition in Eq.\eqref{eq:bccm} now picks up a complex phase factor~\cite{Rummukainen:1995vs,Fu:2011xz, Leskovec:2012gb,Gockeler:2012yj}:
\beq
\psi(\bm r+\bm n L) = e^{i\pi A \widetilde{\bm n}\cdot  \widetilde{\bm d}} \psi(\bm r) \,,
\label{eq:bbc}
\eeq
also known as $\bm d$-periodic boundary condition. The vector $\bm n$ in this equation should be understood as $\bm n=(n_{x}, n_{y}, n_{z}\eta)$.
The condition depends on the boost $\bm d$, as well as the particle masses $m_1$ and $m_2$ via the factor $A$.
This condition is not easy to implement if we work in the CM frame. 
By working in the lab frame, standard periodic boundary conditions can be applied. 

The new condition is not invariant under parity, so the solutions of the Helmholtz equation are a mixture 
of both parities. 
For spinless systems, the irreps now overlap with both even and odd angular momenta $l$, not just the even or odd separately.

Boosting of spinless system in cubic box has been considered in a well-known study in Ref.~\cite{Rummukainen:1995vs} and later extended to unequal masses in Ref.~\cite{Fu:2011xz}.
Boosting of spin-1/2 system in cubic box has been considered in Refs.~\cite{Leskovec:2012gb,Gockeler:2012yj}.
In this work, we reexamine the QCs for cubic boxes and derive new ones for elongated boxes.

For moving frames, the zeta functions in Eq.\eqref{eq:zfun} need to be modified to include the boost $\bm d$,
\beq
\mathcal{Z}_{lm}(q^2,\bm d,\eta) = \sum_{ \widetilde{\bm n}\in P_{\bm d}(\eta)} \frac{\mathcal{Y}_{lm}(\widetilde{\bm n})}{\widetilde{\bm n}^2-q^2}.
\label{eq:zfun_boost_eta}
\eeq
The summation grid changes to
\beq
P_{\bm d}(\eta) =\left\{\widetilde{\bm n}\in\reals^3 \mid \widetilde{\bm n}=\hat{\eta}^{-1}(\bm m-\frac{1}{2}A\, \bm d), \bm m\in \mathds{Z}^3 \right\},
\label{eq:psum_boost_eta}
\eeq
with the projector $\hat{\eta}^{-1}$ acting on a vector $\bm m$
 to mean $\hat{\eta}^{-1} \bm m =(m_x,m_y,m_z/\eta)$.
The evaluation of zeta functions has been described, for example, in Refs.~\cite{Leskovec:2012gb,Gockeler:2012yj,Feng:2004ua,Guo:2016zos}. We implemented a high-precision version that can handle both asymmetric geometry and general moving frames.
Because moving frames single out special directions in space, symmetry in the system is reduced. This is reflected in the proliferation of nonzero matrix elements in the QCs, as summarized in Table~\ref{tab:allW}.
Due to lack of parity in moving frames, there is mixing between odd and even $l$ states within a given irrep. 
This means that the phase shift formulas are generally more complicated for moving states than for the ones at rest.
One consequence is the appearance of zeta functions with odd values of $l$.
\begin{table*}
\caption{Nonzero zeta functions for  rest frame (up to $l=10$),  moving frames (up to $l=8$),  unequal masses,  and $z$ elongated box. 
In the case of cubic box, $\text{w}_{20}$ and $\text{w}_{52}$ vanish for $d= (0,0,0) $ and  $d= (1,1,1) $.
In the case of equal masses, all odd-$l$ functions vanish.
We use the notation $\text{w}_{lm}\equiv \text{wr}_{lm} + I \text{wi}_{lm}$. If the function is real, then it is simply represented by $\text{w}_{lm}$ instead of $\text{wr}_{lm}$; if purely imaginary, then by $\text{wi}_{lm}$. If the function has both nonzero real and imaginary parts, then expressions such as $\text{wi}_{31}\to \text{wr}_{31}$ or $\text{wi}_{33}\to -\text{wr}_{33}$ mean that they have equal magnitude but may differ by a sign, and the function will be represented by its real part. We did not encounter the case where nonzero real and imaginary parts have different magnitudes. All matrix elements of the QCs in Appendix~\ref{sec:ME} are expressed as linear combination of these functions with purely numerical coefficients.}
\label{tab:allW}          
              
$                                
\renewcommand{\arraystretch}{1.2}
\begin{array}{c c  c }
\toprule                         
d & l &  \text{w}_{lm}   \\
\hline           
 \{0,0,0\} & 0 & \left\{\text{w}_{00}\right\} \\
   & 2 & \left\{\text{w}_{20}\right\} \\
   & 4 & \left\{\text{w}_{40},\text{w}_{44}\right\} \\
   & 6 & \left\{\text{w}_{60},\text{w}_{64}\right\} \\
   & 8 & \left\{\text{w}_{80},\text{w}_{84},\text{w}_{88}\right\} \\
   & 10 & \left\{\text{w}_{100},\text{w}_{104},\text{w}_{108}\right\} \\
\hline                           
 \{0,0,1\} & 0 & \left\{\text{w}_{00}\right\} \\
   & 1 & \left\{\text{w}_{10}\right\} \\
   & 2 & \left\{\text{w}_{20}\right\} \\
   & 3 & \left\{\text{w}_{30}\right\} \\
   & 4 & \left\{\text{w}_{40},\text{w}_{44}\right\} \\
   & 5 & \left\{\text{w}_{50},\text{w}_{54}\right\} \\
   & 6 & \left\{\text{w}_{60},\text{w}_{64}\right\} \\
   & 7 & \left\{\text{w}_{70},\text{w}_{74}\right\} \\
   & 8 & \left\{\text{w}_{80},\text{w}_{84},\text{w}_{88}\right\} \\
\hline    
 \{1,1,0\} & 0 & \left\{\text{w}_{00}\right\} \\
   & 1 & \left\{\text{wi}_{11}\to \text{wr}_{11}\right\} \\
   & 2 & \left\{\text{w}_{20},\text{wi}_{22}\right\} \\
   & 3 & \left\{\text{wi}_{31}\to \text{wr}_{31},\text{wi}_{33}\to -\text{wr}_{33}\right\} \\
   & 4 & \left\{\text{w}_{40},\text{wi}_{42},\text{w}_{44}\right\} \\
   & 5 & \left\{\text{wi}_{51}\to \text{wr}_{51},\text{wi}_{53}\to -\text{wr}_{53},\text{wi}_{55}\to \text{wr}_{55}\right\} \\
   & 6 & \left\{\text{w}_{60},\text{wi}_{62},\text{w}_{64},\text{wi}_{66}\right\} \\
   & 7 & \left\{\text{wi}_{71}\to \text{wr}_{71},\text{wi}_{73}\to -\text{wr}_{73},\text{wi}_{75}\to \text{wr}_{75},\text{wi}_{77}\to
   -\text{wr}_{77}\right\} \\
   & 8 & \left\{\text{w}_{80},\text{wi}_{82},\text{w}_{84},\text{wi}_{86},\text{w}_{88}\right\} \\

\hline
 \{1,1,1\} & 0 & \left\{\text{w}_{00}\right\} \\
   & 1 & \left\{\text{w}_{10},\text{wi}_{11}\to \text{wr}_{11}\right\} \\
   & 2 & \left\{\text{w}_{20},\text{wi}_{21}\to \text{wr}_{21},\text{wi}_{22}\right\} \\
   & 3 & \left\{\text{w}_{30},\text{wi}_{31}\to \text{wr}_{31},\text{wi}_{32},\text{wi}_{33}\to -\text{wr}_{33}\right\} \\
   & 4 & \left\{\text{w}_{40},\text{wi}_{41}\to \text{wr}_{41},\text{wi}_{42},\text{wi}_{43}\to -\text{wr}_{43},\text{w}_{44}\right\}
   \\
   & 5 & \left\{\text{w}_{50},\text{wi}_{51}\to \text{wr}_{51},\text{wi}_{52},\text{wi}_{53}\to
   -\text{wr}_{53},\text{w}_{54},\text{wi}_{55}\to \text{wr}_{55}\right\} \\
   & 6 & \left\{\text{w}_{60},\text{wi}_{61}\to \text{wr}_{61},\text{wi}_{62},\text{wi}_{63}\to
   -\text{wr}_{63},\text{w}_{64},\text{wi}_{65}\to \text{wr}_{65},\text{wi}_{66}\right\} \\
   & 7 & \left\{\text{w}_{70},\text{wi}_{71}\to \text{wr}_{71},\text{wi}_{72},\text{wi}_{73}\to
   -\text{wr}_{73},\text{w}_{74},\text{wi}_{75}\to \text{wr}_{75},\text{wi}_{76},\text{wi}_{77}\to -\text{wr}_{77}\right\} \\
   & 8 & \left\{\text{w}_{80},\text{wi}_{81}\to \text{wr}_{81},\text{wi}_{82},\text{wi}_{83}\to
   -\text{wr}_{83},\text{w}_{84},\text{wi}_{85}\to \text{wr}_{85},\text{wi}_{86},\text{wi}_{87}\to
   -\text{wr}_{87},\text{w}_{88}\right\} \\

 \hline
 \{0,1,2\} & 0 & \left\{\text{w}_{00}\right\} \\
   & 1 & \left\{\text{w}_{10},\text{wi}_{11}\right\} \\
   & 2 & \left\{\text{w}_{20},\text{wi}_{21},\text{w}_{22}\right\} \\
   & 3 & \left\{\text{w}_{30},\text{wi}_{31},\text{w}_{32},\text{wi}_{33}\right\} \\
   & 4 & \left\{\text{w}_{40},\text{wi}_{41},\text{w}_{42},\text{wi}_{43},\text{w}_{44}\right\} \\
   & 5 & \left\{\text{w}_{50},\text{wi}_{51},\text{w}_{52},\text{wi}_{53},\text{w}_{54},\text{wi}_{55}\right\} \\
   & 6 & \left\{\text{w}_{60},\text{wi}_{61},\text{w}_{62},\text{wi}_{63},\text{w}_{64},\text{wi}_{65},\text{w}_{66}\right\} \\
   & 7 &
   \left\{\text{w}_{70},\text{wi}_{71},\text{w}_{72},\text{wi}_{73},\text{w}_{74},\text{wi}_{75},\text{w}_{76},\text{wi}_{77}\right\}
   \\
   & 8 &
   \left\{\text{w}_{80},\text{wi}_{81},\text{w}_{82},\text{wi}_{83},\text{w}_{84},\text{wi}_{85},\text{w}_{86},\text{wi}_{87},\text{w}_
   {88}\right\} \\
\bottomrule                      
\end{array}                      
$                                
                    
\end{table*}                     

Further simplification is possible from the closure relation on the zeta functions~\cite{Luscher:1990ux},
\beq
\sum_{m'=-l}^l D^{(l)}_{m'm}({\alpha,\beta,\gamma}) Z_{lm'}=Z_{lm},
\label{eq:closure}
\eeq
where $D$ is the Wigner rotation matrix for a transformation in the little group. This has its origin in the property of spherical harmonics under discrete rotations, 
and holds for all $\bm d$ possibilities and box geometries. A constraint among the zeta functions can be obtained from each group element; not all constraints are independent. 
These introduce further relations between $\text{w}_{lm}$ with the same $l$. We list these relationships in Table~\ref{tab:addW} for
$d=(0,0,0)$ and $d=(1,1,1)$ in cubic box.
They are used to further simplify the matrix elements in the two cases. For all other cases, no new relations are generated by these constraints.
\begin{table}
\caption{Additional relations derived from the closure condition in Eq.\eqref{eq:closure} 
for  rest frame $d= \{0,0,0\} $ and  moving frame $d= \{1,1,1\} $ in cubic box. They are used to further simply the matrix elements in the two cases.}
\label{tab:addW}          
              
$                                
\renewcommand{\arraystretch}{1.4}
\begin{array}{cc}
\toprule                         
 & d= \{0,0,0\}, \text{ group } O_h   \\
\hline           
 \text{} & \text{w}_{44}\to \sqrt{\frac{5}{14}} \text{w}_{40} \\
 \text{} & \text{w}_{64}\to -\sqrt{\frac{7}{2}} \text{w}_{60} \\
 \text{} & \text{w}_{84}\to \frac{1}{3} \sqrt{\frac{14}{11}} \text{w}_{80} \\
 \text{} & \text{w}_{88}\to \frac{1}{3} \sqrt{\frac{65}{22}} \text{w}_{80} \\
 \text{} & \text{w}_{104}\to -\sqrt{\frac{66}{65}} \text{w}_{100} \\
 \text{} & \text{w}_{108}\to -\sqrt{\frac{187}{130}} \text{w}_{100} \\
\hline     
     &  d= \{1,1,1\}, \text{ group } C_{3v}   \\      
  \hline                  
 \text{} & \text{wr}_{11}\to -\frac{\text{w}_{10}}{\sqrt{2}} \\
 \text{} & \text{wi}_{22}\to -\text{wr}_{21} \\
 \text{} & \text{wr}_{31}\to \frac{\sqrt{3} \text{w}_{30}}{4} \\
 \text{} & \text{wr}_{33}\to -\frac{1}{4} \sqrt{5} \text{w}_{30} \\
 \text{} & \text{wi}_{42}\to 2 \sqrt{2} \text{wr}_{41} \\
 \text{} & \text{wr}_{43}\to \sqrt{7} \text{wr}_{41} \\
 \text{} & \text{w}_{44}\to \sqrt{\frac{5}{14}} \text{w}_{40} \\
 \text{} & \text{wr}_{53}\to \sqrt{\frac{5}{7}} \text{w}_{50}+3 \sqrt{\frac{3}{14}} \text{wr}_{51} \\
 \text{} & \text{w}_{54}\to -\sqrt{\frac{5}{14}} \text{w}_{50}-\frac{8 \text{wr}_{51}}{\sqrt{21}} \\
 \text{} & \text{wr}_{55}\to \sqrt{\frac{5}{42}} \text{wr}_{51}-\frac{\text{w}_{50}}{\sqrt{7}} \\
 \text{} & \text{wi}_{62}\to \frac{1}{4} \left(-\sqrt{10} \text{wr}_{61}-6 \text{wr}_{63}\right) \\
 \text{} & \text{wi}_{66}\to \frac{1}{44} \left(2 \sqrt{55} \text{wr}_{63}-9 \sqrt{22} \text{wr}_{61}\right) \\
 \text{} & \text{w}_{64}\to -\sqrt{\frac{7}{2}} \text{w}_{60} \\
 \text{} & \text{wr}_{65}\to \sqrt{\frac{6}{11}} \text{wr}_{61}+\sqrt{\frac{15}{11}} \text{wr}_{63} \\
 \text{} & \text{wi}_{76}\to \frac{363 \text{wi}_{72}-32 i \left(\sqrt{21} \text{w}_{70}-\sqrt{6} \text{wr}_{71}+9
   \sqrt{2} \text{wr}_{73}\right)}{33 \sqrt{143}} \\
 \text{} & \text{w}_{74}\to -\frac{25 \sqrt{42} \text{w}_{70}-248 \sqrt{3} \text{wr}_{71}+120 \text{wr}_{73}}{66
   \sqrt{11}} \\
 \text{} & \text{wr}_{75}\to \frac{14 \sqrt{42} \text{w}_{70}-94 \sqrt{3} \text{wr}_{71}+21 \text{wr}_{73}}{33
   \sqrt{11}} \\
 \text{} & \text{wr}_{77}\to -\frac{1}{11} \sqrt{\frac{13}{11}} \left(2 \sqrt{6} \text{w}_{70}+\sqrt{21}
   \text{wr}_{71}+2 \sqrt{7} \text{wr}_{73}\right) \\
 \text{} & \text{wi}_{82}\to \frac{1}{2} \left(\sqrt{66} \text{wr}_{83}-\sqrt{70} \text{wr}_{81}\right) \\
 \text{} & \text{wi}_{86}\to 3 \sqrt{\frac{33}{26}} \text{wr}_{81}-\sqrt{\frac{35}{26}} \text{wr}_{83} \\
 \text{} & \text{w}_{84}\to \frac{1}{3} \sqrt{\frac{14}{11}} \text{w}_{80} \\
 \text{} & \text{wr}_{85}\to 3 \sqrt{\frac{15}{13}} \text{wr}_{83}-2 \sqrt{\frac{77}{13}} \text{wr}_{81} \\
 \text{} & \text{wr}_{87}\to 2 \sqrt{\frac{21}{13}} \text{wr}_{83}-\sqrt{\frac{55}{13}} \text{wr}_{81} \\
 \text{} & \text{w}_{88}\to \frac{1}{3} \sqrt{\frac{65}{22}} \text{w}_{80} \\
\bottomrule                      
\end{array}                      
$                                
                    
\end{table}                     

For unequal masses, what happens if the two masses are switched? 
This question can be answered by examining the mass dependence in the zeta function 
in Eq.\eqref{eq:zfun_boost_eta}. 
Interchanging $m_1$ and $m_2$ only affects the $A$ factor in the summation grid in  Eq.\eqref{eq:psum_boost_eta}. 
The result is a change in sign of the set of points to be summed over from $\widetilde{\bm n}$ to $- \widetilde{\bm n}$ (the mirror image grid).  
It leads to an overall sign change in the zeta function, which does not affect the QC. 
So the order of $m_1$ and $m_2$ does not matter as far as QC is concerned. 
However, the order matters in terms of total energy of the system: when the higher-mass particle carries more momentum than the lower-mass particle, the system has lower total energy. Examples will be given when we discuss energy spectrum in the box.


\section{Two-particle energies in a periodic box}
\label{sec:scheq}

To check our derivation of the quantization conditions discussed in the previous
section, we want to calculate the spectrum of the two-particle states
in a finite box with periodic boundary conditions. To make the calculation
transparent we will use a nonrelativistic setup with the particles' 
interaction controlled by a rotationally invariant potential. We will
solve the problem numerically using a lattice discretization of the 
Hamiltonian and the associated Schr\"odinger equation. The results
are extrapolated to the continuum limit before comparing them to
the results of the quantization conditions.

\subsection{Lattice Hamiltonian} 

We want to obtain the energy spectrum in a finite box in order to examine the quantization conditions. We achieve this by using discretized lattices of finite lattice spacing $a$, then extrapolating to $a\to 0$ while keeping the physical box size fixed.
Consider the general case $L\times L\times \eta L$ where $\eta$ is the elongation factor in the $z$ direction.
We want to solve the Schr\"{o}dinger equation $H\Psi=E\Psi$ in the box frame (lab frame) with periodic boundary conditions. 
The Hamiltonian of the system is
\beq
H=-\frac{\hbar^2 }{ 2m_1} \nabla_1^2  - \frac{\hbar^2 }{ 2m_2} \nabla_2^2 + V_L(|\bm r_1 - \bm r_2|).
\label{eq:latH}
\eeq
Here $V_L$ is periodic version of the infinite-volume potential $V$,
\beq
V_L(|\bm r_1 - \bm r_2|) = \sum_{\bm n_1, \bm n_2} V(| \bm r_1 +\bm n_1 L- \bm r_2 -\bm n_2 L |).
\label{eq:pot-box}
\eeq
Visually, the continuous space gets tiled into an infinite number of $\eta L^3$ boxes in which the potential is replicated. 
Under this scenario, the potential is no longer rotationally symmetric. Instead, it takes on the symmetry of the box.
The wave functions satisfy periodic boundary conditions
\beq 
\Psi(\bm r_1 +\bm n_1 L, \bm r_2 +\bm n_2 L) =\Psi(\bm r_1, \bm r_2),
\label{eq:bc}
\eeq
where $\bm n_1=(n_{1x}, n_{1y}, n_{1z} \eta)$ and $\bm n_2=(n_{2x}, n_{2y}, n_{2z} \eta)$.
Numerically, the problem can be solved by discretizing the box into a lattice of $N_xN_yN_z$ grid points and an isotropic spacing $a$, so the physical volume is $N_xN_yN_z  a^3$. For the elongated geometry along the $z$ axis we have $N_x=N_y=N$ and $N_z=\eta N$. The Laplacian can be approximated by finite differences on the lattice. However, the dimension of the Hilbert space for the two-particle states grows with $N^6$ and finding the relevant eigenvalues for this Hamiltonian is only practical for very small lattices. 
We seek a method that can reduce it to a $N^3$ problem.
The traditional approach is to separate the problem into the motion of the center of mass plus the relative motion in the CM frame with a reduced mass. The CM motion is constant and is largely decoupled; its presence is only felt in the relative motion through kinematics. Moreover, the periodic boundary condition in the lab frame is so modified in the CM frame that depends on the total CM momentum and the masses of the two particles, as seen in Eq.\eqref{eq:bbc}. The separation of the center-of-mass
motion has to be done carefully since we intend to use the same formalism both
for two-particle states at rest and for the moving case.
We want a formalism that is inducive to the study of moving frames in a natural fashion.
To this end, we project the problem to a new basis consisting of total momentum $\bm P$ and relative coordinates $\bm r$ in the lab frame,
\beq 
\ket{\bm P,\bm r}=\sum_{\bm m} e^{i \bm P \cdot \bm m} \ket{\bm m, \bm m+ \bm r},
\label{eq:Pr}
\eeq
where $\ket{\bm n_1, \bm n_2}$ is the ket in the position representation for two particles.

\begin{widetext} 
In Cartesian coordinates, using a three-point stencil, the Laplacian operator is
\beq
\nabla^2 \psi(\bm r) = \sum_{\mu=1}^3 \frac{\psi(\bm r+a\hat \mu)+\psi(\bm r-a\hat \mu)-2\psi(\bm r)}{a^2} + \mathcal{O}(a^2) \,,
\eeq
and the projection leads to the reduced problem $H\psi(\bm P,\bm r)=E\psi(\bm P,\bm r)$, where the lattice Hamiltonian is given by
\beqs
H\ket{\bm P, \bm r}  =&-\frac{\hbar^2}{ 2} \sum_{\mu} \frac{-1}{ 2 a^2}   
\left[ 
\left( \frac{e^{i P_\mu a} }{ m_1} + \frac{1}{m_2} \right) \ket{\bm P,\bm r + a\hat\mu }
+\left( \frac{e^{-iP_\mu a}}{m_1} + \frac{1}{m_2} \right) \ket{\bm P ,\bm r -a\hat\mu}
+2\left( \frac{1}{ m_1} + \frac{1}{ m_2} \right) \ket{\bm P,\bm r} 
\right] \\
& + V_L(\bm r)\ket{\bm P,\bm r} +  O(a^2).
\label{eq:latH3}
\eeqs
As expected the Hilbert space for fixed total momentum $\bm P$
is invariant under the action of the Hamiltonian and the
eigenvalue problem is more tractable.
The dimension of the space is proportional to $N^3$ and the low-lying spectrum can be obtained quickly for lattices up to $32^3$ on a desktop computer.
For a system at rest ($\bm P=0$), it coincides with the familiar form in the CM frame for relative motion,
\beq
H\ket{\bm 0,\bm r} =-\frac{\hbar^2}{ 2\tilde m} \sum_{\mu} \frac{1}{ 2 a^2}  \left(  \ket{\bm 0, \bm r+a\hat\mu} + \ket{\bm 0, \bm r-a\hat\mu} +2 \ket{\bm 0,\bm r} \right) + V_L(\bm r)\,,
\label{eq:latH3cm}
\eeq
with $\tilde m$ as the reduced mass.
To accelerate the convergence to the continuum we use an improved version with a seven-point stencil,
\beqs
H\ket{\bm P, \bm r} &=-\frac{\hbar^2}{ 2} \sum_{\mu} \frac{-1}{ 180 a^2}  \\ 
& \Big[
-2\left( \frac{e^{3iP_\mu a} }{ m_1} + \frac{1}{ m_2} \right) \ket{\bm P,\bm r+3a\hat\mu}
+27\left( \frac{e^{2iP_\mu a} }{ m_1} + \frac{1}{ m_2} \right) \ket{\bm P, \bm r+2a\hat\mu}
-270\left( \frac{e^{iP_\mu a} }{ m_1} + \frac{1}{ m_2} \right) \ket{\bm P,\bm r+a\hat\mu}
 \\ & 
 -2\left( \frac{e^{-3iP_\mu a} }{ m_1} + \frac{1}{ m_2} \right) \ket{\bm P,\bm r-3a\hat\mu}
+27\left( \frac{e^{-2iP_mu a} }{ m_1} + \frac{1}{ m_2} \right) \ket{\bm P, \bm r-2a\hat\mu}
-270\left( \frac{e^{-iP_\mu a} }{ m_1} + \frac{1}{ m_2} \right) \ket{\bm P,\bm r-a\hat\mu}
 \\ & 
+490\left( \frac{1}{ m_1} + \frac{1}{ m_2} \right) 
\ket{\bm P,\bm r}
\Big]  + V_L(\bm r)\ket{\bm P,\bm r} + O(a^6).
\label{eq:latH7}
\eeqs
\end{widetext} 

To confirm the correctness of the new formalism in the $|\bm P,\bm r\rangle$ basis, we performed the following check.
We solve the original problem in Eq.\eqref{eq:latH} on a $4^3$ lattice to obtain all 4096 eigenvalues and eigenvectors. They contain all 64 sectors of possible discrete total momenta of the system,
\beqs
 \bm P &=\frac{2\pi}{ a} \left( \frac{i }{ N_x}, \frac{j }{ N_y}, \frac{k }{ N_z } \right) \text{ where } \\
 & i=0,\cdots, N_x-1, \\ 
 & j=0,\cdots, N_y-1,  \\
 &k=0,\cdots, N_z-1. 
\label{eq:P}
\eeqs
Note that due to the smallness of the lattice, there are
a lot of accidental degeneracies in this setup and eigenvectors
of different momentum are mixed since they have the same
energy. 
To project out the individual momentum sectors, we lifted
the degeneracy by adding random terms to the Hamiltonian
proportional to the momentum operators. We construct momentum operators on the lattice from the translation operator $T_x=e^{i P_x a}$ (similar in y and z directions). The issue is that it is not Hermitian, so we consider Hermitian alternatives 
\beq 
P_{x1}=\frac{T_x-T_x^\dagger }{ 2I} \text{ and } 
P_{x2}=\frac{T_x+T_x^\dagger }{ 2}.
\eeq 
It turns out that both the ``sine'' and ``cosine'' operators are needed to remove degeneracies. 
Using these operators, we can form a set of commuting Hermitian operators 
\beq 
\hat{O}=\{H, P_{x1}, P_{x2},  P_{y1}, P_{y2}, P_{z1}, P_{z2}\}.
\eeq 
We take a random linear combination of the set and solve for the  eigenfunctions $\ket{\psi}$. 
The eigenvalues of the seven operators in the set can be postcomputed easily: 
$\lambda_i=\opbraket*{\psi}{\hat{O}_i}{\psi}$. 
Then it is just a matter of comparing the momentum eigenvalues with the 64 unique sectors in Eq.\eqref{eq:P} to identify the eigenvalues of H belonging to a certain sector.
The 64 groups of energy levels thus obtained are compared to those computed directly from the projected Hamiltonian in Eq.\eqref{eq:latH3} on the same lattice sector by sector where $\bm P$ is an input.
Perfect agreement is achieved in all 64 sectors, using both the three-point stencil in Eq.\eqref{eq:latH3} and seven-point stencil in Eq.\eqref{eq:latH7} version of the Hamiltonian. 
The same check is carried out for $z$ elongated lattices ($N_x=N_y\neq N_z$).

\subsection{Energy spectrum in the box} 

The spectrum of the Hamiltonian is naturally split into
invariant block with different total momentum $\bm P$.
We need to further consider the effects of the 
rotational symmetry on this spectrum. The relevant
symmetry group is reduced from the infinite volume one
to the lattice group. Furthermore, if we are considering
states with $\bm P\not= 0$, the relevant symmetry group
is further reduced to the {\em little group}, that is
the subgroup of the lattice symmetry group that leaves
the momentum $\bm P$ invariant: a symmetry 
transformation $S_k$ belongs to the little group if
$S_k \bm P = \bm P$.

The reduction method we used to project to 
the total-momentum blocks can be similarly used to further 
reduce the Hamiltonian to the invariant sectors generated 
by the rotational symmetry. To generate the eigenvectors according to the
irreducible representations (irreps) of the relevant lattice 
symmetry group we use two different methods.
The first approach is to compute the low-lying spectrum of H
and then determine which irrep the eigenvectors belong to based on their transformation properties under rotations. Specifically we build the projection operators
\beq
P^{\Gamma,\lambda}\ket{\bm P,\bm r} = \frac{d_\Gamma}{ g}\sum_k [D_{\lambda\lambda}^\Gamma(S_k)]^* \ket{S_k\bm P,S_k\bm r},  
\eeq 
where $\Gamma$ is a given irrep, $\lambda$ the representation row, $d_\Gamma$ the dimension of the irrep, $g$ the total number of elements in the symmetry group, $S_k$ the symmetry transformation, and $D^\Gamma(S_k)$ its representation matrix. If the norm of the rotated eigenvector $P_\Gamma|\bm P,\bm r\rangle$ is nonzero, then the corresponding eigenvalue is classified to belong to the $\lambda$ row of the irrep $\Gamma$ (we assume no accidental degeneracies).
Note that since $S_k$ is a member of the little group, $S_k\bm P$
is $\bm P$ in the equation above.

The second way is to project the Hamiltonian first, 
\beq
H^{\Gamma,\lambda}=P_{\Gamma,\lambda} \,H\, P_{\Gamma,\lambda}.
\eeq 
The projection matrix for row $\lambda$ of irrep $\Gamma$ is constructed as
\beq
P_{\Gamma,\lambda} = \frac{d_\Gamma}{ g} \sum_k [D_{\lambda\lambda}^\Gamma(S_k)]^*  U(S_k)\,.  
\eeq
Matrix $U(S_k)$ represents the action of the symmetry operation:
\beq
U_{\bm r\bm r'}(S_k)=\delta(\bm r-S_k \bm r')\, .
\eeq
Then the spectrum is obtained from the eigenvalues of the projected Hamiltonian.
The two methods produce the same results and serve as a cross check.

Having obtained the lattice Hamiltonian in the reduced $\ket{\bm P,\bm r}$ basis, we need to take the continuum limit to obtain box levels from lattice levels. This is done by increasing the number of grid points and deceasing the lattice spacing simultaneously while keeping the box size fixed, 
\beq
\lim_{\substack{a\to 0\\ N\to \infty }} N a = L.
\eeq 
Since the discretization error is known to behave as $\mathcal{O}(a^6)$, we perform a linear extrapolation 
in $a^6$ using three lattices.

In a later section, we will present the results for this method applied to the cases discussed earlier: rest frame ($\bm P=0$) and four moving frames ($\bm P\not=0$), in both cubic and elongated boxes. But first we discuss the quantization conditions.

\section{Numerical checks}
\label{sec:validation}
In this section, we check our derivation for the QCs in the various scenarios discussed Sec.~\ref{sec:QC}. We seek the simplest way to accomplish this goal: by solving a Schr\"{o}dinger equation with a simple potential in a box with periodic boundary conditions. We compare this spectrum with the one derived from the quantization conditions.

\subsection{Infinite volume phase shifts} 
\label{sec:phases}
\begin{figure}[!htbp]
\includegraphics[scale=0.35]{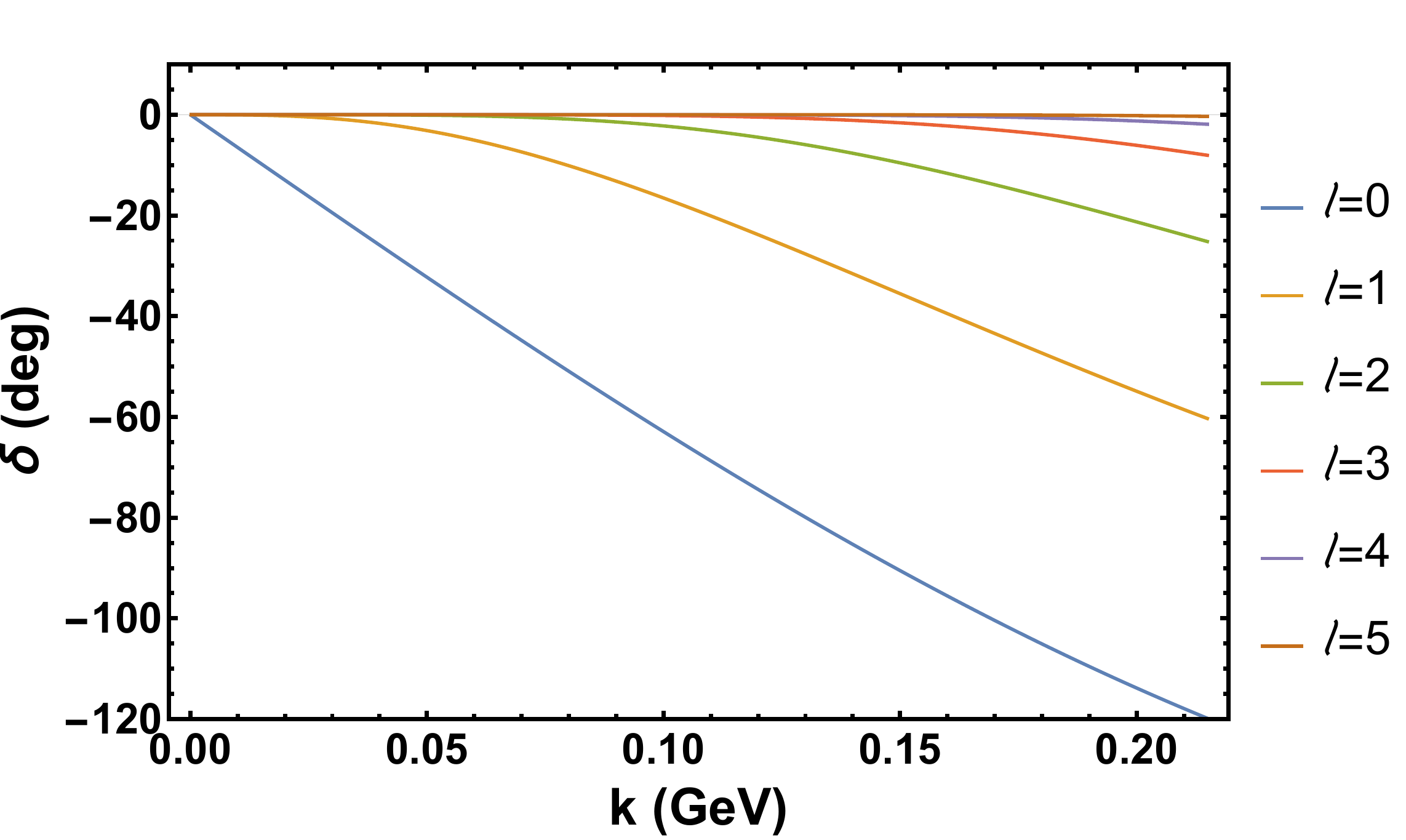}
\caption{Phase shift of the test potential for the lowest six partial waves.}
\label{fig:phase}
\end{figure}

The first step is to compute the phase shifts for a simple potential. 
Consider two particles $m_1=0.138$ GeV and $m_2=0.94$ GeV, interacting through a repulsive potential of Gaussian falloff,
\beq
V(r)=C e^{-0.5(r/R_0)^2}
\label{eq:pot2}
\eeq
where  $C=1.0$ GeV and  $R_0=1.25$ fm. 
The range of the potential is about 4 fm.
The phase shifts can be obtained readily by the variable phase method~\cite{Calogero1967}. 
For partial waves up to $l=5$ and momenta up to about 0.2 GeV, they are shown in Fig.~\ref{fig:phase}.
Some numerical values are given in Table~\ref{tab:phase}.
The phase shifts have the expected $\delta_l(k) \sim k^{2l+1}$ asymptotic behavior. 
The potential is chosen so that in our tests partial waves up to $l=5$ 
can be checked for convergence in the $k$ range we use.
The goal is to check our derivation for the higher order QCs by comparing these energies produced by these phase shifts with the two-particle spectrum in finite volume.
\begin{table}[!htbp]
\caption{Some numerical values of the phase shift (degrees)  at selected values of k (GeV) in Fig.~\ref{fig:phase}.}
\label{tab:phase}          
                                       
\renewcommand{\arraystretch}{1.3}
$
\begin{array}{ccccc}
\toprule   
l & k=0.05 & k=0.1 & k=0.15 & k=0.2 \\                        
 0 & -32.2599 & -62.9184 & -90.5094 & -113.833 \\
 1 & -3.15552 & -16.5361 & -35.5349 & -54.9107 \\
 2 & -0.101947 & -2.22494 & -9.5661 & -21.2908 \\
 3 & -0.00173584 & -0.160577 & -1.61377 & -6.08421 \\
 4 & -0.0000210598 & -0.00781134 & -0.18254 & -1.24548 \\
 5 & -1.817\times10^{-7} & -0.000293742 & -0.0156874 & -0.192916 \\
\bottomrule   
\end{array}                   
$                                                
                    
\end{table}                     
%

\subsection{Results and discussion} 

Since the range of the potential is about 4~fm, a box size of $L=24$~fm is sufficient to make the exponential finite-volume effects negligible.
The large volume is for checking purposes in the quantum-mechanical model. 
We note such a large volume is not accessible in current lattice QCD simulations, although the recent development of the masterfield paradigm makes an important leap in this direction~\protect\cite{Francis_2020,master2021}.
To take the continuum limit we use lattices of $20^3$ with $a=1.2$~fm, $24^3$ with $a=1$~fm, and $30^3$ with $a=0.8$~fm for cubic case. For the elongated case we use the same three lattice spacings and the same size $L=24$~fm in the $x$, and $y$ direction but we elongate the $z$ direction by a factor of $\eta=1.5$. 
The lowest nonzero momentum in the spectrum is controlled by the box size $k_{min}=2\pi/(L\eta)$.
For the higher $k$ values the density of states gets higher. We study states with $k<0.2$ GeV. In this $k$-range only the phase shifts for $\ell\leq 5$ are significantly different from zero (see Fig.~\ref{fig:phase}). Therefore, we expect convergence of the QCs by $l=4$ or $l=5$. The total number of noninteracting levels (with or without degeneracies) under the cutoff are summarized in Table~\ref{tab:kcut} for all the cases considered in this work. We see the number of levels is still fairly large after the $k$ cutoff. In such cases, we apply an additional cut of 40 distinct levels to keep the number manageable, which implies a smaller $k$ range.
Interactions cannot change the number of levels in our model, only shift them, so the noninteracting levels serve as a useful guide.
We provide all the noninteracting levels obtained from kinematics in Sec.~1 of the Supplement Material~\cite{supp}.
For the rest frame of cubic and elongated boxes, we see that the levels are more packed in elongated box (36 vs 14), which also reaches lower in the first nonzero level (0.03 vs 0.05 GeV). The ability to reach lower energy can be regarded as an advantage of the elongated geometry over the cubic.
The integer indices of lab momentum for each particle along with their degeneracy are shown. We see how the particles are arranged to have back-to-back momenta so the total momentum is zero in both lab and CM frames. Level 9 in cubic box is a special case with accidental degeneracy of 6 from $(\pm 3,0,0)$ and 24 from $(\pm 2,\pm 2,\pm 1)$. Its counterpart in elongated box is level 21, but with a different degeneracy of 4 from $(\pm 3,0,0)$ and 16 from $(\pm 3,\pm 2,\pm 1)$. The one with $(\pm 2,\pm 2,\pm 1)$ is level 20 which has lower energy and degeneracy 8.
These $k$ levels are also free-particle poles in the zeta functions in Eqs.\eqref{eq:zfun} and~\eqref{eq:zfun_boost_eta}.
Interactions will cause $k$ to deviate from the free levels. The amount of the deviation is related to phase shifts via the L\"uscher QC. 

\begin{figure*}[b]
\includegraphics[scale=0.9]{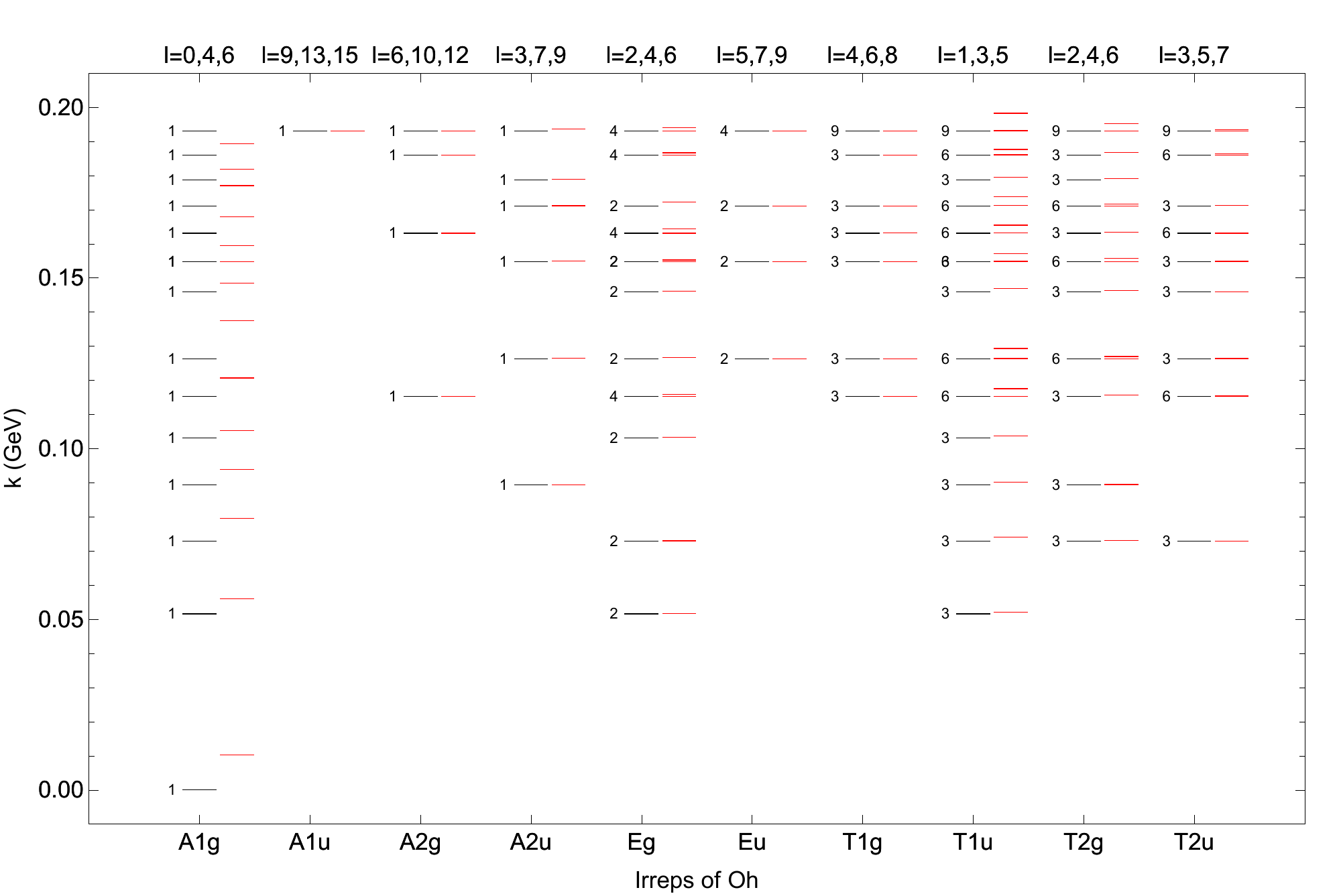}
\caption{Discrete box levels in the rest frame of cubic box of size $L=24$~fm up to $k=0.2$ GeV. The free-particle (black) and interacting (red) levels are shown side by side for comparison. The numbers next to free levels indicate degeneracy. The box energies are obtained in the lab frame from an extrapolation on several lattices, projected into the irreps of $O_h$, then converted to CM momentum $k$. The top labels show the lowest partial waves in each irrep.
Numerical values for both noninteracting and interacting levels can be found in Sec.~1 and Sec.~4 of the Supplemental Material~\protect\cite{supp}.
}
\label{fig:ospecOh}
\end{figure*}

Since the QCs are block diagonalized by irreps, the levels must also be projected into the irreps as discussed earlier. The interacting levels will be given in numerical tables when we study the convergence of QCs.
In Fig.~\ref{fig:ospecOh} we show an example of the projected spectrum for the rest frame in cubic box ($O_h$ group). 
Note that in the noninteracting case levels with the same energy appear in different irreps. Degeneracies in the noninteracting levels are removed by the interactions. 
The strongest interaction is in the $A_{1g}$ channel which is dominated by $l=0$. 
The lowest level (zero) is shifted up by about $10$~MeV.  The next strongest interaction is the $T_{1u}$ channel which is dominated by $l=1$ where the lowest level is shifted up by about $0.5$~MeV. The shift is upwards across the board because the interaction is repulsive everywhere. 
The shifts in other channels are barely visible, but can be resolved numerically.
As an additional check we repeated the entire procedure, including the continuum extrapolations, with the interactions turned off. We compared these results with the expectation from basic kinematics and found perfect agreement.
This also provided us with a straightforward way to determine the multiplicity for each noninteracting level in each irrep.

\begin{figure}[h]
\includegraphics[width=0.5\textwidth]{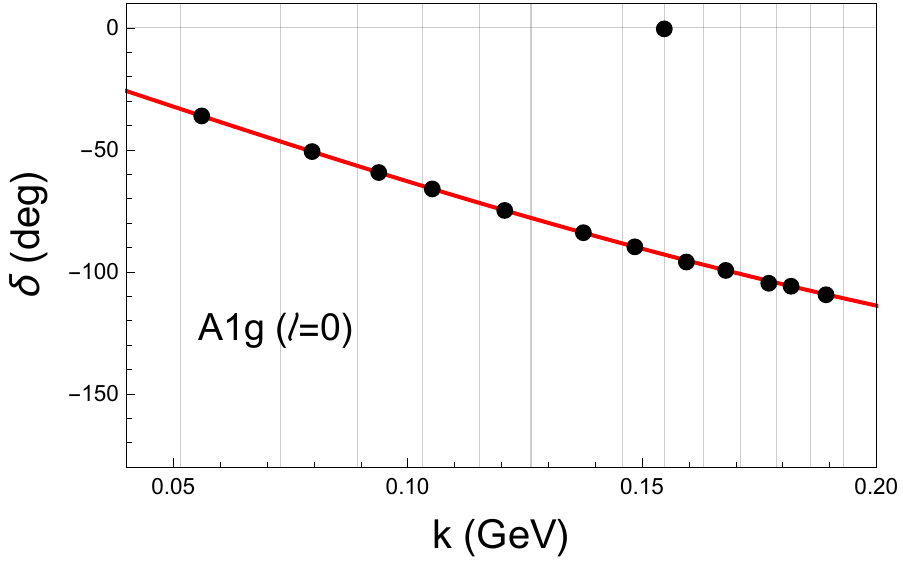}
\caption{Phase shifts reconstruction for the lowest partial wave ($l=0$) in the $A_{1g}$ irrep of rest frame $d=(0,0,0)$ in cubic box. 
The black points are the predicted phase shift via L\"uscher formula. The red curve is the infinite-volume phase shift. 
The faint vertical lines correspond to noninteracting levels in the box. 
}
\label{fig:Oh-A1g}
\end{figure}

\begin{table}[htb]
\caption{Number of distinct and total noninteracting levels for two spinless particles of unequal masses in a box of geometry $L\times L\times \eta L$ with $L=24$ fm and cutoff $k<0.2$ GeV. 
}
\label{tab:kcut}          
                                           
\renewcommand{\arraystretch}{1.3}
$
\begin{array}{c | c c | c c }
\toprule                         
d  & \multicolumn{2}{c |}{\text{cubic } (\eta=1)}  & \multicolumn{2}{c}{\text{$z$ elongated } (\eta=1.5)}     \\
\hline                           
 &  \text{distinct} & \text{total} &  \text{distinct} & \text{total}      \\
\hline                           
(0,0,0) & 14  & 251 & 36 & 359 \\
(0,0,1) & 54 & 252 &  79 & 367 \\
(1,1,0) & 73 & 245 & 108 & 355  \\
(1,1,1)  & 54  & 342 & 204 & 363 \\
(0,1,2)  &107  & 240 &  218 & 366 \\
\bottomrule                      
\end{array}  
$                                                
                    
\end{table}                     
Our objective is to reproduce the finite-volume spectrum using the QCs
and the phase shifts.
 The matrix elements needed to construct the QCs are given in Table~\ref{tab:Oh} and Table~\ref{tab:D4h}  in Appendix~\ref{sec:ME}.
Since the symmetry-adapted QC is a single condition that couples to all possible partial waves in a given irrep, we can compute the phase shift from the spectrum only if the lowest partial wave is retained. 
As an example, we show in Fig.~\ref{fig:Oh-A1g} the phase shift prediction for $l=0$ partial wave in the $A_{1g}$ QC by feeding the interacting energy levels into the L\"uscher formula. 
All other cases can be found in Sec.~2 of the Supplemental Material~\cite{supp}.
We see the reconstruction is excellent up to $k=0.2$~GeV, but with a notable exception point at  $k=0.1548$~GeV (level 9) where the box level seems ``incompatible" with the lowest-order L\"uscher QC. 
A feature of the exception is that it happens at a free-particle pole (faint vertical line). We say the level is ``pinched" at the  free-particle pole.  All other points are in between free-particle poles. The discrepancy is due to the fact that we neglected the higher partial waves in the QC. 

As a general method to assess the effects of higher partial waves, we investigate the convergence of the QC by feeding it the infinite-volume phase shifts and comparing the resulting levels with the box levels. 
We find it more convenient to locate the roots of this equation by solving the real valued QC1 rather than the complex valued QC2. 
After the roots are found from QC1, we pass them through QC2 to double check they are also roots. 
This is a different approach to Ref.\cite{Woss2020} where QC2 is solved by eigenvalue decomposition for coupled channels.
We check the convergence order by order: ``order 1" has only the lowest partial wave, ``order 2" with the next partial wave added, and so on. In the limit that all the partial waves are included, perfect agreement is expected. 
In practice, we check convergence for the lowest five partial waves (up to $l=4$), which is adequate for most applications. In some cases, we consider $l=5$. 
Once convergence is achieved, we consider the QC checked at the given order.

\begin{figure}[h!]
\includegraphics[width=0.5\textwidth]{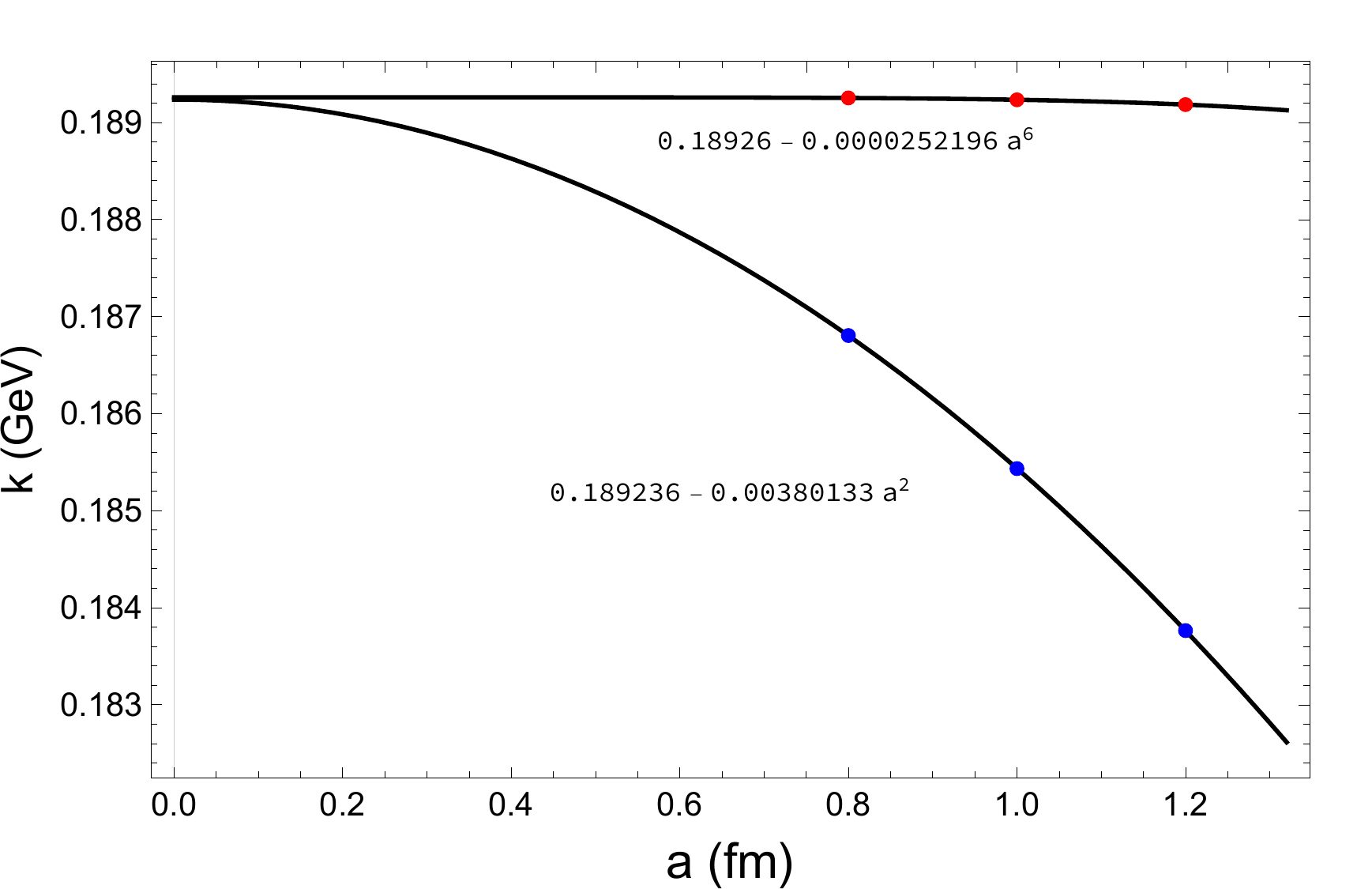}
\caption{Continuum extrapolation of a level in the $A_{1g}$ irrep of $O_h$. Three points are obtained on lattices $20^3$ with $a=1.2$~fm, $24^3$ with $a=1$~fm, and $30^3$ with $a=0.8$~fm. The red points are from the seven-point stencil and the blue points from the three-point stencil. The fitted curves  and forms are also displayed.
}
\label{fig:contra}
\end{figure}

The comparison involves very small differences that are not easily discernible visually. 
To better gauge the quality of the convergence, we introduce a numerical measure 
\beq
\chi^2=\frac{(k_{\rm box} - k_{\rm QC})^2 }{ (k_{\rm box} - k_{\rm lat})^2},
\label{eq:chi2}
\eeq
where $k_{\rm box}$ is the continuum box level extrapolated from the three lattice spacings, $k_{\rm lat}$ the level on the lattice with the finest lattice spacing, 
and $k_{\rm QC}$ the solution from the QC at each order.  The extrapolation is a function of $a^6$, the error present 
in the Hamiltonian from the seven-point stencil approximation in Eq.\eqref{eq:latH7}. 
We show an example of the continuum extrapolation in Fig.~\ref{fig:contra}. For comparison, we also show the result from the standard three-point stencil approximation Eq.\eqref{eq:latH3} which has  $O(a^2)$ discretization errors. We see that they converge to the same result but at different rates.
\begin{figure}[h]
\includegraphics[width=0.5\textwidth]{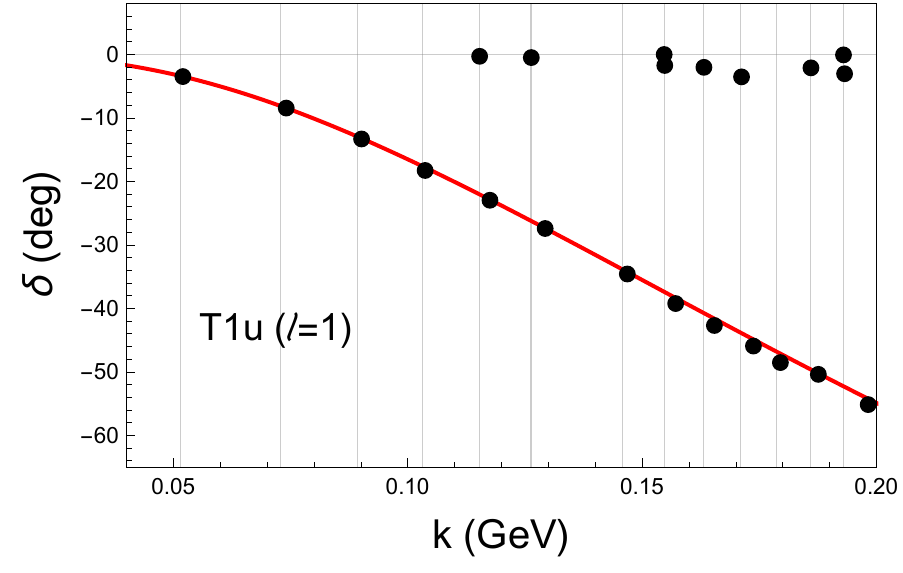}
\caption{Phase shifts reconstruction for the lowest partial wave ($l=1$) in the $T_{1u}$ irrep of rest frame $d=(0,0,0)$ in cubic box. 
The black points are the predicted phase shift via L\"uscher formula. The red curve is the infinite-volume phase shift. 
The faint vertical lines correspond to noninteracting levels in the box. 
}
\label{fig:Oh-T1u}
\end{figure}

Basically, the convergence is measured against the tiny difference 
between the continuum box levels and those from the largest lattice used in the extrapolation (about six decimal places, or 1~eV out of 1~MeV). Note that the $\chi^2$ introduced is not in the standard sense of curve fitting where the best value is around 1. Here the smaller its value, the better the convergence. 
This is a highly sensitive measure: nonconvergence of a single level will have a large contribution to the total $\chi^2$.
In cases of a box level coinciding with a free-particle pole, we indicate it by a red star and replace $k_{\rm QC}$ in Eq.\eqref{eq:chi2} by $k_{\rm free}$. 

We apply the method to the $A_{1g}$ example in Fig.~\ref{fig:Oh-A1g}. 
The results are given in Table~\ref{tab:Oh-A1g}.
The overall quality of convergence  is measured by the total $\chi^2$ of all the levels.
 We see that at order 1 (only $l=0$ $s$ wave is included), it is about 277. The level most responsible for the discrepancy is level 12, 
followed by levels 10, 9, 4, 6. We note that level 9 is the one that is obscured by the noninteracting energy pole, the one that stands out in Fig.~\ref{fig:Oh-A1g}.
At order 2, the QC is enlarged by adding the next partial wave ($l=4$). 
We see that the addition of  $l=4$ causes small changes in the $k$ values that improves the total $\chi^2$. The exception level, level 9, is now approximated well by one of the roots of the QC. 
The quality is indicated by the small $\chi^2$ value of 0.019.  In fact, all the levels at order 2 have small $\chi^2$, leading to a total of 3.73, much smaller than 277 at order 1. This is a prime example of how the higher partial wave impacts the QC and how we check convergence.
\begin{table*}
\caption{The companion table to Fig.~\ref{fig:Oh-A1g} for $A_{1g}$, showing convergence of quantization condition by adding higher partial waves. 
The k levels in GeV are compared directly order by order, using the $\chi^2$ measure defined in Eq.\eqref{eq:chi2}. The red star indicates that the box level is ``pinched" at a free-particle pole.}
\label{tab:Oh-A1g}

\renewcommand{\arraystretch}{1.2}
$
\begin{array}{ccccccc} \toprule
\text{Level} & \text{Box L} & \text{Lattice} & \text{Order 1} & \text{Order 2} & \chi^2 \text{ Order 1} & \chi^2 \text{ Order 2} \\
\hline
 14 & 0.18926 & 0.189253 & 0.189254 & 0.18926 & 0.820197 & 0.00914913 \
\\
 13 & 0.181804 & 0.181799 & 0.181803 & 0.181803 & 0.0205124 & \
0.0146576 \\
 12 & 0.177043 & 0.177041 & 0.177007 & 0.177044 & 174.698 & 0.0110595 \
\\
 11 & 0.167873 & 0.167867 & 0.167872 & 0.167874 & 0.0419179 & \
0.0215601 \\
 10 & 0.159465 & 0.159461 & 0.15944 & 0.159466 & 29.8119 & 0.014445 \
\\
 9 & 0.154755 & 0.154749 &  \textcolor{red}{\ast}0.154723 & 0.154755 & 29.1192 & 0.0191184 \
\\
 8 & 0.148467 & 0.148466 & 0.148466 & 0.148467 & 0.763934 & \
0.000166674 \\
 7 & 0.137488 & 0.137487 & 0.137489 & 0.137489 & 0.0251408 & \
0.0252861 \\
 6 & 0.120712 & 0.120711 & 0.12071 & 0.120712 & 16.368 & 0.0213916 \\
 5 & 0.105257 & 0.105256 & 0.105256 & 0.105257 & 4.20346 & 0.017348 \
\\
 4 & 0.0938371 & 0.0938369 & 0.0938362 & 0.0938372 & 18.3015 & \
0.130459 \\
 3 & 0.0796177 & 0.0796176 & 0.0796178 & 0.0796178 & 0.453209 & \
0.453221 \\
 2 & 0.0560702 & 0.0560701 & 0.0560702 & 0.0560701 & 0.214519 & \
0.623864 \\
 1 & 0.01025 & 0.01025 & 0.0102501 & 0.0102501 & 2.36538 & 2.36787 \\
\bottomrule
&&&& \text{\bf Total }\bm\chi^2 & {\bf 277.206} & {\bf 3.7296} \\
\end{array}
$     

\end{table*}
\begin{table*}
\caption{The companion table to Fig.~\ref{fig:Oh-T1u} for $T_{1u}$, showing convergence of quantization condition by adding higher partial waves. 
The levels are compared directly order by order, using the $\chi^2$ measure defined in Eq.\eqref{eq:chi2}. The red star indicates that the box level is ``pinched" at a free-particle pole.}
\label{tab:Oh-T1u}

\renewcommand{\arraystretch}{1.2}
$
\begin{array}{ccccccccc} \toprule
\text{Level} & \text{Box L} & \text{Lattice} & \text{Order 1} & \text{Order 2} & \text{Order 3} & \chi^2 \text{ Order 1} & \chi^2 \text{ Order 2} & \chi^2 \text{ Order 3} \\
\hline
 22 & 0.198249 & 0.198241 & 0.198146 & 0.198245 & 0.19825 & 167.097 & \
0.249283 & 0.0101593 \\
 21 & 0.193217 & 0.193213 & \textcolor{red}{\ast}0.192974 & 0.193209 & 0.193221 & 2998.88 \
& 3.71223 & 0.470423 \\
 20 & 0.192982 & 0.192978 & \textcolor{red}{\ast}0.192974 & \textcolor{red}{\ast}0.192974 & 0.192975 & \
3.26057 & 3.26057 & 2.49916 \\
 19 & 0.187618 & 0.187613 & 0.187614 & 0.187615 & 0.187619 & 0.747838 \
& 0.42273 & 0.00535232 \\
 18 & 0.186043 & 0.186039 & \textcolor{red}{\ast}0.185954 & 0.186041 & 0.186045 & 389.699 \
& 0.356124 & 0.194373 \\
 17 & 0.179522 & 0.179521 & 0.179489 & 0.179517 & 0.179522 & 807.378 \
& 17.3011 & 0.0125204 \\
 16 & 0.173768 & 0.173763 & 0.173698 & 0.173767 & 0.173769 & 206.536 \
& 0.0359251 & 0.00582583 \\
 15 & 0.171233 & 0.171228 & \textcolor{red}{\ast}0.171053 & 0.171231 & 0.171233 & 1447.64 \
& 0.209609 & 0.00394721 \\
 14 & 0.165438 & 0.165434 & 0.165382 & 0.165438 & 0.165439 & 128.455 \
& 0.0154137 & 0.00520841 \\
 13 & 0.163206 & 0.163201 & \textcolor{red}{\ast}0.163093 & 0.163204 & 0.163206 & 526.933 \
& 0.147073 & 0.0123243 \\
 12 & 0.157179 & 0.157177 & 0.157134 & 0.157178 & 0.157179 & 551.783 \
& 0.156363 & 0.00460883 \\
 11 & 0.154856 & 0.154853 & \textcolor{red}{\ast}0.154723 & 0.154856 & 0.154857 & 1267.98 \
& 0.0467443 & 0.00685293 \\
 10 & 0.154726 & 0.154725 & \textcolor{red}{\ast}0.154723 & \textcolor{red}{\ast}0.154723 & 0.154726 & \
6.59657 & 6.59657 & 0.0176498 \\
 9 & 0.146868 & 0.146867 & 0.14686 & 0.146867 & 0.146868 & 172.894 & \
2.71263 & 0.0020146 \\
 8 & 0.129339 & 0.129339 & 0.129335 & 0.129339 & 0.129339 & 58.3417 & \
0.000653416 & 0.0113939 \\
 7 & 0.126379 & 0.126379 & \textcolor{red}{\ast}0.126331 & 0.126379 & 0.126379 & 30130.9 \
& 0.976735 & 0.00253806 \\
 6 & 0.11756 & 0.117559 & 0.117552 & 0.11756 & 0.11756 & 323.653 & \
0.0133041 & 0.00803915 \\
 5 & 0.115359 & 0.115359 & \textcolor{red}{\ast}0.115324 & 0.115359 & 0.115359 & 13893.6 \
& 0.272455 & 0.002095 \\
 4 & 0.103734 & 0.103734 & 0.10372 & 0.103734 & 0.103734 & 1539.04 & \
0.0089603 & 0.0024343 \\
 3 & 0.0901719 & 0.0901719 & 0.0901658 & 0.0901719 & 0.0901719 & \
11790.7 & 0.0587094 & 0.0204841 \\
 2 & 0.0740985 & 0.0740985 & 0.0740982 & 0.0740985 & 0.0740985 & \
27.9224 & 0.011686 & 0.0285701 \\
 1 & 0.0520501 & 0.0520501 & 0.0520494 & 0.0520501 & 0.0520501 & \
836.425 & 0.0286906 & 0.0308669 \\
\bottomrule
&&&&& \text{\bf Total }\bm\chi^2 & {\bf 67276.5} & {\bf 36.5936} & {\bf 3.35684} \\
\end{array}
$     

\end{table*}

Another interesting example is the $T_{1u}$ irrep in cubic box, shown in Fig.~\ref{fig:Oh-T1u}. 
It poses two challenges. One is there are numerous exceptions in this channel. The other is that some levels are nearly degenerate.
The convergence study is shown in Table~\ref{tab:Oh-T1u}.
This channel is dominated by the $p$ wave, followed by $l=3$ and $l=5$ which also has a twofold multiplicity. 
At order 1, the numerous exceptions are indicated by the red stars. 
At order 2, most are fixed by the addition of $l=3$, but two remain pinched (levels 10 and 20). This demonstrates that the neglected $l=3$ partial wave is largely responsible for the exceptions observed at order 1. A closer examination reveals that the two remaining exceptions are part of nearly degenerate pairs. 
Only one of the two nearly degenerate points (level 11) is a QC solution but the other (level 10) is not.  A similar situation holds for another pair of close-by levels: level 21 is a solution but level 20 is not. This hints at possible influence of the next partial wave $l=5$.  Indeed, at order 3, the two points are both solutions of the QC and we obtain a spectrum that is in complete agreement with the box spectrum. 
The convergence is confirmed by the total $\chi^2$: from 67277 at order 1, to 37 at order 2, and to 3.4 at order 3.

We have confirmed the convergence of all 45 cases in the same manner, as summarized in Table~\ref{tab:all}.
One point to emphasize is that to get agreement for certain cases we needed to consider QC all the way to order 5. The irreps in question (Nos. 10, 14, 18, 21, 33, 38, 42, 44) are, as expected, the ones that allow the most mixing between partial waves.

\begin{table*}
\caption{Summary of the total $\chi^2$ measure showing convergence for all QCs discussed in this work. 
Here $l(n)$ indicates the lowest few partial waves (and multiplicities) that couple to the QC; $N$ is the number of levels under the cutoff of $k=0.2$ GeV or 40 levels. 
Detailed convergence data for every individual energy level can be found in Sec.~4 of the Supplemental Material~\cite{supp}.
}
\label{tab:all}

\renewcommand{\arraystretch}{1.2}
$
\begin{array}{cccclllllll}\toprule     
\multicolumn{11}{c}{\text{\bf Cubic box}}  \\

\text{No.}  & d & \text{Group}    & \text{QC} & l(n)  & \text{N}  & \text{Order 1} & \text{Order 2}  &\text{Order 3}  & \text{Order 4}  & \text{Order 5}   \\
\hline
1  & (0,0,0) & O_h  & A_{1g} & 0, 4, 6, \cdots & 14 & 277.206 & 3.7296 & & &\\
    &             &         & A_{1u} & 9, 13, 15 , \cdots & & &&& &\\
    &             &         & A_{2g} & 6, 10, 12, \cdots  & & &&& &\\
2  &             &         & A_{2u} & 3, 7, 9, \cdots & 6 & 0.0458813  && & &\\
3  &             &         & E_{g} &  2, 4, 6, \cdots & 16 & 2123.07 & 5.44639 & & &\\
4  &             &         & E_{u} & 5, 7, 9, \cdots & 5 & 0.0546959 & & &\\
5  &             &         & T_{1g} &4, 6, 8(2), \cdots &9 &  0.243897 && & &\\
6  &             &         & T_{1u} & 1, 3, 5(2), \cdots & 22 & 67276.5 & 36.5936 & 3.35684 & &\\
7  &             &         & T_{2g} & 2, 4, 6(2), \cdots & 16 &  2485.36 & 2.70125 & & &\\
8  &             &         & T_{2u} & 3, 5, 7(2), \cdots & 14 &  4.55575 & 0.267262 & & &\\

10  & (0,0,1) & C_{4v}  & A_1 & 0, 1, 2, 3, 4(2), \cdots & 40  &  4.08008\times 10^8 & 1.06106\times 10^{10} & 10175.7 & 284.022 & \
6.92792 \\
11  &             &             & A_2 & 4, 5, 6, \cdots & 15 &  7.46202 & 0.145375 & &\\
12  &             &             & B_1 & 2, 3, 4, 5, \cdots & 34 & 276889. & 402.025 & 5.9668 & &\\
13 &             &             & B_2 &  2, 3, 4, 5, \cdots & 27 &  350956. & 779.183 & 3.84012 & &\\
14 &             &             & E &  1, 2, 3(2), 4(2), \cdots & 40 & 1.32751\times 10^7 & 72379.5 & 514.979 & 8.21687  &\\

14  & (1,1,0) & C_{2v}  & A_1 & 0, 1, 2(2), 3(2), 4(3), \cdots &  40 & 1.96654\times 10^8 & 4.77126\times 10^6 & 17298.1 & 190.94 & 5.34113 \\
15  &             &             & A_2 & 2, 3, 4(2), 5(2), \cdots & 40 & 54211.3 & 2998.61 & 3.01327& &\\
16  &             &             & B_1 & 1, 2, 3(2), 4(2), \cdots & 40 &  2.84806\times 10^6 & 47754.8 & 241.517 & 4.62961 &\\
17 &             &             & B_2 &  1, 2, 3(2), 4(2), \cdots & 40 & 9.36978\times 10^6 & 88719.7 & 308.013 & 6.28652  &\\

18  & (1,1,1) & C_{3v}  & A_1 &  0, 1, 2, 3(2), 4(2), \cdots & 40 & 9.16556\times 10^7 & 1.09895\times 10^6 & 22260. & 115.412 & 3.53446 \\
19  &             &             & A_2 & 3, 4, 5, \cdots & 26 & 111.322 & 0.44229 & & &\\
20  &             &             & E & 1, 2(2), 3(2), 4(3), \cdots & 40 & 8.06257\times 10^6 & 23959. & 271.289 & 4.13508&\\

21  & (0,1,2) & C_{1v}  & A_1 &  0, 1(2), 2(3), 3(4), 4(5), \cdots & 40 &  1.01852\times 10^8 & 993629. & 6858.3 & 17.3429 & 2.01417 \\
22  &            &              & A_2 &  1, 2(2), 3(3), 4(4), \cdots & 40 & 8.98645\times 10^6 & 37314.8 & 293.484 & 3.18387  &\\

\hline   
\multicolumn{11}{c}{\text{\bf Elongated box}}   \\
  
\text{No.}  & d & \text{Group}    & \text{QC} & l(n)  & \text{N}  & \text{Order 1} & \text{Order 2}  &\text{Order 3}  & \text{Order 4}  & \text{Order 5}   \\
\hline
23  & (0,0,0) & D_{4h}  & A_{1g} &0, 2, 4(2), 5, \cdots & 40  & 4.06688\times 10^6 & 573.449 & 8.64225  & &\\
24  &             &         & A_{1u} & 5, 7, 9(2) , \cdots & 9  & 0.134224  & &  & &\\
25    &             &         & A_{2g} & 4, 6, 8(2) , \cdots & 3  &  0.00828583 & &  & &\\
26    &             &         & A_{2u} & 1, 3, 5(2), 7(2), \cdots & 34  & 131946. & 16.2795 & 10.7998 & &\\
27  &             &         & E_{g} & 2, 4(2), 6(3), \cdots & 38  &  2009.93 & 0.319341 & & &\\
28  &             &         & E_{u} & 1,3(2),5(3), \cdots & 40  &   221693. & 41.2595 & 8.52709  & &\\
29  &             &         & B_{1g} & 2, 4, 6(2), 8(2),\cdots & 27  & 703.291 & 11.0395 & & &\\
30  &             &         & B_{1u} & 3,5,7(2), 9(2), \cdots & 17  &  7.07159 & 0.168506 & & &\\
31  &             &         & B_{2g} & 2, 4, 6(2), 8(2), \cdots & 22  &  2093.46 & 0.192 & & &\\
32  &             &         & B_{2u} & 3, 5, 7(2), 9(2), \cdots & 21  & 2.06599 & 0.488025 & & &\\

33  & (0,0,1) & C_{4v}  & A_1 & 0, 1, 2, 3, 4(2), \cdots & 40  &  5.70458\times 10^8 & 9.64404\times 10^6 & 14025.4 & 166.777 & \
11.0137 \\
34  &             &             & A_2 & 4, 5, 6, \cdots & 21 & 40.092 & 0.21509 & &\\
35  &             &             & B_1 & 2, 3, 4, 5, \cdots & 40 & 161200. & 274.945 & 5.1077 & &\\
36 &             &             & B_2 &  2, 3, 4, 5, \cdots & 39 & 318459. & 845.868 & 2.82798 & &\\
37 &             &             & E &  1, 2, 3(2), 4(2), \cdots & 40 & 1.66366\times 10^7 & 79648.3 & 353.966 & 7.28402 &\\

38  & (1,1,0) & C_{2v}  & A_1 & 0, 1, 2(2), 3(2), 4(3), \cdots & 40 & 2.19334\times 10^8 & 3.60527\times 10^6 & 9669.56 & 99.3024 & 27.4093\\
39  &             &             & A_2 & 2, 3, 4(2), 5(2), \cdots & 40 & 34211.4 & 295.311 & 0.804797& &\\
40  &             &             & B_1 & 1, 2, 3(2), 4(2), \cdots & 40 & 2.97904\times 10^6 & 28179. & 115.601 & 6.80754  &\\
41 &             &             & B_2 &  1, 2, 3(2), 4(2),  \cdots & 40 & 9.05592\times 10^6 & 71558.8 & 146.885 & 18.7042  &\\

42  & (1,1,1) & C_{1v}  & A_1 &  0, 1(2), 2(3), 3(4), 4(5), \cdots & 40  & 5.45564\times 10^8 & 2.89417\times 10^6 & 15983.3 & 39.1208 & 8.00598 \\
43  &             &             & A_2 & 1, 2(2), 3(3), 4(4), \cdots & 40  &  6.85779\times 10^6 & 23813.9 & 69.434 & 3.24328 &\\

44 & (0,1,2) & C_{1v}  & A_1 &  0, 1(2), 2(3), 3(4), 4(5), \cdots & 40  & 3.79138\times 10^8 & 3.16621\times 10^6 & 6213.69 & 17.0296 & 5.34524\\
45  &            &              & A_2 &  1, 2(2), 3(3), 4(4), \cdots & 40  &  1.27375\times 10^7 & 45256.8 & 146.374 & 2.67257 &\\

\bottomrule
\end{array}
$     

\end{table*}

Finally we want to see that for the ``pinched'' points if we can exploit the sensitivity to the second partial wave to produce predictions for its phase shift based solely on the energy input. Note that this is not in general possible at other points since the equation at next order involves two phase shifts constrained by a single equation. Solving for the second partial wave from a generic QC at order 2 with no multiplicities, we have
\beq
\cos\delta_{\rm 2nd}=M_{22}+\frac{|M_{12}|^2 }{ \cos\delta_{\rm 1st}-M_{11}}.
\label{eq:2nd}
\eeq
The ``pinched'' points occur very near the pole where $|M_{11}| \gg 1$ and
then we can approximate the equation by setting $\cos\delta_{\rm 1st}=0$
and solve for $\delta_{\rm 2nd}$.
Using above-mentioned $A_{1g}$ and $T_{1u}$ as examples, we plot in Fig.~\ref{fig:Oh2nd}  $\cos\delta_{\rm 2nd}$ extracted this way for $l=4$ and $l=3$. We see that the exception points discussed in Figs.~\ref{fig:Oh-A1g} and~\ref{fig:Oh-T1u}  fall on the curve for the infinite-volume phase shift of the second partial wave.
This suggests that $\delta_{\rm 1st}(k)$ and $\delta_{\rm 2nd}(k)$ can be separately isolated by considering the QC at orders 1 and 2 respectively. The values obtained for $\delta_{\rm 2nd}(k)$ can be further checked by comparing with the predictions from other irreps which have it as the lowest partial wave. 
These results demonstrate that exception points at the free-particle poles are sensitive to the second partial wave. If such points are encountered in lattice QCD simulations, then they can be either neglected in the prediction of the lowest partial wave, or be utilized to give an estimate for the second partial wave in the QC.
The recipe is not fail proof. We see that two exception points in $T_{1u}$ do not fall on the curve. It happens when a nearly degenerate pair (levels 10 and 11, or levels 20 and 21) is pinched at the free-particle poles. It indicates the possible influence of the third partial wave $l=5$ in the QC. 
Similar trends are observed in elongated box.
Additional cases can be found in Sec.~3 of the Supplemental Material~\cite{supp}.

\begin{figure}[b]
\includegraphics[scale=0.4]{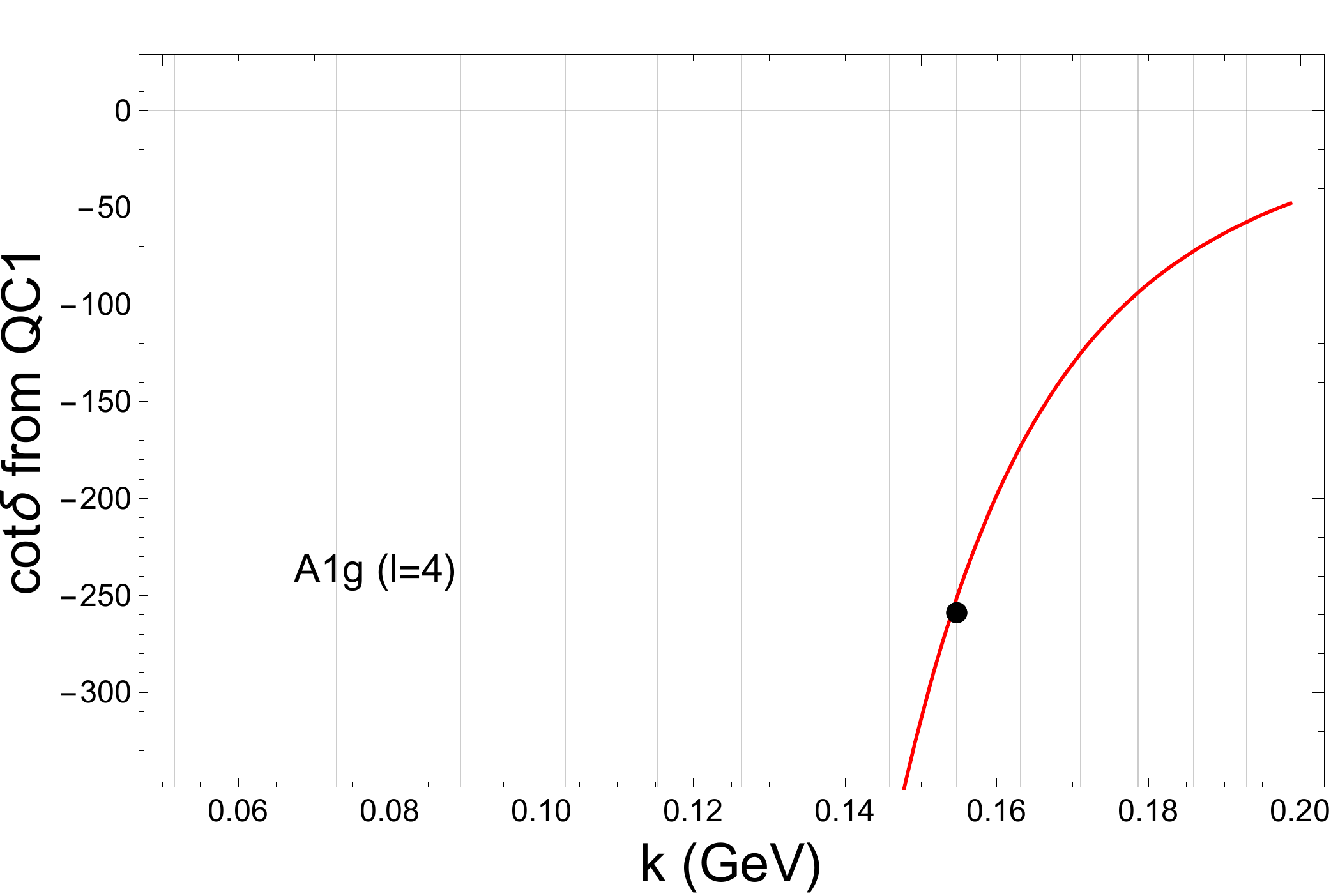}
\includegraphics[scale=0.4]{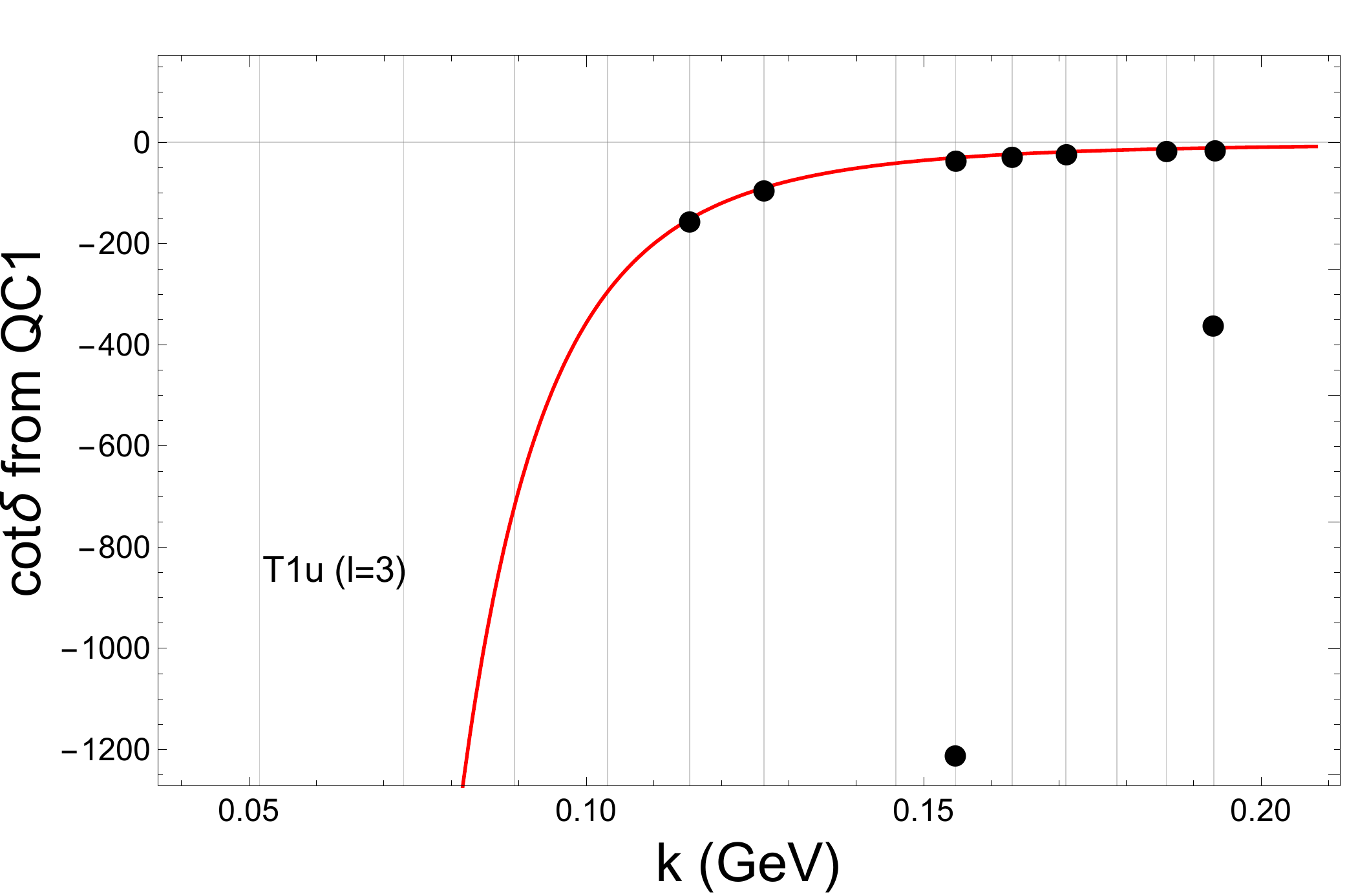}
\caption{Second partial wave in $A_{1g}$ (top) and $T_{1u}$ (bottom) of cubic box. The black points are $\cos\delta_{\rm 2nd}$ in Eq.\eqref{eq:2nd} with $\cos\delta_{\rm 1st}$ neglected, evaluated at the pinched box levels from order 1. The red curve is the infinite-volume $\cos\delta_{\rm 2nd}$. The faint vertical lines are the free-particle poles.
}
\label{fig:Oh2nd}
\end{figure}
%

\section{Conclusion and outlook}
\label{sec:con}

We derived higher-order L\"uscher QCs for scattering of two spinless particles of unequal masses.
Our results were checked numerically by comparing the QC predictions with the spectrum of two-particle states in a box computed
by solving the Schr\"odinger equation.
This is done using a simple potential model in nonrelativistic quantum mechanics. Both the phase shifts in infinite volume and energy levels in finite volume are independently generated in a well-controlled fashion. 
Here is a summary of our findings.

\begin{enumerate}[1)]
\item
We considered a variety of scenarios: rest frame and four moving frames, cubic and elongated geometries. In total, we examined 22 QCs in cubic box and 23 QCs in elongated box. 
The five lowest partial waves in each QC are examined. In some cases, up to $l=5$. 
Matrix elements for all the QCs are given in Appendix~\ref{sec:ME}. Some of the QCs are rederived to include higher partial waves, others are new.
Generically, we expect the QCs to be valid up to terms which vanish exponentially with the box size.

\item
We choose the potential and the box-size so that the systematics associated with finite-volume are negligible, on one hand,
and on the other the results are sensitive to partial waves as high as $\ell=5$. This allows us to
provide very stringent tests for our results.
The numerical checks are done at high precision (to six decimal digits, or differences of 1 eV resolved out of 1 MeV). 
Up to CM momentum $k=0.2$ GeV and up to 40 levels are examined for each of the QCs.
All convergence data, along with noninteracting levels and other information, are provided in the Supplemental Material~\cite{supp}. 

\item
We found sensitivity to the second lowest partial wave in selected QCs through ``pinched" levels which coincide with free-particle poles. The sensitivity can be used to provide an approximate phase shift for the second lowest partial wave despite the presence of the lowest one in a particular channel, but this must be determined  on a case by case basis.
If such levels are encountered in lattice QCD simulations, they can be either ignored or used to estimate the second partial wave.

\item
For the most part, we find elongated boxes work just as well as cubic ones.  This bodes well for using elongated boxes 
as a cost-effective way of varying the kinematic range with a modest increase in the lattice volume. 

\item
Boosting of the two-particle system in both cubic and elongated boxes allows lower energies to be accessed, thus a wider coverage. We considered four basic types of moving frames $d=(0,0,1)$, $(1,1,0)$, $(1,1,1)$, and $(0,1,2)$. Rules for going higher in momentum are given in Table~\ref{tab:boostAll}.
The trade-off for lower energy is the loss of parity which means more mixing of partial waves. 

\item
The effort is already paying dividends. For example, we checked the integer-$J$ QCs in Ref.~\cite{Gockeler:2012yj} for $d=(1,1,0)$ and $d=(1,1,1)$ and found agreement with ours, despite having different forms due to different basis vectors. Those QCs are only given for up to $l=2$. Here we extend up to $l=4$. We also checked against $C_{3v}$ up to $l=4$ from an independent source~\cite{Colin2021} and found agreement.
We also found a few typos in the QCs included in Ref.~\cite{Lee:2017igf}. 
We also checked against all the expressions up to $l=4$ for spinless particles of equal mass at total zero momentum in nonelongated boxes given in Ref.~\cite{Luscher:1990ux} by setting $m_1=m_2$ in our expressions and found agreement.

\item
The numerical check is designed to be transparent and computationally inexpensive.  The entire calculation can be done on a laptop.
\end{enumerate}

For outlook, we envision the following possibilities.
\begin{enumerate}[1)]
\item
The QCs can only be used to extract phase shifts from energy only for the lowest partial waves in each irrep. The predictions are affected by 
cutting off all the higher partial waves. The severity is not known {\it a priori} and it depends on the box geometry and the total momentum of the state. The problem can be turned on its head: can we extract the higher partial waves by 
considering multiple QCs simultaneously?  We have seen in limited cases that higher partial waves can be isolated in a single QC despite the presence of a lower one. Is there a systematic approach, taking advantage of multiple irreps, moving frames, and box size? Thus far this was done using various parametrizations of the scattering amplitude.

\item
Our results are based on a simple repulsive Gaussian potential. This was appropriate for our goal here, which was to check our derivation of the quantization conditions. We note that the same methodology could be easily applied for other potentials, if there is a physical problem that requires calculation of the two-particle spectrum in a finite box.
Any interaction potential can be used in this approach, including potentials given in numerical form or nonlocal potentials $V(r, \bm p)$ where  $\bm p$ can be treated as finite differences on the lattice.
We note also that the QCs are general to any spinless two-body system below the inelastic threshold in finite volume, not just those relevant to nuclear and particle physics.

\item
The formalism can be used to study the finite-volume effects in lattice QCD simulation of physical systems, 
such as the the magnitude of exponential
finite-volume effects ignored by the QCs by considering a smaller box (3.5 to 6 fm); the effect of the range of the model potential; and/or the finite-volume spectrum in the presence
of shallow bound states. Even using the na\"ive $O(a^2)$ discretization to study the influence of cutoff effects on the extracted finite-volume energies could be interesting.

\item
The formalism can be applied with minimal modification to systems with two integer spins.
The same is true for two spin-1/2 particles, such as nucleon-nucleon scattering. 
There is a plethora of NN interaction potentials to work with. 

\item
Another direction is the extension to systems with total half-integer $J$, such as a spin-0 particle and a spin-1/2  particle.
Important physical systems can be studied, a classic example being the delta resonance in pion-nucleon scattering. This is more challenging.  Group theory for double-cover groups are involved for half-integer total angular momentum. 
The QCs for half-integer $J$ should also be checked since they are even more involved than the ones for integer spin. 
For this case the Hamiltonian must be modified to include spin-orbit coupling. 
The spin-orbit interaction has been studied in the continuum but finite-volume using a Fourier basis approach~\cite{Lee:2020fbo}. 
A demonstration on the lattice like the one in this work would be desirable. 
\end{enumerate}

\acknowledgements

We thank Colin Morningstar 
for helpful communications.
This work is supported in part by the U.S. Department of Energy Grant No. DE-FG02-95ER40907. 

\clearpage
\newpage
\appendix

\newpage
\section{Group theory details}
\label{sec:group}
The different scenarios discussed in the main text are classified by their symmetries which can be treated systematically by group theory.
In this appendix, we provide a reasonably self-contained description of 
the group theoretic ingredients needed in this work.

\subsection{Cubic box} 
\begin{figure}
\includegraphics[scale=0.3,angle=0]{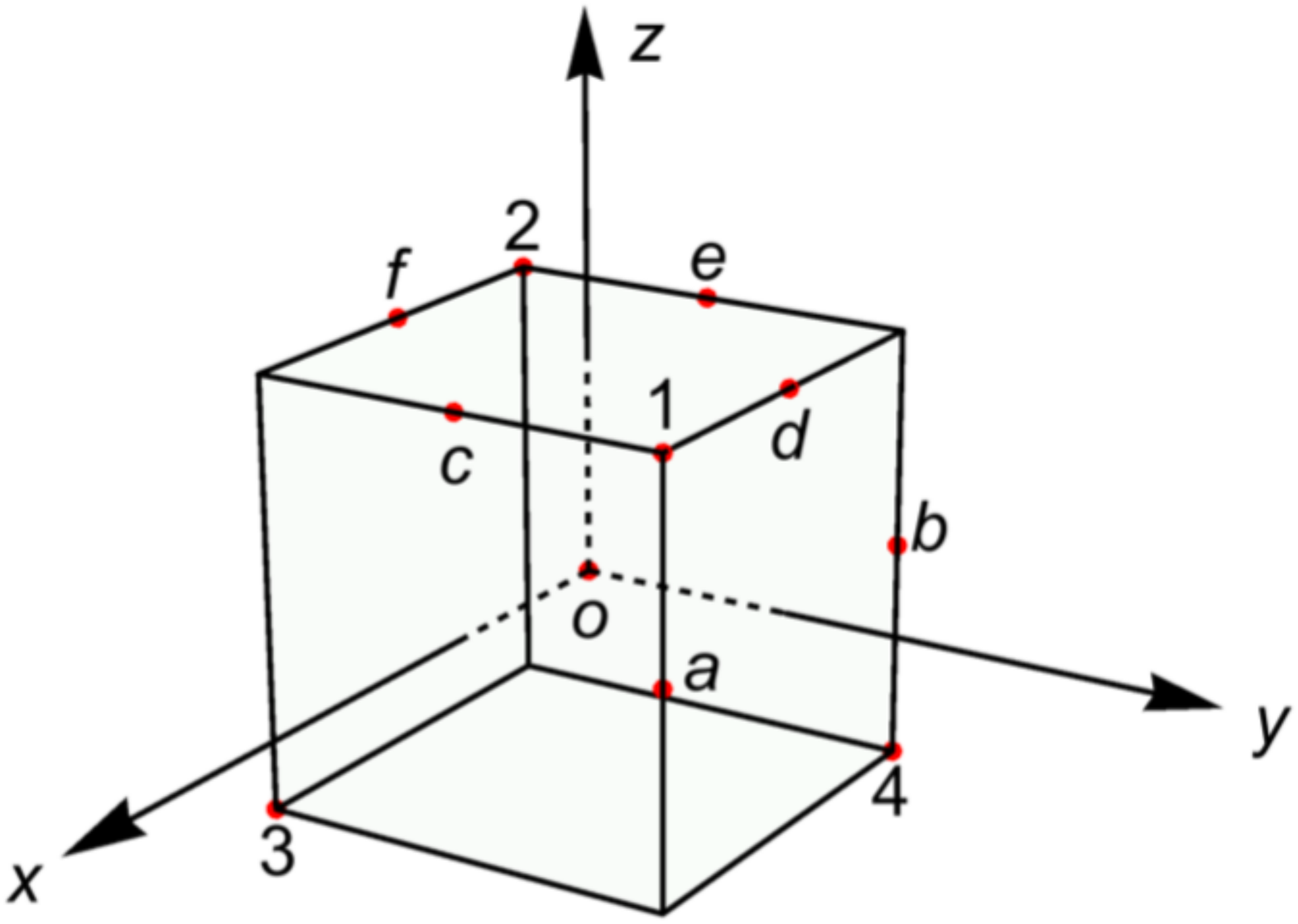}
\caption{The 24 symmetry operations that form the octahedral group $O$ in cubic box.
They are divided into five conjugacy classes: 
the identity ($\mathbb{E}$);  six $\pi/2$ rotations about the three axes (6$C_4$); 
three $\pi$ rotations about the three axes (3$C_2$);  
eight $2\pi/3$ rotations about four body diagonals denoted by 1, 2, 3, 4 (8$C_3$); 
and six $\pi$ rotations about axes  parallel to six face diagonals denoted by $a, b, c, d, e, f$ (6$C'_2$). 
The operations are performed in a right-hand way with the thumb pointing from the center to the various symmetry points. 
}
\label{fig:box_cubic}
\end{figure}

For cubic lattices, the symmetry is depicted in Fig.~\ref{fig:box_cubic}.
The symmetry group is called the octahedral point group $O$ is a finite subgroup of the continuum rotation group $SO(3)$.
It has 24 elements and 5 irreps 
commonly called $A_1$, $A_2$, $E$, $T_1$, and $T_2$ having the respective dimensonality 1,1,2,3, and 3.
The $O$ group is sufficient in describing integral angular momentum in cubic box.
For half-integral angular momentum, its double-covered group $^2O$ is needed.
It is a general group property that the number of irreps is equal to the number of conjugacy classes.
Another property is that the square of irrep dimensions sum to the total number of elements. 

For systems at rest, parity (space inversion) is another symmetry in the box. In this case we need the symmetry group 
$O_h=O\otimes\{ \mathbb{E} , I \}$ which is a direct product with the inversion group (two elements: identity $\mathbb{E}$ and the inversion operation $i$).
The $O_h$ group has 48 elements and 10 irreps that split into two branches:
five with even parity labeled by a plus sign (or {\it g} for {\it gerade}), 
five with odd parity labeled by a minus sign (or {\it u} for {\it ungerade}).
Full details of the $O_h$ group are given in Table~\ref{tab:irrepOh}.
The original $O$ group is embedded in the upper left quadrant of the table.
The table shows how to construct $O_h$ from $O$.
First, double the number of elements by adding a copy of $O$ to the bottom, labelling them as $I1$ to $I24$ so they have one to one correspondence with the original. 
The new elements can also be named by adding the letter $I$ in front of the original names. For example, $C_{2z}$ to  $IC_{2z}$, or one can use the traditional name $\sigma_{z}$ to signify a mirror reflection (rotation about z axis by $\pi/2$ followed by space inversion).
Second, double the number of irrep columns by adding a copy to the right, rename the original as ($A_{1g}$, $A_{2g}$, $E_g$, $T_{1g}$, $T_{2g}$) and the new ones as ($A_{1u}$, $A_{2u}$, $E_u$, $T_{1u}$, and $T_{2u}$). 
Third, change the sign of the rotation matrices $S_k$ and the representations in the added rows.
The rotations in the added rows are also known as improper rotations because the operations also include inversion.
One way to tell whether a rotation matrix is proper or improper is by its determinant: a proper one has $\det[S]=1$, an improper one $\det [S]=-1$.
The $\bm n$, $\omega$, and $ \{\alpha,\beta,\gamma\}$ columns stay the same.
This is practically what it means by $O_h=O\otimes\{ \mathbb{E} , I \}$. 

Operationally, the representations for the $O$ group are as follows.
$A_1$ is the trivial representation, that assigns $1$ to every group element.  The $T_1$ is the three-dimensional representation corresponding to the transformations of vector ($x, y, z$) under rotations whose matrices are  generated via
\beq
t_k=e^{-i(\bm n\cdot\bm J) \omega_k}, \quad k=1,\cdots, 24,
\eeq 
where $(J_k)_{ij}=i\epsilon_{ijk}$. 
For the remaining irreps, a sign change in the character of elements 13 to 24 connects $A_2$ to $A_1$,  $T_2$ to $T_1$, respectively.
The $E$ is a real valued, two-dimensional irrep whose matrices can be obtained from the fact that 
it has Cartesian basis vectors $\sqrt{3}(x^2-y^2)$ and $2z^2-x^2-y^2$.
The five distinct matrices in the table are given by~\cite{Lee:2017igf}. 
\scriptsize
\beqs
& e_1=\left(\begin{array}{cc}
 -1 & 0 \\
 0 & 1 \\
\end{array}\right), 
e_2=\left(\begin{array}{cc}
 -\frac{1}{2} & -\frac{\sqrt{3}}{2} \\
 \frac{\sqrt{3}}{2} & -\frac{1}{2} \\
\end{array}\right), 
 e_3=\left(\begin{array}{cc}
 -\frac{1}{2} & \frac{\sqrt{3}}{2} \\
 -\frac{\sqrt{3}}{2} & -\frac{1}{2} \\
\end{array}\right), \\
& e_4=\left(
\begin{array}{cc}
 \frac{1}{2} & -\frac{\sqrt{3}}{2} \\
 -\frac{\sqrt{3}}{2} & -\frac{1}{2} \\
\end{array}\right), 
 e_5=\left(\begin{array}{cc}
 \frac{1}{2} & \frac{\sqrt{3}}{2} \\
 \frac{\sqrt{3}}{2} & -\frac{1}{2} \\
\end{array}\right).
\eeqs
\normalsize
The character table is implied in the table: for one-dimensional irreps character is simply 1 or -1; for multidimensional irreps character is the trace of the representation matrix.
Note that the rotation angle $\omega$ is defined over $4\pi$ and the Euler angles over
$0\le \alpha < 2\pi$,  $0\le \beta \le \pi$,  $0\le\gamma< 4\pi$, supplemented by the condition that 
$\alpha=0$ when $\beta=0$ or $\pi$ (see~\cite{Chen:2002,Lee:2017igf}). 
This matters when double groups are involved. For the single groups considered in this work, they are equivalent to the traditional Euler angles defined over $2\pi$.  For notational purposes, we denote the operation matrices by $S_k$ and the representation matrices by $D^\Gamma(g)$ for irrep $\Gamma$. 
 Note that $S_k$ and $D^{T_{1u} }(k)$  happen to coincide for the $O$ group.
The full content of the table is used for various aspects to be discussed below.

There are more systematic ways of finding the representation matrices for the irreducible representations in a finite group.
One way is to use the multiplication table for the group (see for example Ref.~\cite{Morningstar:2013bda}).
Another way is known as the Dixon method~\cite{Dixon1970}. A complete set of irreducible unitary representations can be constructed from a single faithful representation.
The method is fairly general: it can find how many conjugacy classes (irreps) and with what dimensionality a new group has, along with its representation matrices. The algorithm is much simpler if the number of irreps and the dimensionalities are known, as is the case for the point groups considered in this work. We applied the simpler version to find the representations in a few cases, taking the $3\times 3$ rotation matrices of group elements as a faithful representation.
After the representation matrices are found, character tables and multiplication tables can be constructed and compared with published versions as a cross check.

\begin{figure}
\includegraphics[scale=0.3,angle=0]{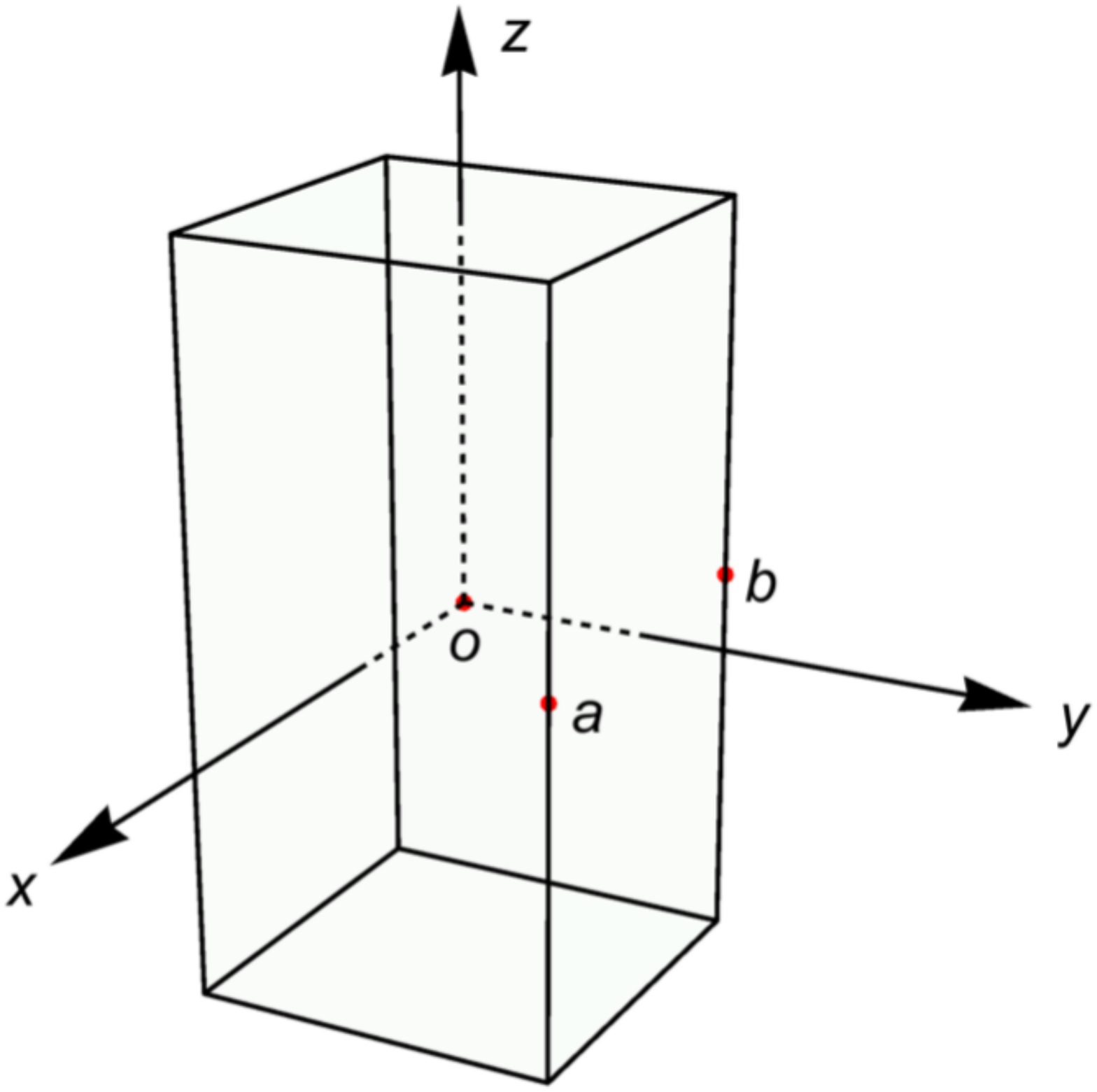}
\caption{The eight symmetry operations that form the dihedral group $D_4$ in elongated box whose dimensions are $L\times L \times \eta L$ where $\eta$ is the elongation factor in the $z$ direction. 
They are divided into five conjugacy classes: 
the identity ($\mathbb{E}$), two $\pi/2$ rotations about the $z$ axis (2$C_4$), 
one $\pi$ rotation about the $z$ axis ($C_2$),  two $\pi$ rotations about $x$ and $y$ axes (2$C'_2$), and  two $\pi$  rotations about the two diagonals in the $xy$ plane denoted by $Oa$ and $Ob$ (2$C^{\prime\prime}_2$). 
}
\label{fig:box_elongated}
\end{figure}
\begin{table*}
\caption{Group table of $O_h$ for rest frame in cubic box. 
The inversion is labeled by preceding with the letter $I$ in the $k$ and $O_{h}$ columns. The matrices $e_k$, $t_k$ are discussed in the text.}
\label{tab:irrepOh}
              
$                                
\renewcommand{\arraystretch}{1.15}
\arraycolsep=6pt
\begin{array}{c | c c c c c c c c c c | c c c c c}

\toprule
k   & O_h          & \bm n       & \omega   & S_k      & \{\alpha,\beta,\gamma\}     
& A_{1g} & A_{2g} & E_g   & T_{1g}    & T_{2g}     & A_{1u} & A_{2u} & E_u   & T_{1u}    & T_{2u}      \\
\hline
 1 & \mathbb{E} & \{0,0,1\} & 4 \pi  & t_1 & \{0,0,0\} & 1 & 1 & \mathds{1}   & t_1 & t_1 & 1 & 1 & \mathds{1}   & t_1 & t_1 \\
 2 & C_{2x} & \{1,0,0\} & \pi  & t_2 & \{0,\pi ,\pi \} & 1 & 1 & \mathds{1}   & t_2 & t_2 & 1 & 1 & \mathds{1}   & t_2 & t_2 \\
 3 & C_{2y} & \{0,1,0\} & \pi  & t_3 & \{0,\pi ,0\} & 1 & 1 & \mathds{1}   & t_3 & t_3 & 1 & 1 & \mathds{1}   & t_3 & t_3 \\
 4 & C_{2z} & \{0,0,1\} & \pi  & t_4 & \{0,0,\pi \} & 1 & 1 & \mathds{1}   & t_4 & t_4 & 1 & 1 & \mathds{1}   & t_4 & t_4 \\
 5 & C_{31}^+ & \{1,1,1\} & \frac{2 \pi }{3} & t_5 & \left\{0,\frac{\pi }{2},\frac{\pi }{2}\right\} & 1 & 1 & e_3 & t_5 & t_5 & 1 & 1
   & e_3 & t_5 & t_5 \\
 6 & C_{32}^+ & \{-1,-1,1\} & \frac{2 \pi }{3} & t_6 & \left\{\pi ,\frac{\pi }{2},\frac{7 \pi }{2}\right\} & 1 & 1 & e_3 & t_6 & t_6 &
   1 & 1 & e_3 & t_6 & t_6 \\
 7 & C_{33}^+ & \{1,-1,-1\} & \frac{2 \pi }{3} & t_7 & \left\{\pi ,\frac{\pi }{2},\frac{5 \pi }{2}\right\} & 1 & 1 & e_3 & t_7 & t_7 &
   1 & 1 & e_3 & t_7 & t_7 \\
 8 & C_{34}^+ & \{-1,1,-1\} & \frac{2 \pi }{3} & t_8 & \left\{0,\frac{\pi }{2},\frac{7 \pi }{2}\right\} & 1 & 1 & e_3 & t_8 & t_8 & 1
   & 1 & e_3 & t_8 & t_8 \\
 9 & C_{31}^- & \{1,1,1\} & \frac{10 \pi }{3} & t_9 & \left\{\frac{\pi }{2},\frac{\pi }{2},3 \pi \right\} & 1 & 1 & e_2 & t_9 & t_9 &
   1 & 1 & e_2 & t_9 & t_9 \\
 10 & C_{32}^- & \{-1,-1,1\} & \frac{10 \pi }{3} & t_{10} & \left\{\frac{3 \pi }{2},\frac{\pi }{2},2 \pi \right\} & 1 & 1 & e_2 &
   t_{10} & t_{10} & 1 & 1 & e_2 & t_{10} & t_{10} \\
 11 & C_{33}^- & \{1,-1,-1\} & \frac{10 \pi }{3} & t_{11} & \left\{\frac{\pi }{2},\frac{\pi }{2},0\right\} & 1 & 1 & e_2 & t_{11} &
   t_{11} & 1 & 1 & e_2 & t_{11} & t_{11} \\
 12 & C_{34}^- & \{-1,1,-1\} & \frac{10 \pi }{3} & t_{12} & \left\{\frac{3 \pi }{2},\frac{\pi }{2},3 \pi \right\} & 1 & 1 & e_2 &
   t_{12} & t_{12} & 1 & 1 & e_2 & t_{12} & t_{12} \\
 13 & C_{4x}^+ & \{1,0,0\} & \frac{\pi }{2} & t_{13} & \left\{\frac{3 \pi }{2},\frac{\pi }{2},\frac{5 \pi }{2}\right\} & 1 & -1 & e_4
   & t_{13} & -t_{13} & 1 & -1 & e_4 & t_{13} & -t_{13} \\
 14 & C_{4y}^+ & \{0,1,0\} & \frac{\pi }{2} & t_{14} & \left\{0,\frac{\pi }{2},0\right\} & 1 & -1 & e_5 & t_{14} & -t_{14} & 1 & -1 &
   e_5 & t_{14} & -t_{14} \\
 15 & C_{4z}^+ & \{0,0,1\} & \frac{\pi }{2} & t_{15} & \left\{0,0,\frac{\pi }{2}\right\} & 1 & -1 & e_1 & t_{15} & -t_{15} & 1 & -1 &
   e_1 & t_{15} & -t_{15} \\
 16 & C_{4x}^- & \{1,0,0\} & \frac{7 \pi }{2} & t_{16} & \left\{\frac{\pi }{2},\frac{\pi }{2},\frac{7 \pi }{2}\right\} & 1 & -1 & e_4
   & t_{16} & -t_{16} & 1 & -1 & e_4 & t_{16} & -t_{16} \\
 17 & C_{4y}^- & \{0,1,0\} & \frac{7 \pi }{2} & t_{17} & \left\{\pi ,\frac{\pi }{2},3 \pi \right\} & 1 & -1 & e_5 & t_{17} & -t_{17} &
   1 & -1 & e_5 & t_{17} & -t_{17} \\
 18 & C_{4z}^- & \{0,0,1\} & \frac{7 \pi }{2} & t_{18} & \left\{0,0,\frac{7 \pi }{2}\right\} & 1 & -1 & e_1 & t_{18} & -t_{18} & 1 &
   -1 & e_1 & t_{18} & -t_{18} \\
 19 & C_{2a} & \{1,1,0\} & \pi  & t_{19} & \left\{0,\pi ,\frac{\pi }{2}\right\} & 1 & -1 & e_1 & t_{19} & -t_{19} & 1 & -1 & e_1 &
   t_{19} & -t_{19} \\
 20 & C_{2b} & \{-1,1,0\} & \pi  & t_{20} & \left\{0,\pi ,\frac{7 \pi }{2}\right\} & 1 & -1 & e_1 & t_{20} & -t_{20} & 1 & -1 & e_1 &
   t_{20} & -t_{20} \\
 21 & C_{2c} & \{1,0,1\} & \pi  & t_{21} & \left\{0,\frac{\pi }{2},\pi \right\} & 1 & -1 & e_5 & t_{21} & -t_{21} & 1 & -1 & e_5 &
   t_{21} & -t_{21} \\
 22 & C_{2d} & \{0,1,1\} & \pi  & t_{22} & \left\{\frac{\pi }{2},\frac{\pi }{2},\frac{\pi }{2}\right\} & 1 & -1 & e_4 & t_{22} &
   -t_{22} & 1 & -1 & e_4 & t_{22} & -t_{22} \\
 23 & C_{2e} & \{-1,0,1\} & \pi  & t_{23} & \left\{\pi ,\frac{\pi }{2},0\right\} & 1 & -1 & e_5 & t_{23} & -t_{23} & 1 & -1 & e_5 &
   t_{23} & -t_{23} \\
  24 & C_{2f} & \{0,-1,1\} & \pi  & t_{24} & \left\{\frac{3 \pi }{2},\frac{\pi }{2},\frac{7 \pi }{2}\right\} & 1 & -1 & e_4 & t_{24} &
   -t_{24} & 1 & -1 & e_4 & t_{24} & -t_{24} \\
     \hline
 I1 & I\mathbb{E} & \{0,0,1\} & 4 \pi  & -t_1 & \{0,0,0\} & 1 & 1 & \mathds{1}   & t_1 & t_1 & -1 & -1 & -\mathds{1}   & -t_1 & -t_1 \\
 I2 & IC_{2x} & \{1,0,0\} & \pi  & -t_2 & \{0,\pi ,\pi \} & 1 & 1 & \mathds{1}   & t_2 & t_2 & -1 & -1 & -\mathds{1}   & -t_2 & -t_2 \\
 I3 & IC_{2y} & \{0,1,0\} & \pi  & -t_3 & \{0,\pi ,0\} & 1 & 1 & \mathds{1}   & t_3 & t_3 & -1 & -1 & -\mathds{1}   & -t_3 & -t_3 \\
 I4 & IC_{2z} & \{0,0,1\} & \pi  & -t_4 & \{0,0,\pi \} & 1 & 1 & \mathds{1}   & t_4 & t_4 & -1 & -1 & -\mathds{1}   & -t_4 & -t_4 \\
 I5 & IC_{31}^+ & \{1,1,1\} & \frac{2 \pi }{3} & -t_5 & \left\{0,\frac{\pi }{2},\frac{\pi }{2}\right\} & 1 & 1 & e_3 & t_5 & t_5 & -1
   & -1 & -e_3 & -t_5 & -t_5 \\
 I6 & IC_{32}^+ & \{-1,-1,1\} & \frac{2 \pi }{3} & -t_6 & \left\{\pi ,\frac{\pi }{2},\frac{7 \pi }{2}\right\} & 1 & 1 & e_3 & t_6 &
   t_6 & -1 & -1 & -e_3 & -t_6 & -t_6 \\
 I7 & IC_{33}^+ & \{1,-1,-1\} & \frac{2 \pi }{3} & -t_7 & \left\{\pi ,\frac{\pi }{2},\frac{5 \pi }{2}\right\} & 1 & 1 & e_3 & t_7 &
   t_7 & -1 & -1 & -e_3 & -t_7 & -t_7 \\
 I8 & IC_{34}^+ & \{-1,1,-1\} & \frac{2 \pi }{3} & -t_8 & \left\{0,\frac{\pi }{2},\frac{7 \pi }{2}\right\} & 1 & 1 & e_3 & t_8 & t_8 &
   -1 & -1 & -e_3 & -t_8 & -t_8 \\
 I9 & IC_{31}^- & \{1,1,1\} & \frac{10 \pi }{3} & -t_9 & \left\{\frac{\pi }{2},\frac{\pi }{2},3 \pi \right\} & 1 & 1 & e_2 & t_9 & t_9
   & -1 & -1 & -e_2 & -t_9 & -t_9 \\
 I10 & IC_{32}^- & \{-1,-1,1\} & \frac{10 \pi }{3} & -t_{10} & \left\{\frac{3 \pi }{2},\frac{\pi }{2},2 \pi \right\} & 1 & 1 & e_2 &
   t_{10} & t_{10} & -1 & -1 & -e_2 & -t_{10} & -t_{10} \\
 I11 & IC_{33}^- & \{1,-1,-1\} & \frac{10 \pi }{3} & -t_{11} & \left\{\frac{\pi }{2},\frac{\pi }{2},0\right\} & 1 & 1 & e_2 & t_{11} &
   t_{11} & -1 & -1 & -e_2 & -t_{11} & -t_{11} \\
 I12 & IC_{34}^- & \{-1,1,-1\} & \frac{10 \pi }{3} & -t_{12} & \left\{\frac{3 \pi }{2},\frac{\pi }{2},3 \pi \right\} & 1 & 1 & e_2 &
   t_{12} & t_{12} & -1 & -1 & -e_2 & -t_{12} & -t_{12} \\
 I13 & IC_{4x}^+ & \{1,0,0\} & \frac{\pi }{2} & -t_{13} & \left\{\frac{3 \pi }{2},\frac{\pi }{2},\frac{5 \pi }{2}\right\} & 1 & -1 &
   e_4 & t_{13} & -t_{13} & -1 & 1 & -e_4 & -t_{13} & t_{13} \\
 I14 & IC_{4y}^+ & \{0,1,0\} & \frac{\pi }{2} & -t_{14} & \left\{0,\frac{\pi }{2},0\right\} & 1 & -1 & e_5 & t_{14} & -t_{14} & -1 & 1
   & -e_5 & -t_{14} & t_{14} \\
 I15 & IC_{4z}^+ & \{0,0,1\} & \frac{\pi }{2} & -t_{15} & \left\{0,0,\frac{\pi }{2}\right\} & 1 & -1 & e_1 & t_{15} & -t_{15} & -1 & 1
   & -e_1 & -t_{15} & t_{15} \\
 I16 & IC_{4x}^- & \{1,0,0\} & \frac{7 \pi }{2} & -t_{16} & \left\{\frac{\pi }{2},\frac{\pi }{2},\frac{7 \pi }{2}\right\} & 1 & -1 &
   e_4 & t_{16} & -t_{16} & -1 & 1 & -e_4 & -t_{16} & t_{16} \\
 I17 & IC_{4y}^- & \{0,1,0\} & \frac{7 \pi }{2} & -t_{17} & \left\{\pi ,\frac{\pi }{2},3 \pi \right\} & 1 & -1 & e_5 & t_{17} &
   -t_{17} & -1 & 1 & -e_5 & -t_{17} & t_{17} \\
 I18 & IC_{4z}^- & \{0,0,1\} & \frac{7 \pi }{2} & -t_{18} & \left\{0,0,\frac{7 \pi }{2}\right\} & 1 & -1 & e_1 & t_{18} & -t_{18} & -1
   & 1 & -e_1 & -t_{18} & t_{18} \\
 I19 & IC_{2a} & \{1,1,0\} & \pi  & -t_{19} & \left\{0,\pi ,\frac{\pi }{2}\right\} & 1 & -1 & e_1 & t_{19} & -t_{19} & -1 & 1 & -e_1 &
   -t_{19} & t_{19} \\
 I20 & IC_{2b} & \{-1,1,0\} & \pi  & -t_{20} & \left\{0,\pi ,\frac{7 \pi }{2}\right\} & 1 & -1 & e_1 & t_{20} & -t_{20} & -1 & 1 &
   -e_1 & -t_{20} & t_{20} \\
 I21 & IC_{2c} & \{1,0,1\} & \pi  & -t_{21} & \left\{0,\frac{\pi }{2},\pi \right\} & 1 & -1 & e_5 & t_{21} & -t_{21} & -1 & 1 & -e_5 &
   -t_{21} & t_{21} \\
 I22 & IC_{2d} & \{0,1,1\} & \pi  & -t_{22} & \left\{\frac{\pi }{2},\frac{\pi }{2},\frac{\pi }{2}\right\} & 1 & -1 & e_4 & t_{22} &
   -t_{22} & -1 & 1 & -e_4 & -t_{22} & t_{22} \\
 I23 & IC_{2e} & \{-1,0,1\} & \pi  & -t_{23} & \left\{\pi ,\frac{\pi }{2},0\right\} & 1 & -1 & e_5 & t_{23} & -t_{23} & -1 & 1 & -e_5
   & -t_{23} & t_{23} \\
 I24 & C_{2f} & \{0,-1,1\} & \pi  & -t_{24} & \left\{\frac{3 \pi }{2},\frac{\pi }{2},\frac{7 \pi }{2}\right\} & 1 & -1 & e_4 & t_{24}
   & -t_{24} & -1 & 1 & -e_4 & -t_{24} & t_{24} \\
 \bottomrule
\end{array}
$  

\end{table*}
\begin{table*}
\caption{Group table of  $D_{4h}$ for rest frame in elongated box. The inversion is labeled by preceding with the letter $I$ in the $k$ and $O_{h}$ columns. The representations of $E$ are given in terms of Pauli matrices.}
\label{tab:irrepD4h}
              
$                                
\renewcommand{\arraystretch}{1.15}
\arraycolsep=6pt
\begin{array}{c | c c c c c c c c c c | c c c c c}

\toprule
k   & D_{4h}          & \bm n       & \omega   & S_k      & \{\alpha,\beta,\gamma\}     
& A_{1g} & A_{2g} & B_{1g}   & B_{2g}    & E_{g}     & A_{1u} & A_{2u} & B_{1u}   & B_{2u}    & E_{u}      \\
\hline
 1 & \mathbb{E} & \{0,0,1\} & 4 \pi  & t_1 & \{0,0,0\} & 1 & 1 & 1 & 1 & \mathds{1}  & 1 & 1 & 1 & 1 & \mathds{1}  \\
 2 & C_{4z}^+ & \{0,0,1\} & \frac{\pi }{2} & t_{15} & \left\{0,0,\frac{\pi }{2}\right\} & 1 & 1 & -1 & -1 & -i \sigma _2 & 1 &
   1 & -1 & -1 & -i \sigma _2 \\
 3 & C_{4z}^- & \{0,0,1\} & \frac{7 \pi }{2} & t_{18} & \left\{0,0,\frac{7 \pi }{2}\right\} & 1 & 1 & -1 & -1 & i \sigma _2 &
   1 & 1 & -1 & -1 & i \sigma _2 \\
 4 & C_{2z} & \{0,0,1\} & \pi  & t_4 & \{0,0,\pi \} & 1 & 1 & 1 & 1 & -\mathds{1}  & 1 & 1 & 1 & 1 & -\mathds{1}  \\
 5 & C_{2x} & \{1,0,0\} & \pi  & t_{2} & \{0,\pi ,\pi \} & 1 & -1 & 1 & -1 & \sigma _3 & 1 & -1 & 1 & -1 & \sigma _3 \\
 6 & C_{2y} & \{0,1,0\} & \pi  & t_{3} & \{0,\pi ,0\} & 1 & -1 & 1 & -1 & -\sigma _3 & 1 & -1 & 1 & -1 & -\sigma _3 \\
 7 & C_{2a} & \{1,1,0\} & \pi  & t_{19} & \left\{0,\pi ,\frac{\pi }{2}\right\} & 1 & -1 & -1 & 1 & \sigma _1 & 1 & -1 & -1
   & 1 & \sigma _1 \\
 8 & C_{2b} & \{-1,1,0\} & \pi  & t_{20} & \left\{0,\pi ,\frac{7 \pi }{2}\right\} & 1 & -1 & -1 & 1 & -\sigma _1 & 1 & -1 &
   -1 & 1 & -\sigma _1 \\
   \hline
 I1 & I\mathbb{E} & \{0,0,1\} & 4 \pi  & -t_1 & \{0,0,0\} & 1 & 1 & 1 & 1 & \mathds{1}  & -1 & -1 & -1 & -1 & -\mathds{1}  \\
 I2 & IC_{4z}^+ & \{0,0,1\} & \frac{\pi }{2} & -t_{15} & \left\{0,0,\frac{\pi }{2}\right\} & 1 & 1 & -1 & -1 & -i \sigma _2 &
   -1 & -1 & 1 & 1 & i \sigma _2 \\
 I3 & IC_{4z}^- & \{0,0,1\} & \frac{7 \pi }{2} & -t_{18} & \left\{0,0,\frac{7 \pi }{2}\right\} & 1 & 1 & -1 & -1 & i \sigma _2
   & -1 & -1 & 1 & 1 & -i \sigma _2 \\
 I4 & IC_{2z} & \{0,0,1\} & \pi  & -t_4 & \{0,0,\pi \} & 1 & 1 & 1 & 1 & -\mathds{1}  & -1 & -1 & -1 & -1 & \mathds{1}  \\
 I5 & IC_{2x} & \{1,0,0\} & \pi  & -t_{2} & \{0,\pi ,\pi \} & 1 & -1 & 1 & -1 & \sigma _3 & -1 & 1 & -1 & 1 & -\sigma _3
   \\
 I6 & IC_{2y} & \{0,1,0\} & \pi  & -t_{3} & \{0,\pi ,0\} & 1 & -1 & 1 & -1 & -\sigma _3 & -1 & 1 & -1 & 1 & \sigma _3 \\
 I7 & IC_{2a} & \{1,1,0\} & \pi  & -t_{19} & \left\{0,\pi ,\frac{\pi }{2}\right\} & 1 & -1 & -1 & 1 & \sigma _1 & -1 & 1 &
   1 & -1 & -\sigma _1 \\
 I8 & IC_{2b} & \{-1,1,0\} & \pi  & -t_{20} & \left\{0,\pi ,\frac{7 \pi }{2}\right\} & 1 & -1 & -1 & 1 & -\sigma _1 & -1 &
   1 & 1 & -1 & \sigma _1 \\
 \bottomrule
\end{array}
$  

\end{table*}
\begin{table}
\caption{Group table of $C_{4v}$ for moving frame $d=(0,0,1)$ in both cubic and elongated boxes.
The corresponding elements in $O_h$ and $D_{4h}$ that preserve the direction are indicated. 
}
\label{tab:C4vgroup}          
              
$                                
\renewcommand{\arraystretch}{1.2}
\arraycolsep=2pt
\begin{array}{ccccccccccc}
\toprule
  C_{4v} &O_h & D_{4h} &\bm n & \omega & \{\alpha,\beta,\gamma\} & A_1 & A_2 & B_1 & B_2 & E\\
  \hline
 \mathbb{E} & 1 & 1 & \{0,0,1\} & 4 \pi  & \{0,0,0\} & 1 & 1 & 1 & 1 & \mathcal{I} \\
 C_{2z} & 4 & 4 & \{0,0,1\} & \pi  & \{0,0,\pi \} & 1 & 1 & 1 & 1 & -\mathcal{I} \\
 C_{4z}^+ & 15 & 2 & \{0,0,1\} & \frac{\pi }{2} & \left\{0,0,\frac{\pi }{2}\right\} & 1 & 1 & -1 & -1 & i \sigma _2 \\
 C_{4z}^- & 18 & 3 & \{0,0,1\} & \frac{7 \pi }{2} & \left\{0,0,\frac{7 \pi }{2}\right\} & 1 & 1 & -1 & -1 & -i \sigma _2
   \\
 IC_{2x} & I2 & I5 & \{1,0,0\} & \pi  & \{0,\pi ,\pi \} & 1 & -1 & 1 & -1 & \sigma _3 \\
 IC_{2y}  & I3 & I6 & \{0,1,0\} & \pi  & \{0,\pi ,0\} & 1 & -1 & 1 & -1 & -\sigma _3 \\
  IC_{2a}  & I19 & I7 & \{1,1,0\} & \pi  & \left\{0,\pi ,\frac{\pi }{2}\right\} & 1 & -1 & -1 & 1 & -\sigma _1 \\
 IC_{2b}  & I20 & I8 & \{-1,1,0\} & \pi  & \left\{0,\pi ,\frac{7 \pi }{2}\right\} & 1 & -1 & -1 & 1 & \sigma _1 \\
     \bottomrule
\end{array}$
                    
\end{table}                     
\begin{table}
\caption{Group table of $C_{3v}$ for moving frame $d=(1,1,1)$ in cubic box only. 
The corresponding elements in $O_h$ that preserve the direction are indicated. 
This moving frame is not allowed in $z$ elongated box.
}
\label{tab:C3vgroup}          
              
$                                
\renewcommand{\arraystretch}{1.2}
\arraycolsep=2pt
\begin{array}{ccccccccc}
\toprule
  C_{3v} &O_h &\bm n & \omega & \{\alpha,\beta,\gamma\} & A_1 & A_2 & E \\
 \hline
 \mathbb{E} & 1 & \{0,0,1\} & 4 \pi  & \{0,0,0\} & 1 & 1 & \mathcal{I} \\
 C_{31}^+ & 5 & \{1,1,1\} & \frac{2 \pi }{3} & \left\{0,\frac{\pi }{2},\frac{\pi }{2}\right\} & 1 & 1 & \frac{1}{2}
   \left(-\mathcal{I}+i \sqrt{3} \sigma _2\right) \\
 C_{31}^- & 9 & \{1,1,1\} & \frac{10 \pi }{3} & \left\{\frac{\pi }{2},\frac{\pi }{2},3 \pi \right\} & 1 & 1 &
   \frac{1}{2} \left(-\mathcal{I}-i \sqrt{3} \sigma _2\right) \\
  IC_{2b}  & I20 & \{-1,1,0\} & \pi  & \left\{0,\pi ,\frac{7 \pi }{2}\right\} & 1 & -1 & \frac{1}{2} \left(\sigma
   _3-\sqrt{3} \sigma _1\right) \\
  IC_{2e}  & I23 & \{-1,0,1\} & \pi  & \left\{\pi ,\frac{\pi }{2},0\right\} & 1 & -1 & -\sigma _3 \\
  IC_{2f}  & I24 & \{0,-1,1\} & \pi  & \left\{\frac{3 \pi }{2},\frac{\pi }{2},\frac{7 \pi }{2}\right\} & 1 & -1 &
   \frac{1}{2} \left(\sqrt{3} \sigma _1+\sigma _3\right) \\
    \bottomrule
\end{array}$                                
                    
\end{table}                     
\begin{table}   
\caption{Group table of $C_{2v}$ for moving frame $d=(1,1,0)$ in both cubic and elongated boxes. 
The corresponding elements in $O_h$ and $D_{4h}$ that preserve the direction are indicated. 
}
\label{tab:C2vgroup}          
              
$                                
\renewcommand{\arraystretch}{1.2}
\arraycolsep=2pt
\begin{array}{cccccccccc}
\toprule
  C_{2v} &O_h & D_{4h} &\bm n & \omega & \{\alpha,\beta,\gamma\} & A_1 & A_2 & B_1 & B_2\\
 \hline
 \mathbb{E} & 1 & 1 & \{0,0,1\} & 4 \pi  & \{0,0,0\} & 1 & 1 & 1 & 1 \\
 C_{2a} & 19 & 7 & \{1,1,0\} & \pi  & \left\{0,\pi ,\frac{\pi }{2}\right\} & 1 & 1 & -1 & -1 \\
  IC_{2b}  & I20 & I8 & \{-1,1,0\} & \pi  & \left\{0,\pi ,\frac{7 \pi }{2}\right\} & 1 & -1 & -1 & 1 \\
  IC_{2z}  & I4 & I4 & \{0,0,1\} & \pi  & \{0,0,\pi \} & 1 & -1 & 1 & -1 \\
  \bottomrule
\end{array}$                                
                    
\end{table}               
\begin{table}[h]
\caption{Group table of $C_{1v}$ for moving frame $d=(0,1,2)$ in cubic and elongated box (top), and $d=(1,1,1)$ in the $z$ elongated box (bottom). The corresponding elements in $O_h$ and $D_{4h}$ that preserve the directions are indicated. 
}
\label{tab:C1vgroup}          
              
$                                
\renewcommand{\arraystretch}{1.2}
\arraycolsep=2pt
\begin{array}{ccccccc}
\toprule
  C_{1v} &O_h\, (D_{4h}) &\bm n & \omega & \{\alpha,\beta,\gamma\} & A_1 & A_2\\
  \mathbb{E} & 1\,(1)  & \{0,0,1\} & 4 \pi  & \{0,0,0\} & 1 & 1 \\
 IC_{2x}  & I2 \,(I5) & \{1,0,0\} & \pi  & \{0,\pi ,\pi \} & 1 & -1 \\[3pt]
\hline
  C_{1v} & D_{4h} &\bm n & \omega & \{\alpha,\beta,\gamma\} & A_1 & A_2\\
  \mathbb{E}  & 1 & \{0,0,1\} & 4 \pi  & \{0,0,0\} & 1 & 1 \\
 IC_{2b}   & I8 &  \{-1,1,0\} & \pi  & \left\{0,\pi ,\frac{7 \pi }{2}\right\}  & 1 & -1 \\
  \bottomrule
\end{array}$                                
                    
\end{table}                     

\subsection{Elongated box} 
For elongated lattices, the symmetry is depicted in Fig.~\ref{fig:box_elongated}.
The discussion is similar to the cubic case, except that the symmetry is reduced
from the octahedral group  $O$ to the dihedral group $D_{4}$ which has eight elements and five irreps.
As far as group operations are concerned, the $D_4$ group is isomorphic to the symmetry of a square.
Including parity the group is $D_{4h}=D_4\otimes\{ \mathbb{E} , I \}$ which has 26 elements and 10 irreps (five even parity, five odd parity)
Full details of the $D_{4h}$ group are given in Table~\ref{tab:irrepD4h}. The construction of $D_{4h}$ from $D_4$ is similar to $O_h$ from $O$. The two-dimensional irrep $E$ is represented by Pauli matrices. It is interesting to note from the rotation elements in $S_k$ that $D_{4h}$ is a subgroup of $O_h$ with 16 common elements ($1,15,18,4, 2,3,19,20, I1,I15,I18,I4, I2,I3,I19,I20$).

\subsection{Moving frames} 
If the two-particle system has nonzero total momentum in the box,
then we say such systems are in moving frames or boosted relative to the box frame (lab frame).
We use $\bm d=(n_x,n_y,n_z)$ with $n_i \in Z^3$ to denote the moving frame in both cubic and elongated boxes, see Eq.\eqref{eq:momenta}. 
A moving frame singles out a special direction $\bm d$ in space so the group symmetry is reduced to the so-called little groups $C_{nv}$. 
Their elements $S_k$ are derived from those of the  $O_h$ and $D_{4h}$ by the general requirement that they must preserve the moving direction $S_k \bm d = \bm d$.
 In this work, we consider four distinct moving frames: $d=(0,0,1)$, $d=(1,1,0)$, $d=(1,1,1)$, and $d=(0,1,2)$, in both cubic and elongated geometries. 
 The little group tables are given in \cref{tab:C4vgroup,tab:C3vgroup,tab:C2vgroup,tab:C1vgroup}.
Moving frames with higher total momentum are covered by the four basic types. The allowed moving frames and the little groups they belong to are summarized in Table~\ref{tab:boostAll} in the main text. If integer multiples of the four distinct $\bm d$ are considered, the little group tables apply without modifications.
On the other hand, if permutations are considered, such as $d=(0,1,1)$ in cubic box, the irreps and representations remain the same, but the elements change to new subsets of $O_h$. Consequently, basis vectors change, thus leading to quantization conditions that look very different. However, physical results (angular momentum, energy levels, and phase shift) from $d=(0,1,1)$ and $d=(1,1,0)$ QCs must be the same due to cubic symmetry, since the two momenta are related by a change of the coordinate system.
This is effectively a rotation of the zeta functions by the Euler angles $(\alpha,\beta,\gamma)=(\pi/2,0,0)$.  
The QC takes a new form, but the roots are the same as the original one. So the same original QC applies to all equivalent permutations.

\subsection{Angular momentum decomposition}
\label{sec:angdecomp}
As an application of the group tables given above, we discuss how angular momentum quantum number is affected by the group symmetries. It also serves as a consistency check of the group properties.

For spherically symmetric interactions the eigenstates of the Hamiltonian in the
infinite volume form multiplets that furnish bases for the irreps of the rotational group $SO(3)$. 
These multiplets are labeled by the angular momentum $l=0, 1, 2,\ldots$.
For both cubic and elongated boxes, these multiplets split into smaller sets that mix under the action of rotations that leave the box invariant, forming the bases for one of the irreps of the group. 
Then the question is: for a given $l$, what irreps are coupled to it? 
To answer this we can decompose the irrep $l$ of the rotation group   
into a direct sum of the irreps of the group,  $l=\bigoplus_\Gamma n(\Gamma, l) \Gamma$,  
where the coefficient is called the multiplicity, which tells how many times irrep $\Gamma$ appears in the given $l$.
This can be calculated by
\beq
n(\Gamma,l) = \frac{1}{g} \sum_k \chi(k,\Gamma) \,  \chi(\omega_k,l),
\label{eq:decom}
\eeq
where $k$ runs through all the elements of the symmetry group and $g$ is the total number of elements in the group. 
Here $\chi(k,\Gamma)$ is the character of element $k$ in irrep $\Gamma$, and 
$\chi(\omega_k,l)$ the character of the rotation group for angular momentum $J$ and rotation angle $\omega_k$.
This can be computed as follows~\cite{Tinkham:1992}.
Any rotation is characterized by a rotation axis and the rotation angle $\omega$. 
Since the character (trace) of the matrix is invariant under similarity transformations the result
will be equal to an equivalent rotation around the $z$ axis (the similarity matrix in this
case is simply a rotation that takes the rotation axis into the $z$ axis).
The character is then the trace of this diagonal matrix
\beq
\chi(\omega,l) =\sum_{m=-l}^{l} e^{-i m \omega} = \frac{\sin[(l+1/2) \omega]}{\sin(\omega/2)}.
\label{eq:charSO3}
\eeq
Note that limits must be taken if division by zero is encountered in evaluating this equation.
Finally, because of mixed parities in the little groups $C_{nv}$, a factor $(-1)^l$ must be inserted into the sum in Eq.\eqref{eq:decom} for elements that contain inversions (or improper rotations). This is so because inversion induces $(-1)^l$ on the spherical harmonics.
The decompositions thus obtained for all the groups considered in this work are summarized in Table~\ref{tab:all}.

\subsection{Basis vectors}
\label{sec:basis}

The irreps of the continuum rotation group with $l = 0, 1, 2, \cdots, \infty$  are defined in the
$(2l+1)$-dimensional space spanned on the basis vectors $|lm\rangle$, which are the standard spherical harmonics.
These representations are reducible under the symmetry group into its irreps $\Gamma$.  
In other words, certain subspaces in the space spanned by $|lm\rangle$ are invariant under the symmetry transformations, 
furnishing irreps for the symmetry group. We construct the basis vectors with the following projection operator,
\beq
P_\Gamma^{\lambda' \lambda}  |lm\rangle = \frac{d_\Gamma}{|G|} \sum_{g\in G} 
[D^\Gamma_{\lambda'\lambda}(g)]^* \sum_{m'=-l}^l D^l_{m'm}(g) |lm'\rangle
\label{eq:wignerD}
\eeq
where $g$ a group element, $d_\Gamma$ the dimensionality of irrep $\Gamma$, $|G|$ the order of the group, $D_\Gamma(g)$ the irreducible representation for irrep $\Gamma$, and $D^l(g)$ the Wigner $D$ matrix evaluated at the Euler angles $\alpha, \beta,\gamma$ of each group element. 
Note that the projected vector coefficients are on the columns of $(P_\Gamma^{\lambda' \lambda} )_{m'm}$.
If we rotate a projected vector from row $\lambda$, then we get
\beq
S(g) P_\Gamma^{\lambda\lambda} |lm\rangle = D^\Gamma_{\lambda'\lambda}(g) P_\Gamma^{\lambda' \lambda} |lm\rangle.
\eeq
So the procedure to find all the basis vectors is as follows.
For a given $l$ and irrep $\Gamma$, we construct the projection matrix for row 1, and then perform a QR decomposition,
$P_\Gamma^{11}=Q^\dagger R$ with $Q$ unitary and $R$ upper triangular.
If matrix $R$ is zero, then  no basis vectors exist for this $l$.
Otherwise, the decomposition has the matrix structure  
$\left[(2l+1)\times(2l+1)\right]=\left[(2l+1)\times k\right] \times \left[k\times (2l+1)\right] $, 
where the rank $k$ reveals the number of linearly independent vectors (multiplicity) for this $l$.  
The $k$ columns of $Q^\dagger\equiv |\Gamma 1 l n \rangle$ with $n=1,\cdots,k$ are the basis vectors corresponding to row 1. If irrep $\Gamma$ is onedimensional, then this is the final vector.
Otherwise, the basis vectors for the remaining rows  are obtained from the first row by
\beq
|\Gamma \lambda' l n \rangle=(P_\Gamma^{\lambda'1}R^{-1})^T \text{ with } \lambda'=2,\cdots,d_\Gamma,
\eeq
where $R^{-1}$ is the pseudoinverse of $R$, that is, $R R^{-1}=I_{k\times k}$.

If a group element contains inversion (or improper rotation), a factor $(-1)^l$ must be inserted in front of the Wigner $D$ function in Eq.\eqref{eq:wignerD} due to the parity of spherical harmonics $|lm\rangle$.
There is freedom to choose the overall phase factor for each $l$ in each irrep.
The basis vectors obtained in this procedure are orthonormal,
\beq
\langle \Gamma' \lambda' l' n' | \Gamma \lambda l n \rangle
= \delta_{\Gamma' \Gamma} \delta_{\lambda'\lambda} \delta_{l'l}\delta_{n'n}.
\eeq
They also satisfy the following property,
\beq
D^\Gamma_{\lambda'\lambda}(g) =\langle \Gamma \lambda' l n | S(g) | \Gamma \lambda l n \rangle
= \left[Q_{\lambda'} D^l(g) Q^\dagger_\lambda\right]_{nn},
\eeq
for all group elements $g$, angular momentum $l$, multiplicity $n$, and irrep $\Gamma$. 
This relation can serve as a strong consistency check of both the irrep representation matrices and the basis vectors.

All the basis vectors used in this work (up to $l=5$) are listed 
in Table~\ref{tab:basisOh} for $O_h$ group in cubic box, 
Table~\ref{tab:basisD4h} for $D_{4h}$ group in elongated box,  
Table~\ref{tab:basisC4v} for the little group $C_{4v}$ for both box geometries, 
Table~\ref{tab:basisC3v}  for the little group $C_{3v}$,  for cubic box only,
Table~\ref{tab:basisC2v}  for the little group $C_{2v}$, for both geometries,
Table~\ref{tab:basisC1v111} for the little group $C_{1v}$ in elongated box only, 
and Table~\ref{tab:basisC1v012} for the little group $C_{1v}$ for both geometries.
The basis vectors are needed to project the quantization condition into block-diagonalized sectors by irrep as explained in Sec.~\ref{sec:block}.

\begin{table*}
\caption{Basis vectors of group $O_h$ in terms of spherical harmonics $Y[l,m]$ up to $l=5$ for rest frame in cubic box. 
Multiplicities are indicated by $n$. Multimultidimensional components are indicated by $\alpha$.}
\label{tab:basisOh}

$      
\renewcommand{\arraystretch}{1.5}
\begin{array}{ccccl}
\toprule
\Gamma & l          & n & \alpha & \text{Basis vectors of  } O_h                       \\
\hline 
 A_{1g} & 0 & 1 & 1 & \text{Y[0,0]} \\
 A_{1g} & 4 & 1 & 1 & \frac{1}{2} \sqrt{\frac{7}{3}} \text{Y[4,0]}+\frac{1}{2} \sqrt{\frac{5}{6}} \text{Y[4,-4]}+\frac{1}{2}
   \sqrt{\frac{5}{6}} \text{Y[4,4]} \\
 A_{2u} & 3 & 1 & 1 & \frac{\text{Y[3,-2]}}{\sqrt{2}}-\frac{\text{Y[3,2]}}{\sqrt{2}} \\
 E_g & 2 & 1 & 1 & \frac{\text{Y[2,-2]}}{\sqrt{2}}+\frac{\text{Y[2,2]}}{\sqrt{2}} \\
 E_g & 2 & 1 & 2 & \text{Y[2,0]} \\
 E_g & 4 & 1 & 1 & \frac{\text{Y[4,-2]}}{\sqrt{2}}+\frac{\text{Y[4,2]}}{\sqrt{2}} \\
 E_g & 4 & 1 & 2 & -\frac{1}{2} \sqrt{\frac{5}{3}} \text{Y[4,0]}+\frac{1}{2} \sqrt{\frac{7}{6}} \text{Y[4,-4]}+\frac{1}{2}
   \sqrt{\frac{7}{6}} \text{Y[4,4]} \\
\bottomrule
\end{array}
$      

\end{table*}
\begin{table}
\caption{Basis vectors of group $D_{4h}$ in terms of spherical harmonics $Y[l,m]$ up to $l=5$ for rest frame $d=(0,0,0)$  in $z$ elongated box. 
Multiplicities are indicated by $n$. Multimultidimensional components are indicated by $\alpha$. }
\label{tab:basisD4h}

$      
\renewcommand{\arraystretch}{1.5}
\begin{array}{ccccl}
\toprule
\Gamma & l          & n & \alpha & \text{Basis of  } D_{4h}                   \\
\hline 
 A_{1g} & 0 & 1 & 1 & \text{Y[0,0]} \\
 A_{1g} & 2 & 1 & 1 & \text{Y[2,0]} \\
 A_{1g} & 4 & 1 & 1 & \frac{\text{Y[4,-4]}+\text{Y[4,4]}}{\sqrt{2}} \\
 A_{1g} & 4 & 2 & 1 & \text{Y[4,0]} \\
 A_{1u} & 5 & 1 & 1 & \frac{\text{Y[5,-4]}-\text{Y[5,4]}}{\sqrt{2}} \\
 A_{2g} & 4 & 1 & 1 & \frac{\text{Y[4,-4]}-\text{Y[4,4]}}{\sqrt{2}} \\
 A_{2u} & 1 & 1 & 1 & \text{Y[1,0]} \\
\bottomrule
\end{array}
$      

\end{table}
\begin{table}
\caption{Basis vectors of group $C_{4v}$ in terms of spherical harmonics $Y[l,m]$ up to $l=5$ for for  moving frame $d=(0,0,1)$ in both cubic and $z$ elongated boxes. 
Multiplicities are indicated by $n$. Multidimensional components are indicated by $\alpha$. }
\label{tab:basisC4v}

$      
\renewcommand{\arraystretch}{1.5}
\begin{array}{ccccl }
\toprule
\Gamma  & l & n & \alpha & \text{Basis vectors of } C_{4v}    \\
\hline
 A_1 & 0 & 1 & 1 & \text{Y[0,0]} \\
 A_1 & 1 & 1 & 1 & \text{Y[1,0]} \\
 A_1 & 2 & 1 & 1 & \text{Y[2,0]} \\
 A_1 & 3 & 1 & 1 & \text{Y[3,0]} \\
 A_1 & 4 & 1 & 1 & \frac{\text{Y[4,-4]}+\text{Y[4,4]}}{\sqrt{2}} \\
 A_1 & 4 & 2 & 1 & \text{Y[4,0]} \\
 A_1 & 5 & 1 & 1 & \frac{\text{Y[5,-4]}+\text{Y[5,4]}}{\sqrt{2}} \\
\bottomrule
\end{array}
$      

\end{table}
\begin{table*}
\caption{Basis vectors of little group $C_{3v}$ used for moving frame $d=(1,1,1)$  in the cubic box only.
Multiplicities are indicated by $n$. Multidimensional components are indicated by $\alpha$.
 }
 \label{tab:basisC3v}

$      
\renewcommand{\arraystretch}{1.8}
\begin{array}{cccc | l}
\toprule
\Gamma& l & n & \alpha &\text{Basis vectors of } C_{3v}     \\
\hline
 A_1 & 0 & 1 & 1 & \text{Y[0,0]} \\
 A_1 & 1 & 1 & 1 & \frac{(1-i) \text{Y[1,0]}}{\sqrt{6}}+\frac{\text{Y[1,-1]}}{\sqrt{3}}+\frac{i \text{Y[1,1]}}{\sqrt{3}} \\
 A_1 & 2 & 1 & 1 & \frac{(1-i) \text{Y[2,-1]}}{\sqrt{6}}+\frac{(1+i)
   \text{Y[2,1]}}{\sqrt{6}}+\frac{\text{Y[2,-2]}}{\sqrt{6}}-\frac{\text{Y[2,2]}}{\sqrt{6}} \\
 A_1 & 3 & 1 & 1 & \frac{(1+i) \text{Y[3,0]}}{\sqrt{6}}-\frac{i \text{Y[3,-1]}}{2 \sqrt{2}}+\frac{\text{Y[3,1]}}{2
   \sqrt{2}}+\frac{1}{2} \sqrt{\frac{5}{6}} \text{Y[3,-3]}-\frac{1}{2} i \sqrt{\frac{5}{6}} \text{Y[3,3]} \\
 A_1 & 3 & 2 & 1 & \frac{\text{Y[3,-2]}}{\sqrt{2}}-\frac{\text{Y[3,2]}}{\sqrt{2}} \\
 A_1 & 4 & 1 & 1 & \frac{1}{2} \sqrt{\frac{7}{3}} \text{Y[4,0]}+\frac{1}{2} \sqrt{\frac{5}{6}} \text{Y[4,-4]}+\frac{1}{2}
   \sqrt{\frac{5}{6}} \text{Y[4,4]} \\
 A_1 & 4 & 2 & 1 & \frac{i \text{Y[4,-1]}}{2 \sqrt{6}}-\frac{\text{Y[4,1]}}{2 \sqrt{6}}+\frac{\left(\frac{1}{2}-\frac{i}{2}\right)
   \text{Y[4,-2]}}{\sqrt{3}}-\frac{\left(\frac{1}{2}-\frac{i}{2}\right) \text{Y[4,2]}}{\sqrt{3}}+\frac{1}{2} \sqrt{\frac{7}{6}}
   \text{Y[4,-3]}-\frac{1}{2} i \sqrt{\frac{7}{6}} \text{Y[4,3]} \\
\bottomrule
\end{array}
$      

\end{table*}
\begin{table*}
\caption{Basis vectors of little group $C_{2v}$  used for moving frame $d=(1,1,0)$  in both cubic and $z$ elongated boxes. 
 }
 \label{tab:basisC2v}

$      
\renewcommand{\arraystretch}{2.0}
\begin{array}{ccccc}
\toprule
\text{Irrep} & l & n & \alpha & \text{Basis vectors of } C_{2v}     \\
\hline
 A_1 & 0 & 1 & 1 & \text{Y[0,0]} \\
 A_1 & 1 & 1 & 1 & \frac{\text{Y[1,-1]}+i \text{Y[1,1]}}{\sqrt{2}} \\
 A_1 & 2 & 1 & 1 & \frac{\text{Y[2,-2]}-\text{Y[2,2]}}{\sqrt{2}} \\
 A_1 & 2 & 2 & 1 & \text{Y[2,0]} \\
 A_1 & 3 & 1 & 1 & \frac{\text{Y[3,-3]}-i \text{Y[3,3]}}{\sqrt{2}} \\
 A_1 & 3 & 2 & 1 & \frac{\text{Y[3,-1]}+i \text{Y[3,1]}}{\sqrt{2}} \\
 A_1 & 4 & 1 & 1 & \frac{\text{Y[4,-4]}+\text{Y[4,4]}}{\sqrt{2}} \\
 A_1 & 4 & 2 & 1 & \frac{\text{Y[4,-2]}-\text{Y[4,2]}}{\sqrt{2}} \\
 A_1 & 4 & 3 & 1 & \text{Y[4,0]} \\
 A_2 & 2 & 1 & 1 & \frac{\text{Y[2,-1]}-i \text{Y[2,1]}}{\sqrt{2}} \\
 A_2 & 3 & 1 & 1 & \frac{\text{Y[3,-2]}+\text{Y[3,2]}}{\sqrt{2}} \\
 A_2 & 4 & 1 & 1 & \frac{\text{Y[4,-3]}+i \text{Y[4,3]}}{\sqrt{2}} \\
 A_2 & 4 & 2 & 1 & \frac{\text{Y[4,-1]}-i \text{Y[4,1]}}{\sqrt{2}} \\
 B_1 & 1 & 1 & 1 & \frac{\text{Y[1,-1]}-i \text{Y[1,1]}}{\sqrt{2}} \\
 B_1 & 2 & 1 & 1 & \frac{\text{Y[2,-2]}+\text{Y[2,2]}}{\sqrt{2}} \\
 B_1 & 3 & 1 & 1 & \frac{\text{Y[3,-3]}+i \text{Y[3,3]}}{\sqrt{2}} \\
 B_1 & 3 & 2 & 1 & \frac{\text{Y[3,-1]}-i \text{Y[3,1]}}{\sqrt{2}} \\
 B_1 & 4 & 1 & 1 & \frac{\text{Y[4,-4]}-\text{Y[4,4]}}{\sqrt{2}} \\
 B_1 & 4 & 2 & 1 & \frac{\text{Y[4,-2]}+\text{Y[4,2]}}{\sqrt{2}} \\
 B_2 & 1 & 1 & 1 & \text{Y[1,0]} \\
 B_2 & 2 & 1 & 1 & \frac{\text{Y[2,-1]}+i \text{Y[2,1]}}{\sqrt{2}} \\
 B_2 & 3 & 1 & 1 & \frac{\text{Y[3,-2]}-\text{Y[3,2]}}{\sqrt{2}} \\
 B_2 & 3 & 2 & 1 & \text{Y[3,0]} \\
 B_2 & 4 & 1 & 1 & \frac{\text{Y[4,-3]}-i \text{Y[4,3]}}{\sqrt{2}} \\
 B_2 & 4 & 2 & 1 & \frac{\text{Y[4,-1]}+i \text{Y[4,1]}}{\sqrt{2}} \\
\bottomrule
\end{array}
$      

\end{table*}
\begin{table}
\caption{Basis vectors of little group $C_{1v}$  used for  moving frame $d=(1,1,1)$ in the $z$ elongated box only.  }
\label{tab:basisC1v111}

$      
\renewcommand{\arraystretch}{1.5}
\begin{array}{ccccc}
\toprule
\text{Irrep} & l & n & \alpha & \text{Basis vectors of } C_{1v}     \\
\hline
 A_1 & 0 & 1 & 1 & \text{Y[0,0]} \\
 A_1 & 1 & 1 & 1 & \frac{\text{Y[1,-1]}+i \text{Y[1,1]}}{\sqrt{2}} \\
 A_1 & 1 & 2 & 1 & \text{Y[1,0]} \\
 A_1 & 2 & 1 & 1 & \frac{\text{Y[2,-2]}-\text{Y[2,2]}}{\sqrt{2}} \\
 A_1 & 2 & 2 & 1 & \frac{\text{Y[2,-1]}+i \text{Y[2,1]}}{\sqrt{2}} \\
 A_1 & 2 & 3 & 1 & \text{Y[2,0]} \\
 A_1 & 3 & 1 & 1 & \frac{\text{Y[3,-3]}-i \text{Y[3,3]}}{\sqrt{2}} \\
 A_1 & 3 & 2 & 1 & \frac{\text{Y[3,-2]}-\text{Y[3,2]}}{\sqrt{2}} \\
 A_1 & 3 & 3 & 1 & \frac{\text{Y[3,-1]}+i \text{Y[3,1]}}{\sqrt{2}} \\
 A_1 & 3 & 4 & 1 & \text{Y[3,0]} \\
 A_1 & 4 & 1 & 1 & \frac{\text{Y[4,-4]}+\text{Y[4,4]}}{\sqrt{2}} \\
 A_1 & 4 & 2 & 1 & \frac{\text{Y[4,-3]}-i \text{Y[4,3]}}{\sqrt{2}} \\
 A_1 & 4 & 3 & 1 & \frac{\text{Y[4,-2]}-\text{Y[4,2]}}{\sqrt{2}} \\
 A_1 & 4 & 4 & 1 & \frac{\text{Y[4,-1]}+i \text{Y[4,1]}}{\sqrt{2}} \\
 A_1 & 4 & 5 & 1 & \text{Y[4,0]} \\
 A_2 & 1 & 1 & 1 & \frac{\text{Y[1,-1]}-i \text{Y[1,1]}}{\sqrt{2}} \\
 A_2 & 2 & 1 & 1 & \frac{\text{Y[2,-2]}+\text{Y[2,2]}}{\sqrt{2}} \\
 A_2 & 2 & 2 & 1 & \frac{\text{Y[2,-1]}-i \text{Y[2,1]}}{\sqrt{2}} \\
 A_2 & 3 & 1 & 1 & \frac{\text{Y[3,-3]}+i \text{Y[3,3]}}{\sqrt{2}} \\
 A_2 & 3 & 2 & 1 & \frac{\text{Y[3,-2]}+\text{Y[3,2]}}{\sqrt{2}} \\
 A_2 & 3 & 3 & 1 & \frac{\text{Y[3,-1]}-i \text{Y[3,1]}}{\sqrt{2}} \\
 A_2 & 4 & 1 & 1 & \frac{\text{Y[4,-4]}-\text{Y[4,4]}}{\sqrt{2}} \\
 A_2 & 4 & 2 & 1 & \frac{\text{Y[4,-3]}+i \text{Y[4,3]}}{\sqrt{2}} \\
 A_2 & 4 & 3 & 1 & \frac{\text{Y[4,-2]}+\text{Y[4,2]}}{\sqrt{2}} \\
 A_2 & 4 & 4 & 1 & \frac{\text{Y[4,-1]}-i \text{Y[4,1]}}{\sqrt{2}} \\
\bottomrule
\end{array}
$      

\end{table}
\begin{table}
\caption{Basis vectors of little group $C_{1v}$  used for  moving frame $d=(0,1,2)$ in both cubic and $z$ elongated boxes.}
\label{tab:basisC1v012}

$      
\renewcommand{\arraystretch}{1.5}
\begin{array}{ccccc}
\toprule
\text{Irrep} & l & n & \alpha & \text{Basis vectors of } C_{1v}     \\
\hline
 A_1 & 0 & 1 & 1 & \text{Y[0,0]} \\
 A_1 & 1 & 1 & 1 & \frac{\text{Y[1,-1]}+\text{Y[1,1]}}{\sqrt{2}} \\
 A_1 & 1 & 2 & 1 & \text{Y[1,0]} \\
 A_1 & 2 & 1 & 1 & \frac{\text{Y[2,-2]}+\text{Y[2,2]}}{\sqrt{2}} \\
 A_1 & 2 & 2 & 1 & \frac{\text{Y[2,-1]}+\text{Y[2,1]}}{\sqrt{2}} \\
 A_1 & 2 & 3 & 1 & \text{Y[2,0]} \\
 A_1 & 3 & 1 & 1 & \frac{\text{Y[3,-3]}+\text{Y[3,3]}}{\sqrt{2}} \\
 A_1 & 3 & 2 & 1 & \frac{\text{Y[3,-2]}+\text{Y[3,2]}}{\sqrt{2}} \\
 A_1 & 3 & 3 & 1 & \frac{\text{Y[3,-1]}+\text{Y[3,1]}}{\sqrt{2}} \\
 A_1 & 3 & 4 & 1 & \text{Y[3,0]} \\
 A_1 & 4 & 1 & 1 & \frac{\text{Y[4,-4]}+\text{Y[4,4]}}{\sqrt{2}} \\
 A_1 & 4 & 2 & 1 & \frac{\text{Y[4,-3]}+\text{Y[4,3]}}{\sqrt{2}} \\
 A_1 & 4 & 3 & 1 & \frac{\text{Y[4,-2]}+\text{Y[4,2]}}{\sqrt{2}} \\
 A_1 & 4 & 4 & 1 & \frac{\text{Y[4,-1]}+\text{Y[4,1]}}{\sqrt{2}} \\
 A_1 & 4 & 5 & 1 & \text{Y[4,0]} \\
 A_2 & 1 & 1 & 1 & \frac{\text{Y[1,-1]}-\text{Y[1,1]}}{\sqrt{2}} \\
 A_2 & 2 & 1 & 1 & \frac{\text{Y[2,-2]}-\text{Y[2,2]}}{\sqrt{2}} \\
 A_2 & 2 & 2 & 1 & \frac{\text{Y[2,-1]}-\text{Y[2,1]}}{\sqrt{2}} \\
 A_2 & 3 & 1 & 1 & \frac{\text{Y[3,-3]}-\text{Y[3,3]}}{\sqrt{2}} \\
 A_2 & 3 & 2 & 1 & \frac{\text{Y[3,-2]}-\text{Y[3,2]}}{\sqrt{2}} \\
 A_2 & 3 & 3 & 1 & \frac{\text{Y[3,-1]}-\text{Y[3,1]}}{\sqrt{2}} \\
 A_2 & 4 & 1 & 1 & \frac{\text{Y[4,-4]}-\text{Y[4,4]}}{\sqrt{2}} \\
 A_2 & 4 & 2 & 1 & \frac{\text{Y[4,-3]}-\text{Y[4,3]}}{\sqrt{2}} \\
 A_2 & 4 & 3 & 1 & \frac{\text{Y[4,-2]}-\text{Y[4,2]}}{\sqrt{2}} \\
 A_2 & 4 & 4 & 1 & \frac{\text{Y[4,-1]}-\text{Y[4,1]}}{\sqrt{2}} \\
\bottomrule
\end{array}
$      

\end{table}
%

\section{Matrix elements for quantization conditions}
\label{sec:ME}

Here we collect all matrix elements for the QCs discussed in the main text.
They apply to rest frame and four moving frames in both cubic and elongated boxes, and unequal masses.
Up to five partial waves ($l=4$) are considered in each QC. The only exception is rest frame in cubic and elongated boxes where up to $l=5$ partial waves are  considered. 
The matrix elements are linear combinations of nonzero elements given in Table~\ref{tab:allW} supplemented by Table~\ref{tab:addW}. To construct the QC corresponding to a particular irrep from a table, we use the following scheme. 
Since the QC matrices are Hermitian we only display the upper triangular elements, 
in the regular order of starting at the top left, then zigzagging to the bottom right. 
The matrices have varying dimensions depending on the multiplicity $n$ associated with $l$. The dimension can be inferred by adding up the multiplicities partial wave by partial wave. For example, if a QC has content $l(n)=0(1), 1(1), 2(2), 3(2), 4(3), \cdots$ (such as the $A_1$ irrep of $C_{2v}$, see Table~\ref{tab:all}), then QC of order 5 includes all partial waves up to $l=4$ and its dimension is $9\times 9$.  One can reconstruct the full $M$ matrix of 81 elements for the QC by reading in the 45 ordered upper-triangular elements in Table~\ref{tab:C2vA1}. Once the full matrix is recovered, going to lower orders is a simple matter of keeping the relevant rows and columns. For example, for order 4, delete the 3 outer rows and columns to reduce the matrix to $6\times 6$. For order 2, keeping only the lower $2\times 2$ of the full matrix.  And so on.

In all the tables, we use 
\beq \text{w}_{lm}\equiv \text{wr}_{lm} + I \text{wi}_{lm},
\eeq
with the following notation.
If the function is real, we represent it simply by $\text{w}_{lm}$ (without the $r$ label).
If the function is purely imaginary, we represent it by $\text{wi}_{lm}$. 
If the function has both parts nonzero, then expressions such as $\text{wi}_{31}\to \text{wr}_{31}$ or $\text{wi}_{33}\to -\text{wr}_{33}$ mean they have equal magnitude but may differ by a sign, and we represent the function by its real part $\text{wr}_{lm}$ (with the $r$ label). We did not encounter the case where nonzero real and imaginary parts have different magnitudes. 
Note that due to angular momentum coupling, $\text{w}_{lm}$ functions up to $l=8$ are involved for partial waves up to $l=4$ in the QC.

Matrix elements for rest frame $d=(0,0,0)$ in cubic box are listed in Table~\ref{tab:Oh}.

Matrix elements for rest frame $d=(0,0,0)$ in elongated box are listed in Table~\ref{tab:D4h}.

Matrix elements moving frame $d=(0,0,1)$ in both cubic and elongated boxes  are listed in Tables~\ref{tab:C4v}.

Matrix elements for moving frame $d=(1,1,0)$ in both cubic and elongated boxes are listed 
in~\cref{tab:C2vA1,tab:C2vA2,tab:C2vB1,tab:C2vB2}.

Matrix elements for moving frame $d=(1,1,1)$ in cubic box only are listed in 
in~\cref{tab:C3vA1,tab:C3vA2,tab:C3vE}.

Matrix elements for moving frame $d=(0,1,2)$ in both cubic and elongated boxes are listed 
in~\cref{tab:C1v012A1,tab:C1v012A2}.

Matrix elements for moving frame $d=(1,1,1)$ in elongated box only are listed 
in~\cref{tab:C1v111A1,tab:C1v111A2}.

\begin{table*}
\caption{Matrix elements for rest frame in cubic box (group $O_{h}$) up to $l=5$.  
}
\label{tab:Oh}

$      
\renewcommand{\arraystretch}{2.0}
\begin{array}{cccccl}
\toprule
\Gamma & l  & n & l'  & n'  &M^\Gamma_{ln, l'n'}   \\
\hline 
A_{1g} & 0 & 1 & 0 & 1 & w_{00} \\
   & 0 & 1 & 4 & 1 & 2 \sqrt{\frac{3}{7}} w_{40} \\
   & 4 & 1 & 4 & 1 & w_{00}+\frac{108 w_{40}}{143}+\frac{80 w_{60}}{11 \sqrt{13}}+\frac{560 w_{80}}{143 \sqrt{17}} \\
A_{2u}& 3 & 1 & 3 & 1 & w_{00}-\frac{12 w_{40}}{11}+\frac{80 w_{60}}{11 \sqrt{13}} \\
E_g& 2 & 1 & 2 & 1 & w_{00}+\frac{6 w_{40}}{7} \\
   & 2 & 1 & 4 & 1 & -\frac{40}{77} \sqrt{3} w_{40}-\frac{30}{11} \sqrt{\frac{3}{13}} w_{60} \\
   & 4 & 1 & 4 & 1 & w_{00}+\frac{108 w_{40}}{1001}-\frac{64 w_{60}}{11 \sqrt{13}}+\frac{392 w_{80}}{143 \sqrt{17}} \\
\bottomrule
\end{array}
$      

\end{table*}
\clearpage\newpage
\renewcommand{\arraystretch}{1.8} 
\begin{longtable*}{CCCCCL} 
\caption{Matrix elements for rest frame in the elongated box (group $D_{4h}$) up to $l=5$.}
\label{tab:D4h} \\

\hline \endfirsthead
\multicolumn{6}{c} {{\tablename\ \thetable{:} \it Continued}} \\
\hline \endhead
\hline \\ \endfoot
\hline \\   \endlastfoot

\hline \\
\Gamma & l  & n  &  l'    & n'  &M^\Gamma_{l n , l' n'}  \\
\hline 
 A_{1g} & 0 & 1 & 0 & 1 & \text{w}_{00} \\
   & 0 & 1 & 2 & 1 & -\text{w}_{20} \\
   & 0 & 1 & 4 & 1 & \sqrt{2} \text{w}_{44} \\
   & 0 & 1 & 4 & 2 & \text{w}_{40} \\
   & 2 & 1 & 2 & 1 & \text{w}_{00}+\frac{2 \sqrt{5} \text{w}_{20}}{7}+\frac{6 \text{w}_{40}}{7} \\
   & 2 & 1 & 4 & 1 & \frac{4 \sqrt{10} \text{w}_{44}}{11}-\frac{15}{11} \sqrt{\frac{2}{13}}
   \text{w}_{64} \\
\hline
\end{longtable*}
\begin{table*}
\caption{Matrix elements for moving frame $d=(0,0,1)$ in both cubic and elongated boxes (group $C_{4v}$).
The $B_{1/2}$ irreps are combined by upper/lower signs. 
}
\label{tab:C4v}

$      
\renewcommand{\arraystretch}{0.8}
\fontsize{8}{9}
\begin{array}{cccccl}
\toprule
\text{Irrep} & l  & n  &  l'    & n'  & M^\Gamma_{l n , l' n'}  \\
\hline 
 A_1 & 0 & 1 & 0 & 1 & \text{w}_{00} \\
   & 0 & 1 & 1 & 1 & i \text{w}_{10} \\
   & 0 & 1 & 2 & 1 & -\text{w}_{20} \\
   & 0 & 1 & 3 & 1 & -i \text{w}_{30} \\
   & 0 & 1 & 4 & 1 & \sqrt{2} \text{w}_{44} \\
   & 0 & 1 & 4 & 2 & \text{w}_{40} \\
   & 1 & 1 & 1 & 1 & \text{w}_{00}+\frac{2 \text{w}_{20}}{\sqrt{5}} \\
\bottomrule
\end{array}
$      

\end{table*}
\begin{table*}
\caption{Matrix elements for moving frame $d=(1,1,0)$ in both cubic and elongated boxes,  group $C_{2v}$, and irrep $A_{1}$.
}
\label{tab:C2vA1}

$ 
\fontsize{8}{9}     
\renewcommand{\arraystretch}{0.0}
\begin{array}{c c c c l }
\toprule
 l  & n  &  l'    & n'  & M^\Gamma_{l n , l' n'}  \\
\hline 
 0 & 1 & 0 & 1 & \text{w}_{00} \\
 0 & 1 & 1 & 1 & (-1-i) \sqrt{2} \text{wr11} \\
 0 & 1 & 2 & 1 & i \sqrt{2} \text{wi}_{22} \\
 0 & 1 & 2 & 2 & -\text{w}_{20} \\
 0 & 1 & 3 & 1 & (-1+i) \sqrt{2} \text{wr33} \\
 0 & 1 & 3 & 2 & (1+i) \sqrt{2} \text{wr31} \\
 0 & 1 & 4 & 1 & \sqrt{2} \text{w}_{44} \\
\bottomrule
\end{array}
$      

\end{table*}
\begin{table*}
\caption{Matrix elements for moving frame $d=(1,1,0)$ in both cubic and elongated boxes,  group $C_{2v}$, and irrep $A_{2}$.
}
\label{tab:C2vA2}

$      
\renewcommand{\arraystretch}{1.9}
\begin{array}{c c c c l }
\toprule
 l  & n  &  l'    & n'  & M^\Gamma_{l n , l' n'}  \\
\hline 
 2 & 1 & 2 & 1 & \text{w}_{00}+\frac{\sqrt{5} \text{w}_{20}}{7}-\frac{\sqrt{30} \text{wi}_{22}}{7}-\frac{4
   \text{w}_{40}}{7}-\frac{2 \sqrt{10} \text{wi}_{42}}{7} \\
 2 & 1 & 3 & 1 & (-1-i) \sqrt{\frac{6}{7}} \text{wr11}-\frac{(1+i)
   \text{wr31}}{\sqrt{3}}-\left(\frac{1}{3}+\frac{i}{3}\right) \sqrt{5} \text{wr33}+(2+2 i) \sqrt{\frac{10}{231}}
   \text{wr51}-\left(\frac{4}{3}+\frac{4 i}{3}\right) \sqrt{\frac{5}{11}} \text{wr53} \\
 2 & 1 & 4 & 1 & i \sqrt{\frac{5}{7}} \text{wi}_{22}+\frac{5}{11} i \sqrt{\frac{15}{7}} \text{wi}_{42}+\frac{2}{11} i
   \sqrt{15} \text{w}_{44}-\frac{4}{11} i \sqrt{\frac{10}{13}} \text{wi}_{62}+\frac{20}{11} i \sqrt{\frac{3}{13}}
   \text{w}_{64} \\
 2 & 1 & 4 & 2 & -\frac{1}{7} \sqrt{30} \text{w}_{20}-\frac{\sqrt{5} \text{wi}_{22}}{7}-\frac{5 \sqrt{6}
   \text{w}_{40}}{77}+\frac{9 \sqrt{15} \text{wi}_{42}}{77}+\frac{10}{11} \sqrt{\frac{6}{13}} \text{w}_{60}+\frac{4}{11}
   \sqrt{\frac{70}{13}} \text{wi}_{62} \\
 3 & 1 & 3 & 1 & \text{w}_{00}-\frac{7 \text{w}_{40}}{11}+\frac{\sqrt{70} \text{w}_{44}}{11}+\frac{10 \text{w}_{60}}{11
   \sqrt{13}}+\frac{10}{11} \sqrt{\frac{14}{13}} \text{w}_{64} \\
 3 & 1 & 4 & 1 & (-1-i) \text{wr11}-\left(\frac{1}{11}+\frac{i}{11}\right) \sqrt{14}
   \text{wr31}+\left(\frac{1}{13}+\frac{i}{13}\right) \sqrt{55} \text{wr51}\\&&&& -\left(\frac{5}{13}+\frac{5 i}{13}\right)
   \sqrt{\frac{42}{11}} \text{wr55}-\left(\frac{4}{143}+\frac{4 i}{143}\right) \sqrt{35}
   \text{wr71}-\left(\frac{4}{13}+\frac{4 i}{13}\right) \sqrt{\frac{105}{11}} \text{wr75} \\
\bottomrule
\end{array}
$      

\end{table*}
\begin{table*}
\caption{Matrix elements for moving frame $d=(1,1,0)$ in both cubic and elongated boxes,  group $C_{2v}$, and irrep $B_{1}$.
}
\label{tab:C2vB1}

$      
\renewcommand{\arraystretch}{1.9}
\begin{array}{c c c c l }
\toprule
 l  & n  &  l'    & n'  & M^\Gamma_{l n , l' n'}   \\
\hline 
 1 & 1 & 1 & 1 & \text{w}_{00}-\frac{\text{w}_{20}}{\sqrt{5}}-\sqrt{\frac{6}{5}} \text{wi}_{22} \\
 1 & 1 & 2 & 1 & (-1-i) \sqrt{\frac{6}{5}} \text{wr11}+(1+i) \sqrt{\frac{3}{35}} \text{wr31}-\frac{(3+3 i)
   \text{wr33}}{\sqrt{7}} \\
 1 & 1 & 3 & 1 & \frac{3 i \text{wi}_{22}}{\sqrt{7}}-\frac{i \text{wi}_{42}}{\sqrt{21}}+\frac{2 i
   \text{w}_{44}}{\sqrt{3}} \\
 1 & 1 & 3 & 2 & -3 \sqrt{\frac{2}{35}} \text{w}_{20}-\sqrt{\frac{3}{35}} \text{wi}_{22}+\sqrt{\frac{2}{7}}
   \text{w}_{40}+\sqrt{\frac{5}{7}} \text{wi}_{42} \\
 1 & 1 & 4 & 1 & -\frac{(2-2 i) \text{wr33}}{\sqrt{3}}+\frac{(1-i) \text{wr53}}{\sqrt{33}}+(1-i) \sqrt{\frac{15}{11}}
   \text{wr55} \\
 1 & 1 & 4 & 2 & (1+i) \sqrt{\frac{5}{7}} \text{wr31}-\frac{(1+i) \text{wr33}}{\sqrt{21}}-(1+i) \sqrt{\frac{2}{11}}
   \text{wr51}+(2+2 i) \sqrt{\frac{7}{33}} \text{wr53} \\
 2 & 1 & 2 & 1 & \text{w}_{00}-\frac{2 \sqrt{5} \text{w}_{20}}{7}+\frac{\text{w}_{40}}{7}+\sqrt{\frac{10}{7}}
   \text{w}_{44} \\
\bottomrule
\end{array}
$      

\end{table*}
\begin{table*}
\caption{Matrix elements for moving frame $d=(1,1,0)$ in both cubic and elongated boxes,  group $C_{2v}$, and irrep $B_{2}$.
}
\label{tab:C2vB2}

$      
\renewcommand{\arraystretch}{1.9}
\begin{array}{c c c c l }
\toprule
 l  & n  &  l'    & n'  & M^\Gamma_{l n , l' n'}    \\
\hline 
 1 & 1 & 1 & 1 & \text{w}_{00}+\frac{2 \text{w}_{20}}{\sqrt{5}} \\
 1 & 1 & 2 & 1 & (-1-i) \sqrt{\frac{6}{5}} \text{wr11}-(4+4 i) \sqrt{\frac{3}{35}} \text{wr31} \\
 1 & 1 & 3 & 1 & i \sqrt{\frac{6}{7}} \text{wi}_{22}+2 i \sqrt{\frac{2}{7}} \text{wi}_{42} \\
 1 & 1 & 3 & 2 & -3 \sqrt{\frac{3}{35}} \text{w}_{20}-\frac{4 \text{w}_{40}}{\sqrt{21}} \\
 1 & 1 & 4 & 1 & (-1+i) \sqrt{\frac{2}{3}} \text{wr33}-(4-4 i) \sqrt{\frac{2}{33}} \text{wr53} \\
 1 & 1 & 4 & 2 & (1+i) \sqrt{\frac{10}{7}} \text{wr31}+\frac{(4+4 i) \text{wr51}}{\sqrt{11}} \\
 2 & 1 & 2 & 1 & \text{w}_{00}+\frac{\sqrt{5} \text{w}_{20}}{7}+\frac{\sqrt{30} \text{wi}_{22}}{7}-\frac{4
   \text{w}_{40}}{7}+\frac{2 \sqrt{10} \text{wi}_{42}}{7} \\
\bottomrule
\end{array}
$      

\end{table*}

\clearpage\newpage
\renewcommand{\arraystretch}{1.6} 
\begin{longtable*}{CCCCCL} 
\caption{Matrix elements for moving frame $d=(1,1,1)$ in cubic box only, group $C_{3v}$, and irrep $A_{1}$.}
\label{tab:C3vA1} \\

\hline \endfirsthead
\multicolumn{6}{c} {{\tablename\ \thetable{:} \it Continued}} \\
\hline \endhead
\hline \\ \endfoot
\hline \\   \endlastfoot

\hline \\

 l  & n  &  l'    & n'  & M^\Gamma_{l n , l' n'}  \\
\hline 
 0 & 1 & 0 & 1 & \text{w}_{00} \\
 0 & 1 & 1 & 1 & (1+i) \sqrt{\frac{3}{2}} \text{w}_{10} \\
 0 & 1 & 2 & 1 & -i \sqrt{6} \text{wr21} \\
 0 & 1 & 3 & 1 & (1-i) \sqrt{\frac{3}{2}} \text{w}_{30} \\
 0 & 1 & 3 & 2 & -\sqrt{2} \text{wi}_{32} \\
 0 & 1 & 4 & 1 & 2 \sqrt{\frac{3}{7}} \text{w}_{40} \\
\hline
\end{longtable*}
\begin{table*}
\caption{Matrix elements for moving frame $d=(1,1,1)$ in cubic box only, group $C_{3v}$, and irrep $A_{2}$.
}
\label{tab:C3vA2}

$      
\renewcommand{\arraystretch}{2.0}
\begin{array}{c c  c c  l }
\toprule
 l  & n  &  l'    & n'  & M^\Gamma_{l n , l' n'}    \\
\hline 
 3 & 1 & 3 & 1 & \text{w}_{00}-\frac{2 \text{w}_{40}}{11}-\frac{60 \text{w}_{60}}{11 \sqrt{13}}+\sqrt{\frac{10}{3}}
   \text{wr21}+\frac{8 \sqrt{5} \text{wr41}}{11}+\frac{5}{11} \sqrt{\frac{14}{39}} \text{wr61}+\frac{10}{11}
   \sqrt{\frac{105}{13}} \text{wr63} \\
 3 & 1 & 4 & 1 & \frac{(1+i) \text{w}_{10}}{\sqrt{2}}+\left(\frac{1}{11}+\frac{i}{11}\right) \sqrt{42}
   \text{w}_{30}-\left(\frac{1}{11}+\frac{i}{11}\right) \sqrt{35} \text{wi}_{32}+\left(\frac{5}{13}+\frac{5 i}{13}\right)
   \sqrt{\frac{6}{11}} \text{w}_{50}-\left(\frac{1253}{4719}+\frac{917 i}{4719}\right) \sqrt{10}
   \text{w}_{70}-\left(\frac{6}{143}+\frac{6 i}{143}\right) \sqrt{210} \text{wi}_{72} \\
&&&& +\left(\frac{20}{13}+\frac{20 i}{13}\right)
   \sqrt{\frac{5}{11}} \text{wr51}+\left(\frac{226}{4719}+\frac{10 i}{363}\right) \sqrt{35}
   \text{wr71}-\left(\frac{50}{1573}-\frac{46 i}{1573}\right) \sqrt{105} \text{wr73} \\
 4 & 1 & 4 & 1 & \text{w}_{00}+\frac{54 \text{w}_{40}}{143}-\frac{4 \text{w}_{60}}{11 \sqrt{13}}-\frac{448 \text{w}_{80}}{143
   \sqrt{17}}+\frac{\sqrt{30} \text{wr21}}{11}-\frac{216 \sqrt{5} \text{wr41}}{143}+\frac{7}{11} \sqrt{\frac{42}{13}}
   \text{wr61}+\frac{10}{11} \sqrt{\frac{105}{13}} \text{wr63}+\frac{168}{11} \sqrt{\frac{2}{17}} \text{wr81}-\frac{56}{13}
   \sqrt{\frac{210}{187}} \text{wr83} \\
\bottomrule
\end{array}
$      

\end{table*}

\clearpage\newpage
\renewcommand{\arraystretch}{2.1} 
\begin{longtable*}{CCCCL} 
\caption{Matrix elements for moving frame $d=(1,1,1)$ in cubic box only, group $C_{3v}$, and irrep $E$.}
\label{tab:C3vE} \\

\hline \endfirsthead
\multicolumn{5}{c} {{\tablename\ \thetable{:} \it Continued}} \\
\hline \endhead
\hline \\ \endfoot
\hline \\   \endlastfoot

\hline \\

 l  & n  &  l'    & n'  & M^\Gamma_{l n , l' n'}   \\
\hline 
 1 & 1 & 1 & 1 & \text{w}_{00}+\sqrt{\frac{6}{5}} \text{wr21} \\
 1 & 1 & 2 & 1 & (1+i) \sqrt{\frac{2}{5}} \text{w}_{10}-(2-3 i) \sqrt{\frac{3}{70}} \text{w}_{30}-\frac{2
   \text{wi}_{32}}{\sqrt{7}} \\
 1 & 1 & 2 & 2 & -\frac{(1-2 i) \text{w}_{10}}{\sqrt{5}}+(1+3 i) \sqrt{\frac{3}{35}} \text{w}_{30}+\sqrt{\frac{2}{7}}
   \text{wi}_{32} \\
 1 & 1 & 3 & 1 & -2 \sqrt{\frac{2}{21}} \text{w}_{40}-\frac{6 i \text{wr21}}{\sqrt{35}}-(14+18 i) \sqrt{\frac{2}{105}}
   \text{wr41} \\
 1 & 1 & 3 & 2 & -(2-4 i) \sqrt{\frac{2}{105}} \text{w}_{40}+\frac{\left(\frac{18}{5}-\frac{6 i}{5}\right)
   \text{wr21}}{\sqrt{7}}+(8+2 i) \sqrt{\frac{2}{21}} \text{wr41} \\
 1 & 1 & 4 & 1 & (1-i) \sqrt{\frac{10}{29}} \text{w}_{30}-(4+14 i) \sqrt{\frac{10}{2233}} \text{w}_{50}+(16-28 i)
   \sqrt{\frac{3}{2233}} \text{wr51} \\
\hline
\end{longtable*}

\clearpage\newpage
\renewcommand{\arraystretch}{1.0} 
\begin{longtable*}{CCCCL} 
\caption{Matrix elements for moving frame $d=(0,1,2)$ in both cubic and elongated boxes,  group $C_{1v}$, and irrep $A_{1}$.}
\label{tab:C1v012A1} \\

\hline \endfirsthead
\multicolumn{5}{c} {{\tablename\ \thetable{:} \it Continued}} \\
\hline \endhead
\hline \\ \endfoot
\hline \\   \endlastfoot

\hline \\

 l  & n  &  l'    & n'  & M^\Gamma_{l n , l' n'}   \\
\hline 
 0 & 1 & 0 & 1 & \text{w}_{00} \\
 0 & 1 & 1 & 1 & -\sqrt{2} \text{wi}_{11} \\
 0 & 1 & 1 & 2 & i \text{w}_{10} \\
 0 & 1 & 2 & 1 & -\sqrt{2} \text{w}_{22} \\
 0 & 1 & 2 & 2 & -i \sqrt{2} \text{wi}_{21} \\
 0 & 1 & 2 & 3 & -\text{w}_{20} \\
\hline
\end{longtable*}

\onecolumngrid\clearpage 
\renewcommand{\arraystretch}{2.05} 
\begin{longtable*}{CCCCL} 
\caption{Matrix elements for moving frame $d=(0,1,2)$ in both cubic and elongated boxes,  group $C_{1v}$, and irrep $A_{2}$.}
\label{tab:C1v012A2} \\

\hline \endfirsthead
\multicolumn{5}{c} {{\tablename\ \thetable{:} \it Continued}} \\
\hline \endhead
\hline \\ \endfoot
\hline \\   \endlastfoot

\hline \\

 l  & n  &  l'    & n'  & M^\Gamma_{l n , l' n'}    \\
\hline 
 1 & 1 & 1 & 1 & \text{w}_{00}-\frac{\text{w}_{20}}{\sqrt{5}}+\sqrt{\frac{6}{5}} \text{w}_{22} \\
 1 & 1 & 2 & 1 & -\sqrt{\frac{6}{5}} \text{wi}_{11}+\sqrt{\frac{3}{35}} \text{wi}_{31}-\frac{3 \text{wi}_{33}}{\sqrt{7}} \\
 1 & 1 & 2 & 2 & i \sqrt{\frac{3}{5}} \text{w}_{10}-\frac{3 i \text{w}_{30}}{\sqrt{35}}+i \sqrt{\frac{6}{7}} \text{w}_{32} \\
 1 & 1 & 3 & 1 & -\frac{3 \text{w}_{22}}{\sqrt{7}}+\frac{\text{w}_{42}}{\sqrt{21}}-\frac{2 \text{w}_{44}}{\sqrt{3}} \\
 1 & 1 & 3 & 2 & -i \sqrt{\frac{6}{7}} \text{wi}_{21}+\frac{i \text{wi}_{41}}{\sqrt{7}}-i \text{wi}_{43} \\
 1 & 1 & 3 & 3 & -3 \sqrt{\frac{2}{35}} \text{w}_{20}+\sqrt{\frac{3}{35}} \text{w}_{22}+\sqrt{\frac{2}{7}}
   \text{w}_{40}-\sqrt{\frac{5}{7}} \text{w}_{42} \\
\hline
\end{longtable*}

\onecolumngrid\clearpage
\renewcommand{\arraystretch}{1.85} 
\begin{longtable*}{CCCCL} 
\caption{Matrix elements for moving frame $d=(1,1,1)$ in elongated box only,  group $C_{1v}$, and irrep $A_{1}$.}
\label{tab:C1v111A1} \\

\hline \endfirsthead
\multicolumn{5}{c} {{\tablename\ \thetable{:} \it Continued}} \\
\hline \endhead
\hline \\ \endfoot
\hline \\   \endlastfoot

\hline \\

 l  & n  &  l'    & n'  & M^\Gamma_{l n , l' n'}   \\
\hline 
 0 & 1 & 0 & 1 & \text{w}_{00} \\
 0 & 1 & 1 & 1 & (-1-i) \sqrt{2} \text{wr11} \\
 0 & 1 & 1 & 2 & i \text{w}_{10} \\
 0 & 1 & 2 & 1 & i \sqrt{2} \text{wi}_{22} \\
 0 & 1 & 2 & 2 & (1-i) \sqrt{2} \text{wr21} \\
 0 & 1 & 2 & 3 & -\text{w}_{20} \\
\hline
\end{longtable*}

\onecolumngrid\clearpage
\renewcommand{\arraystretch}{1.85} 
\begin{longtable*}{CCCCL} 
\caption{Matrix elements for moving frame $d=(1,1,1)$ in elongated box only,  group $C_{1v}$, and irrep $A_{2}$.}
\label{tab:C1v111A2} \\

\hline \endfirsthead
\multicolumn{5}{c} {{\tablename\ \thetable{:} \it Continued}} \\
\hline \endhead
\hline \\ \endfoot
\hline \\   \endlastfoot

\hline \\

 l  & n  &  l'    & n'  & M^\Gamma_{l n , l' n'}   \\
\hline 
 1 & 1 & 1 & 1 & \text{w}_{00}-\frac{\text{w}_{20}}{\sqrt{5}}-\sqrt{\frac{6}{5}} \text{wi}_{22} \\
 1 & 1 & 2 & 1 & (-1-i) \sqrt{\frac{6}{5}} \text{wr11}+(1+i) \sqrt{\frac{3}{35}} \text{wr31}-\frac{(3+3 i)
   \text{wr33}}{\sqrt{7}} \\
 1 & 1 & 2 & 2 & i \sqrt{\frac{3}{5}} \text{w}_{10}-\frac{3 i \text{w}_{30}}{\sqrt{35}}-i \sqrt{\frac{6}{7}}
   \text{wi}_{32} \\
 1 & 1 & 3 & 1 & \frac{3 i \text{wi}_{22}}{\sqrt{7}}-\frac{i \text{wi}_{42}}{\sqrt{21}}+\frac{2 i
   \text{w}_{44}}{\sqrt{3}} \\
 1 & 1 & 3 & 2 & (1-i) \sqrt{\frac{6}{7}} \text{wr21}-\frac{(1-i) \text{wr41}}{\sqrt{7}}+(1-i) \text{wr43} \\
 1 & 1 & 3 & 3 & -3 \sqrt{\frac{2}{35}} \text{w}_{20}-\sqrt{\frac{3}{35}} \text{wi}_{22}+\sqrt{\frac{2}{7}}
   \text{w}_{40}+\sqrt{\frac{5}{7}} \text{wi}_{42} \\
\hline
\end{longtable*}
%

\onecolumngrid\clearpage
\twocolumngrid
\bibliography{xvalidation}

\begin{supplement}
\clearpage
\appendix
\begin{widetext}
\section*{\large Supplemental Material} \label{sec:sup} 
It consists of four parts: noninteracting levels, phase shift reconstruction for the lowest partial wave from the L\"uscher formula, sensitivity to the second partial wave, and convergence data for every energy level examined.
All material (tables, figures, equations, and citations) is  cross-referenced with the main text.
\end{widetext}

\renewcommand{\arraystretch}{1.4} 
%

\subsection{Noninteracting levels}
\label{sup:freek}

This section contains the noninteracting levels for the ten cases considered in this work:
rest frame plus four moving frames in cubic and elongated geometries.
They are obtained from kinematics in Eq.\eqref{eq:kinP}, Eq.\eqref{eq:momenta}, Eq.\eqref{eq:elab}, Eq.\eqref{eq:ecm}, and Eq.\eqref{eq:ecm2} for unequal masses of $m_1=0.138$ GeV and $m_2=0.94$ GeV. The box has the geometry $L\times L\times \eta L$ with $L=24$ fm and $\eta=1$ for cubic and $\eta=1.5$ for $z$ elongated. The spectrum is bounded from below. The total number of levels anticipated in various scenarios are numerous, as summarized in Table~\ref{tab:kcut}.  We apply upper cuts to keep the number of levels manageable. First a cutoff of $k<0.2$ GeV is applied. If the number of levels is still too large after the $k$ cutoff, we limit it to 40.
The noninteracting levels serve multiple purposes: as a guide on how the two particles are moving in the lab frame by their n indices and degeneracies; as a baseline to judge the interacting levels given below in the convergence section; as poles of the zeta functions and plotted as vertical lines in graphs; 
as a cross check of the spectrum calculation from the lattice Hamiltonian
(the spectrum calculation with the interaction turned off should reproduce the noninteracting levels given here after extrapolating to the continuum).

\begin{table}[b]
\caption{Noninteracting levels for rest frame $d=(0,0,0)$ in cubic box. }
\label{tab:freekOh}          
\begin{tabular}{ccccccccc}                                   
\toprule
 Level & $n_{1x}$ & $n_{1y}$ & $n_{1z}$ & $n_{2x}$ & $n_{2y}$ & $n_{2z}$ & k (GeV) & Deg.\\
  \hline
 14 & -3 & -2 & -1 & 3 & 2 & 1 & 0.192974 & 12 \\
 13 & -3 & -2 & 0 & 3 & 2 & 0 & 0.185954 & 12 \\
 12 & -2 & -2 & -2 & 2 & 2 & 2 & 0.178659 & 2 \\
 11 & -3 & -1 & -1 & 3 & 1 & 1 & 0.171053 & 6 \\
 10 & -3 & -1 & 0 & 3 & 1 & 0 & 0.163093 & 12 \\
 9 & -3 & 0 & 0 & 3 & 0 & 0 & 0.154723 & 12 \\
 8 & -2 & -2 & 0 & 2 & 2 & 0 & 0.145875 & 6 \\
 7 & -2 & -1 & -1 & 2 & 1 & 1 & 0.126331 & 6 \\
 6 & -2 & -1 & 0 & 2 & 1 & 0 & 0.115324 & 12 \\
 5 & -2 & 0 & 0 & 2 & 0 & 0 & 0.103149 & 6 \\
 4 & -1 & -1 & -1 & 1 & 1 & 1 & 0.0893296 & 2 \\
 3 & -1 & -1 & 0 & 1 & 1 & 0 & 0.0729373 & 6 \\
 2 & -1 & 0 & 0 & 1 & 0 & 0 & 0.0515745 & 6 \\
 1 & 0 & 0 & 0 & 0 & 0 & 0 & 0. & 1 \\
\bottomrule                     
\end{tabular}                    
\end{table}                     
\begin{table}[b]
\caption{Noninteracting levels for rest frame $d=(0,0,0)$ in $z$ elongated box with $\eta=1.5$. }
\label{tab:freekD4h}          
\begin{tabular}{ccccccccc}                                   
\toprule
 Level & $n_{1x}$ & $n_{1y}$ & $n_{1z}$ & $n_{2x}$ & $n_{2y}$ & $n_{2z}$ & k (GeV) & Deg.\\
   \hline
 36 & -3 & -2 & -2 & 3 & 2 & 2 & 0.198262 & 16 \\
 35 & -3 & -1 & -3 & 3 & 1 & 3 & 0.192974 & 16 \\
 34 & -3 & -2 & -1 & 3 & 2 & 1 & 0.189106 & 16 \\
 33 & -1 & -1 & -5 & 1 & 1 & 5 & 0.186747 & 8 \\
 32 & -3 & -2 & 0 & 3 & 2 & 0 & 0.185954 & 16 \\
 31 & -2 & -1 & -4 & 2 & 1 & 4 & 0.179484 & 24 \\
 30 & -2 & -2 & -3 & 2 & 2 & 3 & 0.178659 & 8 \\
 29 & -3 & -1 & -2 & 3 & 1 & 2 & 0.176997 & 16 \\
 28 & -2 & 0 & -4 & 2 & 0 & 4 & 0.171915 & 10 \\
 27 & -3 & 0 & -2 & 3 & 0 & 2 & 0.169317 & 8 \\
 26 & -3 & -1 & -1 & 3 & 1 & 1 & 0.166678 & 16 \\
 25 & -3 & -1 & 0 & 3 & 1 & 0 & 0.163093 & 8 \\
 24 & -2 & -2 & -2 & 2 & 2 & 2 & 0.161271 & 8 \\
 23 & -3 & 0 & -1 & 3 & 0 & 1 & 0.158498 & 8 \\
 22 & -1 & -1 & -4 & 1 & 1 & 4 & 0.155676 & 8 \\
 21 & -3 & 0 & 0 & 3 & 0 & 0 & 0.154723 & 20 \\
 20 & -2 & -2 & -1 & 2 & 2 & 1 & 0.149872 & 8 \\
 19 & -1 & 0 & -4 & 1 & 0 & 4 & 0.146884 & 8 \\
 18 & -2 & -2 & 0 & 2 & 2 & 0 & 0.145875 & 12 \\
 17 & 0 & 0 & -4 & 0 & 0 & 4 & 0.137532 & 2 \\
 16 & -2 & -1 & -2 & 2 & 1 & 2 & 0.13427 & 16 \\
 15 & -1 & -1 & -3 & 1 & 1 & 3 & 0.126331 & 8 \\
 14 & -2 & 0 & -2 & 2 & 0 & 2 & 0.12397 & 8 \\
 13 & -2 & -1 & -1 & 2 & 1 & 1 & 0.12034 & 16 \\
 12 & -2 & -1 & 0 & 2 & 1 & 0 & 0.115324 & 16 \\
 11 & -2 & 0 & -1 & 2 & 0 & 1 & 0.108729 & 8 \\
 10 & -2 & 0 & 0 & 2 & 0 & 0 & 0.103149 & 6 \\
 9 & -1 & -1 & -2 & 1 & 1 & 2 & 0.100243 & 8 \\
 8 & -1 & 0 & -2 & 1 & 0 & 2 & 0.0859575 & 8 \\
 7 & -1 & -1 & -1 & 1 & 1 & 1 & 0.0806353 & 8 \\
 6 & -1 & -1 & 0 & 1 & 1 & 0 & 0.0729373 & 4 \\
 5 & 0 & 0 & -2 & 0 & 0 & 2 & 0.068766 & 2 \\
 4 & -1 & 0 & -1 & 1 & 0 & 1 & 0.0619848 & 8 \\
 3 & -1 & 0 & 0 & 1 & 0 & 0 & 0.0515745 & 4 \\
 2 & 0 & 0 & -1 & 0 & 0 & 1 & 0.034383 & 2 \\
 1 & 0 & 0 & 0 & 0 & 0 & 0 & 0. & 1 \\
\bottomrule                     
\end{tabular}                   
\end{table}                     
\begin{table}
\caption{Noninteracting levels for moving frame $d=(0,0,1)$ in cubic box. }
\label{tab:freekOh001}          
\begin{tabular}{ccccccccc}                                   
\toprule
 Level & $n_{1x}$ & $n_{1y}$ & $n_{1z}$ & $n_{2x}$ & $n_{2y}$ & $n_{2z}$ & k (GeV) & Deg.\\
  \hline
 40 & -2 & -2 & 2 & 2 & 2 & -1 & 0.17493 & 4 \\
 39 & -3 & -1 & -1 & 3 & 1 & 2 & 0.173158 & 8 \\
 38 & -1 & 0 & -3 & 1 & 0 & 4 & 0.169369 & 4 \\
 37 & -3 & -1 & 1 & 3 & 1 & 0 & 0.16918 & 8 \\
 36 & -3 & 0 & -1 & 3 & 0 & 2 & 0.165299 & 4 \\
 35 & -1 & -1 & 3 & 1 & 1 & -2 & 0.165105 & 4 \\
 34 & -3 & -1 & 0 & 3 & 1 & 1 & 0.163226 & 8 \\
\bottomrule                     
\end{tabular}                      
\end{table}                     
\begin{table}
\caption{Noninteracting levels for moving frame $d=(0,0,1)$ in $z$ elongated box with $\eta=1.5$. }
\label{tab:freekD4h001}          
\begin{tabular}{ccccccccc}                                   
\toprule
 Level & $n_{1x}$ & $n_{1y}$ & $n_{1z}$ & $n_{2x}$ & $n_{2y}$ & $n_{2z}$ & k (GeV) & Deg.\\
  \hline
 40 & -1 & 0 & -4 & 1 & 0 & 5 & 0.151013 & 4 \\
 39 & -2 & -2 & -1 & 2 & 2 & 2 & 0.150943 & 4 \\
 38 & -2 & 0 & -3 & 2 & 0 & 4 & 0.14902 & 4 \\
 37 & -2 & -2 & 1 & 2 & 2 & 0 & 0.148924 & 4 \\
 36 & -2 & -2 & 0 & 2 & 2 & 1 & 0.145941 & 4 \\
 35 & -2 & 0 & 3 & 2 & 0 & -2 & 0.142796 & 4 \\
 34 & -1 & 0 & 4 & 1 & 0 & -3 & 0.142771 & 4 \\
\bottomrule                     
\end{tabular}                      
\end{table}                     
\begin{table}
\caption{Noninteracting levels for moving frame $d=(1,1,0)$ in cubic box.}
\label{tab:freekOh110}          
\begin{tabular}{ccccccccc}                                   
\toprule
 Level & $n_{1x}$ & $n_{1y}$ & $n_{1z}$ & $n_{2x}$ & $n_{2y}$ & $n_{2z}$ & k (GeV) & Deg.\\
  \hline
 40 & -2 & 1 & -2 & 3 & 0 & 2 & 0.157186 & 4 \\
 39 & 0 & 3 & -1 & 1 & -2 & 1 & 0.156982 & 4 \\
 38 & -2 & -2 & 0 & 3 & 3 & 0 & 0.155212 & 1 \\
 37 & -2 & 2 & -1 & 3 & -1 & 1 & 0.155005 & 6 \\
 36 & 1 & 3 & 0 & 0 & -2 & 0 & 0.154798 & 2 \\
 35 & -1 & 2 & -2 & 2 & -1 & 2 & 0.152792 & 4 \\
 34 & -2 & 0 & -2 & 3 & 1 & 2 & 0.15076 & 4 \\
\bottomrule                     
\end{tabular}                      
\end{table}                     
\begin{table}
\caption{Noninteracting levels for moving frame $d=(1,1,0)$ in $z$ elongated box with $\eta=1.5$. }
\label{tab:freekD4h110}          
\begin{tabular}{ccccccccc}                                   
\toprule
 Level & $n_{1x}$ & $n_{1y}$ & $n_{1z}$ & $n_{2x}$ & $n_{2y}$ & $n_{2z}$ & k (GeV) & Deg.\\
  \hline
 40 & -2 & 1 & -2 & 3 & 0 & 2 & 0.137101 & 4 \\
 39 & 2 & 2 & 0 & -1 & -1 & 0 & 0.136538 & 1 \\
 38 & -1 & 2 & -2 & 2 & -1 & 2 & 0.13204 & 4 \\
 37 & -1 & -1 & -3 & 2 & 2 & 3 & 0.131942 & 2 \\
 36 & -2 & 0 & -2 & 3 & 1 & 2 & 0.129683 & 4 \\
 35 & -2 & -1 & -1 & 3 & 2 & 1 & 0.128888 & 4 \\
 34 & 1 & 2 & -2 & 0 & -1 & 2 & 0.126777 & 4 \\
\bottomrule                     
\end{tabular}                      
\end{table}                     
\begin{table}
\caption{Noninteracting levels for moving frame $d=(1,1,1)$ in cubic box. }
\label{tab:freekOh111}          
\begin{tabular}{ccccccccc}                                   
\toprule
 Level & $n_{1x}$ & $n_{1y}$ & $n_{1z}$ & $n_{2x}$ & $n_{2y}$ & $n_{2z}$ & k (GeV) & Deg.\\
  \hline
 40 & -2 & 2 & 2 & 3 & -1 & -1 & 0.175179 & 3 \\
 39 & -3 & 1 & 1 & 4 & 0 & 0 & 0.17341 & 3 \\
 38 & -3 & -1 & 0 & 4 & 2 & 1 & 0.171622 & 6 \\
 37 & -1 & -1 & 3 & 2 & 2 & -2 & 0.169437 & 3 \\
 36 & -3 & 0 & 1 & 4 & 1 & 0 & 0.167607 & 6 \\
 35 & 2 & 2 & 2 & -1 & -1 & -1 & 0.167224 & 1 \\
 34 & -2 & -2 & -1 & 3 & 3 & 2 & 0.165756 & 3 \\
\bottomrule                     
\end{tabular}                      
\end{table}                     
\begin{table}
\caption{Noninteracting levels for moving frame $d=(1,1,1)$ in $z$ elongated box with $\eta=1.5$. }
\label{tab:freekD4h111}    
\begin{tabular}{ccccccccc}                                   
\toprule
 Level & $n_{1x}$ & $n_{1y}$ & $n_{1z}$ & $n_{2x}$ & $n_{2y}$ & $n_{2z}$ & k (GeV) & Deg.\\
  \hline
 40 & -1 & 2 & 0 & 2 & -1 & 1 & 0.112806 & 2 \\
 39 & 1 & 2 & 1 & 0 & -1 & 0 & 0.110646 & 2 \\
 38 & -1 & -1 & -2 & 2 & 2 & 3 & 0.110102 & 1 \\
 37 & -2 & 0 & 0 & 3 & 1 & 1 & 0.110038 & 2 \\
 36 & 0 & 1 & 3 & 1 & 0 & -2 & 0.108707 & 2 \\
 35 & 0 & 0 & -3 & 1 & 1 & 4 & 0.107955 & 1 \\
 34 & 1 & 2 & 0 & 0 & -1 & 1 & 0.106598 & 2 \\
\bottomrule                     
\end{tabular}                      
\end{table}                     
\begin{table}
\caption{Noninteracting levels for moving frame $d=(0,1,2)$ in cubic box. }
\label{tab:freekOh012}   
\begin{tabular}{ccccccccc}                                   
\toprule
 Level & $n_{1x}$ & $n_{1y}$ & $n_{1z}$ & $n_{2x}$ & $n_{2y}$ & $n_{2z}$ & k (GeV) & Deg.\\
  \hline
 40 & -2 & -1 & -1 & 2 & 2 & 3 & 0.134984 & 4 \\
 39 & 0 & 2 & 2 & 0 & -1 & 0 & 0.131952 & 1 \\
 38 & 0 & -1 & -2 & 0 & 2 & 4 & 0.130087 & 1 \\
 37 & -2 & 1 & -1 & 2 & 0 & 3 & 0.12984 & 2 \\
 36 & -1 & 0 & -2 & 1 & 1 & 4 & 0.127443 & 3 \\
 35 & -1 & -2 & 1 & 1 & 3 & 1 & 0.127191 & 4 \\
 34 & 0 & 1 & -2 & 0 & 0 & 4 & 0.124742 & 1 \\
\bottomrule                     
\end{tabular}                      
\end{table}                     
\begin{table}
\caption{Noninteracting levels for moving frame $d=(0,1,2)$ in $z$ elongated box with $\eta=1.5$. }
\label{tab:freekD4h012} 
\begin{tabular}{ccccccccc}                                   
\toprule
 Level & $n_{1x}$ & $n_{1y}$ & $n_{1z}$ & $n_{2x}$ & $n_{2y}$ & $n_{2z}$ & k (GeV) & Deg.\\
  \hline
 40 & 0 & -2 & 0 & 0 & 3 & 2 & 0.110104 & 1 \\
 39 & -1 & -1 & -2 & 1 & 2 & 4 & 0.109824 & 2 \\
 38 & -1 & 2 & 0 & 1 & -1 & 2 & 0.109812 & 2 \\
 37 & -1 & 0 & 3 & 1 & 1 & -1 & 0.107725 & 2 \\
 36 & -2 & 0 & 1 & 2 & 1 & 1 & 0.106478 & 2 \\
 35 & 0 & 2 & -1 & 0 & -1 & 3 & 0.105765 & 1 \\
 34 & 0 & 1 & 3 & 0 & 0 & -1 & 0.104516 & 1 \\
\bottomrule                     
\end{tabular}                      
\end{table}                     
%

\clearpage
\subsection{Phase shift reconstruction for the lowest partial wave }
\label{sup:low}

The L\"uscher method has predictive power if only the lowest partial wave is retained in the quantization conditions (QC) and the box levels are used as input. The truncated QC is often referred to as the ``L\"uscher formula".
This section contains predictions from the L\"uscher formula for all QCs in this work. 
They are given in the norm order $|\bm d|^2$. Within each d, results for the two geometries are presented side-by-side to facilitate comparison.

In the case of rest frame $d=(0,0,0)$, the group symmetry is $O_h$ in cubic box, $D_{4h}$ in $z$ elongated box. 
The results are shown in Fig.~\ref{fig:Oh-all-irreps}.
We have confirmed in all cases that deviation from the curve is due to mixing with higher partial waves.
In other words, all box levels (black points) in the figures can be accommodated by the QC if 
it is enlarged to include higher partial-wave infinite-volume phase shifts up to the designed cutoff $l=5$.

\begin{figure*}
\includegraphics[scale=0.75]{Oh-all-irreps.pdf}
\includegraphics[scale=0.75]{D4h-all-irreps.pdf}
\caption{Phase shifts reconstruction for the rest frame $d=(0,0,0)$. 
The left column is from a cubic box of $L\times L\times L$ with size L=24 fm.  The right column is from a $z$ elongated box $L\times L\times (\eta L)$ with the same $L$ and $\eta=1.5$. The red curve is the infinite-volume phase shift. The black points are the predictions from the L\"uscher formula using the box levels for k.
The faint vertical lines correspond to noninteracting levels in the boxes. The labels indicate the irreps and the lowest partial wave in that irrep. Two placeholders ($A_{1u}$ and $A_{2g}$) are inserted on the left to facilitate a side-by-side comparison. 
}
\label{fig:Oh-all-irreps}
\end{figure*}
%

In the case of moving frame $d=(0,0,1)$,
the group symmetry is $C_{4v}$ in both box geometries. In group theory language, the same operations that form the little group $C_{4v}$ can be found in both $O_h$ and $D_{4h}$, see Table~\ref{tab:C4vgroup}. They share the same QCs in form. The difference is in the evaluation of zeta functions ($\eta=1$ vs. $\eta =1.5$), and energy levels.
 The matrix elements needed to construct the QCs are given in Table~\ref{tab:C4v}.  
The lowest-order phase shift reconstruction is shown side by side in Fig.~\ref{fig:C4v}.
We see that the levels are more packed than in the rest frame; both cutoffs ($k=0.2$ GeV or 40 levels) are in effect.
The numerous points that fall outside the the curves indicate strong mixing with higher partial waves.
This is confirmed by the convergence of the QCs in the supplement.
It suggests that if this moving frame is relied upon to predict the lowest partial waves ($l=0,1,2,4$) in the five irreps,
only the points that fall on the curves are useful.  Furthermore, the prediction is most reliable in the lower energy region.
If all the data points are used for the reconstruction of phase shifts, then all partial waves up to $l=4$ are needed in the QCs.
Unlike the lowest-order QC, predicting the phase shifts is more challenging if multiple partial waves are involved in a single QC~\cite{Morningstar_2017}. 
It may be possible to use a combination of multiple QCs, interpolation techniques, multiple volumes.
The noninteracting levels can be found in Table~\ref{tab:freekOh001} for cubic box, 
and Table~\ref{tab:freekD4h001} for $z$ elongated box.
In the first two levels in cubic box, we see an example of dependence on the ordering of the two masses mentioned earlier. 
When the heavier particle 2 carries the momentum of the system, the energy is lower than when particle 1 does it (0.0066 vs. 0.045). The same is true in elongated box (0.0044 vs. 0.03).
It is also an example that moving frames lead to lower CM energy in the system than rest frames. (0.0066 vs. 0.05 in cubic and 0.0044 vs. 0.034 in elongated). 

\begin{figure*}
\includegraphics[height=0.55\textheight]{Oh-C4v-all-irreps.pdf}
\includegraphics[height=0.55\textheight]{D4h-C4v-all-irreps.pdf}
\caption{Phase shift reconstruction for the lowest partial wave in each irrep for moving frame $d=(0,0,1)$ in cubic box (left) and elongated box (right). The symmetry group is $C_{4v}$ in both cases. Deviation from the curve is due to mixing with  higher partial waves.
}
\label{fig:C4v}
\end{figure*}
%

In the case of moving frame $d=(1,1,0)$,
the group symmetry is $C_{2v}$ in both geometries (see Table~\ref{tab:C2vgroup}). 
All four irreps ($A_1$, $A_2$, $B_1$, $B_2$) are one dimensional.
They share the same QCs in form. The difference is in the evaluation of zeta functions ($\eta=1$ vs. $\eta =1.5$), and energy levels. The matrix elements needed to construct the QCs are given in Tables~\ref{tab:C2vA1} to~\ref{tab:C2vB2}.  
The lowest-order phase shift reconstruction is shown side by side in Fig.~\ref{fig:C2v}.
We have confirmed that all the points in the figure are accounted for if higher partial waves up to $l=4$ are 
included in the QC. 
The noninteracting levels are given in Table~\ref{tab:freekOh110} for cubic box, 
and Table~\ref{tab:freekD4h110} for $z$ elongated box.
The mass-order pair is level 1 and level 5 in cubic box, and level 1 and level 6 in $z$ elongated box. There are other configurations in between the two levels. Another feature is that whenever there is a level in cubic box that does not involve z direction, one can find a level in elongated box that has the same energy (1 and 1, 2 and 3, 4 and 5, 7 and 10, 10 and 12, 12 and 17, and so on). The feature still holds when interaction is turned on since the two geometries are the same as far as x and y directions are concerned for these pairs. They shift by the same mount by the interaction.

\begin{figure*}
\includegraphics[scale=0.9]{Oh-C2v-all-irreps.pdf}
\includegraphics[scale=0.9]{D4h-C2v-all-irreps.pdf}
\caption{Phase shift reconstruction for the lowest partial wave in each irrep for moving frame $d=(1,1,0)$ in cubic box (left) and elongated box (right). The symmetry group is $C_{2v}$ in both cases. Deviation from the curve is due to mixing with  higher partial waves.
}
\label{fig:C2v}
\end{figure*}
%

In the case of moving frame $d=(1,1,1)$,
it is allowed in both geometries, but in terms of symmetry, there is a distinction. 
The symmetry group is $C_{3v}$ in cubic box,  but  $C_{1v}$ in $z$ elongated box. The group tables can be found in Table~\ref{tab:C3vgroup} and bottom half of Table~\ref{tab:C1vgroup}. Consequently the QCs are different. 
For cubic box, they are $A_1$, $A_2$, and $E$ whose matrix elements are in Tables~\ref{tab:C3vA1} to~\ref{tab:C3vE}. 
They have access to the lowest partial waves $l=0$ ($A_1$), $l=1$ ($E$), and $l=3$ ($A_2$).
For elongated box, they are $A_1$, $A_2$ whose matrix elements are in Tables~\ref{tab:C1v111A1} to~\ref{tab:C1v111A2}. 
Due to reduced symmetry, they only have $l=0$ ($A_1$) and $l=1$ ($A_2$), and high degree of mixing and multiplicities.
Nonetheless, the QCs converge up to $l=4$.
The noninteracting levels are given in Table~\ref{tab:freekOh111} for cubic box, 
and Table~\ref{tab:freekD4h111} for $z$ elongated box.
The mass-order pair is level 1 and level 6 in cubic box, and 1 and 12 in elongated box.

\begin{figure*}
\includegraphics[scale=0.9]{Oh-C3v-all-irreps.pdf}
\includegraphics[scale=0.9]{D4h-C1v111-all-irreps.pdf}
\caption{Phase shift reconstruction for the lowest partial wave in each irrep for moving frame $d=(1,1,1)$ in cubic box (left) and elongated box (right). The symmetry group is $C_{3v}$ in cubic case and $C_{1v}$ in elongated case. Deviation from the curve is due to mixing with  higher partial waves.
}
\label{fig:C3v}
\end{figure*}
%

In the case of moving frame $d=(0,1,2)$, it has the least symmetry.
Although it is not used very often in practice, we want to check its convergence nonetheless.
The group symmetry is $C_{1v}$ in both geometries as given in the top half of Table~\ref{tab:C1vgroup}. 
They share the same QCs in form. 
The difference is in the evaluation of zeta functions ($\eta=1$ vs. $\eta =1.5$), and energy levels. 
The matrix elements needed to construct the QCs are given in Tables~\ref{tab:C1v012A1} to~\ref{tab:C1v012A2}.  
The noninteracting levels are given in Table~\ref{tab:freekOh012} for cubic box, 
and Table~\ref{tab:freekD4h012} for $z$ elongated box.
The mass order pair is level 1 and level 18 in cubic box, 1 and 16 in elongated box.
For $A_1$, The partial wave content with associated multiplicities is $0 (1), 1(2), 2(3), 3(4), 4(5),\cdots$.
The QC has the matrix dimension of 15x15 at order 5.
For $A_2$, The partial wave content with associated multiplicities is $1 (1), 2(2), 3(3), 4(4),\cdots$.
The QC has the matrix dimension of 10x10 at order 4.
Despite the high degree of partial wave mixing, convergence is found.

\begin{figure*}
\includegraphics[scale=0.9]{Oh-C1v-all-irreps.pdf}
\includegraphics[scale=0.9]{D4h-C1v012-all-irreps.pdf}
\caption{Phase shift reconstruction for the lowest partial wave in each irrep for moving frame $d=(0,1,2)$ in cubic box (left) and elongated box (right). The symmetry group is $C_{1v}$ in both cases. Deviation from the curve is due to mixing with  higher partial waves.
}
\label{fig:C1v}
\end{figure*}
%

\clearpage
\subsection{Sensitivity to second partial wave}
\label{sup:2nd}
Additional cases in Fig.~\ref{fig:Oh2nd} in cubic box and Fig.~\ref{fig:D4h2nd} for elongated box.
The sensitivity comes from the `pinched' points in Fig.~\ref{fig:Oh-all-irreps}. Due to strong partial wave mixing,  sensitivity for moving frames is not considered.
\begin{figure}
\includegraphics[scale=0.4]{Oh-A1g-l4-pinched.pdf}
\includegraphics[scale=0.4]{Oh-T1u-l3-pinched.pdf}
\includegraphics[scale=0.4]{Oh-T2g-l4-pinched.pdf}
\includegraphics[scale=0.4]{Oh-Eg-l4-pinched.pdf}
\includegraphics[scale=0.4]{Oh-T2u-l5-pinched.pdf}
\caption{Second partial waves in rest frame of cubic box.
The black points are $\cos\delta_{2nd}$ in Eq.\eqref{eq:2nd} with $\cos\delta_{1st}$ neglected, evaluated at the pinched box levels from order 1. The red curves are the infinite-volume  values for $\cos\delta_{2nd}$. The faint vertical lines are the free-particle levels.
}
\label{fig:Oh2nd}
\end{figure}
\begin{figure}
\includegraphics[width=0.5\textwidth]{D4h-A1g-l2-pinched.pdf}
\includegraphics[width=0.5\textwidth]{D4h-A2u-l3-pinched.pdf}
\includegraphics[width=0.5\textwidth]{D4h-Eu-l3-pinched.pdf}
\includegraphics[width=0.5\textwidth]{D4h-Eg-l4-pinched.pdf}
\includegraphics[width=0.5\textwidth]{D4h-B1g-l4-pinched.pdf}
\caption{Second partial waves in rest frame of elongated box.
}
\label{fig:D4h2nd}
\end{figure}
%

\clearpage
\begin{widetext}
\subsection{Convergence data}
\label{sup:data}

This section has  information on the convergence of every level in every quantization condition (QC) considered in this work.  
We provide two forms to demonstrate convergence. One is a graph of absolute value of QC2 in Eq.\eqref{eq:QC2}. Since QC2 is complex valued, a solution must be satisfied by both real and imaginary parts. The absolute value gives a cleaner look.
Using the infinite volume phase shifts as input, we plot the determinant $|QC2|$ as a function $k$ (blue curve). It is a well-behaved function bounded from both above and below. It gives a full view of the QC.  Next we locate the solutions (roots) of the quantization condition. We find QC1 is more sensitive for root-finding purposes than QC2. Once the roots are located they are passed to QC2 and displayed as empty red circles. Last, we plot the the determinant with both infinite-volume phase shift and box k levels as input (black dots). Deviations between the red and black points are a measure of effects from higher partial waves. 
This is done by enlarging the QC order by order. Convergence is achieved when the red and black points agree. In other words, the box levels are accounted for by the QC with all the partials waves at the order considered.
Another form is numerical using the high-precision $\chi^2$-measure defined in Eq.\eqref{eq:chi2}. 
The two forms are presented together as a figure, one combo for each QC. Interesting features are pointed out in the captions.
A couple of them ($A_{1g}$ and $T_{1u}$) were used as examples in the main text. 
The full collection has 45 QCs. The breakdown is $8+5+4+3+2=22$ in cubic box, and $10+5+4+2+2=23$ in elongated box.
They are given below by the order 1 to 45 in the summary Table~\ref{tab:all}.

%
\begin{figure*}[!b]
\includegraphics[width=.6\textwidth]{Oh-A1g-2orders.pdf}
\includegraphics[width=.6\textwidth]{Oh-A1g-2orders-table.png}
\caption{Convergence data for $A_{1g}$ irrep of the $O_h$ group. Top: graphical view by QC2. The red open circles are the roots of the QC when infinite-volume phase shifts are fed to it. The black dots are the residue of the QC when both the infinite-volume phase shifts and the box levels are used as input. Bottom: numerical table using a $\chi^2$ measure defined in the text. The red star is the free-particle pole near the box level. The partial waves $l$ involved in each order along with their multiplicities can be found in the title of the top graph. This one is used as an example in the main text.
}
\label{fig:Oh-A1g2}
\end{figure*}
\begin{figure*}
\includegraphics[width=.6\textwidth]{Oh-A2u-1order.pdf}
\includegraphics[width=.5\textwidth]{Oh-A2u-1order-table.png}
\caption{Convergence data for cubic, rest frame, group $O_h$, and irrep $A_{2u}$. The convergence is excellent due to the large gap to the next partial wave ($l=7$). This suggests $A_{2u}$ is good at predicting the $l=3$ phase shift in cubic box.}
\label{fig:Oh-A2u}
\end{figure*}
\begin{figure*}
\includegraphics[width=.6\textwidth]{Oh-Eg-2orders.pdf}
\includegraphics[width=.6\textwidth]{Oh-Eg-2orders-table.png}
\caption{Convergence data for cubic, rest frame, group $O_h$, and irrep $E_g$. There are 5 exception levels that are part of the 5 close-by pairs at order 1. They are all accounted for at order 2.}
\label{fig:Oh-Eg}
\end{figure*}
\begin{figure*}
\includegraphics[width=.7\textwidth]{Oh-Eu-1order.pdf}
\includegraphics[width=.5\textwidth]{Oh-Eu-1order-table.png}
\caption{Convergence data for cubic, rest frame, group $O_h$, and irrep $E_u$.}
\label{fig:Oh-Eu}
\end{figure*}
\begin{figure*}
\includegraphics[width=.7\textwidth]{Oh-T1g-1order.pdf}
\includegraphics[width=.5\textwidth]{Oh-T1g-1order-table.png}
\caption{Convergence data for cubic, rest frame, group $O_h$, and irrep $T_{1g}$.}
\label{fig:Oh-T1g}
\end{figure*}

\begin{figure*}
\includegraphics[width=.8\textwidth]{Oh-T1u-3orders.pdf}
\includegraphics[width=.8\textwidth]{Oh-T1u-3orders-table.png}
\caption{Convergence data for cubic, rest frame, group $O_h$, and irrep $T_{1u}$. This one is used an example and discussed in detail in the main text.
}
\label{fig:Oh-T1u2}
\end{figure*}
\begin{figure*}
\includegraphics[width=.8\textwidth]{Oh-T2g-2orders.pdf}
\includegraphics[width=.7\textwidth]{Oh-T2g-2orders-table.png}
\caption{Convergence data for cubic, rest frame, group $O_h$, and irrep $T_{2g}$.}
\label{fig:Oh-T2g}
\end{figure*}
\begin{figure*}
\includegraphics[width=.8\textwidth]{Oh-T2u-2orders.pdf}
\includegraphics[width=.7\textwidth]{Oh-T2u-2orders-table.png}
\caption{Convergence data for cubic, rest frame, group $O_h$, and irrep $T_{2u}$.}
\label{fig:Oh-T2u}
\end{figure*}
%
\begin{figure*}
\includegraphics[width=.5\textwidth]{Oh-C4vA1-5orders.pdf}
\includegraphics[width=.8\textwidth]{Oh-C4vA1-5orders-table.png}
\caption{Convergence data for cubic, moving frame $d=(0,0,1)$, group $C_{4v}$, and irrep $A_1$.}
\label{fig:Oh-C4vA1}
\end{figure*}
\begin{figure*}
\includegraphics[width=.8\textwidth]{Oh-C4vA2-1order.pdf}
\includegraphics[width=.7\textwidth]{Oh-C4vA2-1order-table.png}
\caption{Convergence data for cubic, moving frame $d=(0,0,1)$, group $C_{4v}$, and irrep $A_2$.}
\label{fig:Oh-C4vA2}
\end{figure*}
\begin{figure*}
\includegraphics[width=.7\textwidth]{Oh-C4vB1-3orders.pdf}
\includegraphics[width=.7\textwidth]{Oh-C4vB1-3orders-table.png}
\caption{Convergence data for cubic, moving frame $d=(0,0,1)$, group $C_{4v}$, and irrep $B_1$.}
\label{fig:Oh-C4vB1}
\end{figure*}

\begin{figure*}
\includegraphics[width=.7\textwidth]{Oh-C4vB2-3orders.pdf}
\includegraphics[width=.7\textwidth]{Oh-C4vB2-3orders-table.png}
\caption{Convergence data for cubic, moving frame $d=(0,0,1)$, group $C_{4v}$, and irrep $B_2$.}
\label{fig:Oh-C4vB2}
\end{figure*}
\begin{figure*}
\includegraphics[width=.5\textwidth]{Oh-C4vE-4orders.pdf}
\includegraphics[width=.8\textwidth]{Oh-C4vE-4orders-table.png}
\caption{Convergence data for cubic, moving frame $d=(0,0,1)$, group $C_{4v}$, and irrep $E$.}
\label{fig:Oh-C4vE}
\end{figure*}
%
\begin{figure*}
\includegraphics[width=.5\textwidth]{Oh-C2vA1-5orders.pdf}
\includegraphics[width=.8\textwidth]{Oh-C2vA1-5orders-table.png}
\caption{Convergence data for cubic, moving frame $d=(1,1,0)$, group $C_{2v}$, and irrep $A_1$.}
\label{fig:Oh-C2vA1}
\end{figure*}

\begin{figure*}
\includegraphics[width=.6\textwidth]{Oh-C2vA2-3orders.pdf}
\includegraphics[width=.65\textwidth]{Oh-C2vA2-3orders-table.png}
\caption{Convergence data for cubic, moving frame $d=(1,1,0)$, group $C_{2v}$, and irrep $A_2$.}
\label{fig:Oh-C2vA2}
\end{figure*}
\begin{figure*}
\includegraphics[width=.5\textwidth]{Oh-C2vB1-4orders.pdf}
\includegraphics[width=.8\textwidth]{Oh-C2vB1-4orders-table.png}
\caption{Convergence data for cubic, moving frame $d=(1,1,0)$, group $C_{2v}$, and irrep $B_1$.}
\label{fig:Oh-C2vB1}
\end{figure*}
\begin{figure*}
\includegraphics[width=.5\textwidth]{Oh-C2vB2-4orders.pdf}
\includegraphics[width=.8\textwidth]{Oh-C2vB2-4orders-table.png}
\caption{Convergence data for cubic, moving frame $d=(1,1,0)$, group $C_{2v}$, and irrep $B_2$.}
\label{fig:Oh-C2vB2}
\end{figure*}
%
\begin{figure*}
\includegraphics[width=.5\textwidth]{Oh-C3vA1-5orders.pdf}
\includegraphics[width=.8\textwidth]{Oh-C3vA1-5orders-table.png}
\caption{Convergence data for cubic, moving frame $d=(1,1,1)$, group $C_{3v}$, and irrep $A_1$.}
\label{fig:Oh-C3vA1}
\end{figure*}
\begin{figure*}
\includegraphics[width=.8\textwidth]{Oh-C3vA2-2orders.pdf}
\includegraphics[width=.7\textwidth]{Oh-C3vA2-2orders-table.png}
\caption{Convergence data for cubic, moving frame $d=(1,1,1)$, group $C_{3v}$, and irrep $A_2$.}
\label{fig:Oh-C3vA2}
\end{figure*}
\begin{figure*}
\includegraphics[width=.55\textwidth]{Oh-C3vE-4orders.pdf}
\includegraphics[width=.75\textwidth]{Oh-C3vE-4orders-table.png}
\caption{Convergence data for cubic, moving frame $d=(1,1,1)$, group $C_{3v}$, and irrep $E$.}
\label{fig:Oh-C3vE}
\end{figure*}
%
\begin{figure*}
\includegraphics[width=.5\textwidth]{Oh-C1vA1-5orders.pdf}
\includegraphics[width=.8\textwidth]{Oh-C1vA1-5orders-table.png}
\caption{Convergence data for cubic, moving frame $d=(0,1,2)$, group $C_{1v}$, and irrep $A_1$.}
\label{fig:Oh-C1vA1}
\end{figure*}
\begin{figure*}
\includegraphics[width=.5\textwidth]{Oh-C1vA2-4orders.pdf}
\includegraphics[width=.8\textwidth]{Oh-C1vA2-4orders-table.png}
\caption{Convergence data for cubic, moving frame $d=(0,1,2)$, group $C_{1v}$, and irrep $A_2$.}
\label{fig:Oh-C1vA2}
\end{figure*}
%


%
\begin{figure*}
\includegraphics[width=.7\textwidth]{D4h-A1g-3orders.pdf}
\includegraphics[width=.7\textwidth]{D4h-A1g-3orders-table.png}
\caption{Convergence data for elongated, rest frame, group $D_{4h}$, and irrep $A_{1g}$.}
\label{fig:D4h-A1g}
\end{figure*}
\begin{figure*}
\includegraphics[width=.7\textwidth]{D4h-A1u-1order.pdf}
\includegraphics[width=.5\textwidth]{D4h-A1u-1order-table.png}
\caption{Convergence data for elongated, rest frame, group $D_{4h}$, and irrep $A_{1u}$.}
\label{fig:D4h-A1u}
\end{figure*}
\begin{figure*}
\includegraphics[width=.7\textwidth]{D4h-A2g-1order.pdf}
\includegraphics[width=.6\textwidth]{D4h-A2g-1order-table.png}
\caption{Convergence data for elongated, rest frame, group $D_{4h}$, and irrep $A_{2g}$.}
\label{fig:D4h-A2g}
\end{figure*}
\begin{figure*}
\includegraphics[width=.7\textwidth]{D4h-A2u-3orders.pdf}
\includegraphics[width=.7\textwidth]{D4h-A2u-3orders-table.png}
\caption{Convergence data for elongated, rest frame, group $D_{4h}$, and irrep $A_{2u}$.}
\label{fig:D4h-A2u}
\end{figure*}
\begin{figure*}
\includegraphics[width=.6\textwidth]{D4h-Eg-2orders.pdf}
\includegraphics[width=.6\textwidth]{D4h-Eg-2orders-table.png}
\caption{Convergence data for elongated, rest frame, group $D_{4h}$, and irrep $E_g$.}
\label{fig:D4h-Eg}
\end{figure*}
\begin{figure*}
\includegraphics[width=.6\textwidth]{D4h-Eu-3orders.pdf}
\includegraphics[width=.7\textwidth]{D4h-Eu-3orders-table.png}
\caption{Convergence data for elongated, rest frame, group $D_{4h}$, and irrep $E_u$.}
\label{fig:D4h-Eu}
\end{figure*}
\begin{figure*}
\includegraphics[width=.7\textwidth]{D4h-B1g-2orders.pdf}
\includegraphics[width=.6\textwidth]{D4h-B1g-2orders-table.png}
\caption{Convergence data for elongated, rest frame, group $D_{4h}$, and irrep $B_{1g}$.}
\label{fig:D4h-B1g}
\end{figure*}
\begin{figure*}
\includegraphics[width=.8\textwidth]{D4h-B1u-1order.pdf}
\includegraphics[width=.7\textwidth]{D4h-B1u-1order-table.png}
\caption{Convergence data for elongated, rest frame, group $D_{4h}$, and irrep $B_{1u}$. The convergence is as good as in cubic box. More levels are available  in the elongated box.}
\label{fig:D4h-B1u}
\end{figure*}
\begin{figure*}
\includegraphics[width=.8\textwidth]{D4h-B2g-2orders.pdf}
\includegraphics[width=.7\textwidth]{D4h-B2g-2orders-table.png}
\caption{Convergence data for elongated, rest frame, group $D_{4h}$, and irrep $B_{2g}$.}
\label{fig:D4h-B2g}
\end{figure*}
\begin{figure*}
\includegraphics[width=.8\textwidth]{D4h-B2u-2orders.pdf}
\includegraphics[width=.7\textwidth]{D4h-B2u-2orders-table.png}
\caption{Convergence data for elongated, rest frame, group $D_{4h}$, and irrep $B_{2u}$.}
\label{fig:D4h-B2u}
\end{figure*}
%
%
\begin{figure*}
\includegraphics[width=.5\textwidth]{D4h-C4vA1-5orders.pdf}
\includegraphics[width=.85\textwidth]{D4h-C4vA1-5orders-table.png}
\caption{Convergence data for elongated, moving frame $d=(0,0,1)$, group $C_{4v}$, and irrep $A_1$.}
\label{fig:D4h-C4vA1}
\end{figure*}
\begin{figure*}
\includegraphics[width=.8\textwidth]{D4h-C4vA2-1order.pdf}
\includegraphics[width=.7\textwidth]{D4h-C4vA2-1order-table.png}
\caption{Convergence data for elongated, moving frame $d=(0,0,1)$, group $C_{4v}$, and irrep $A_2$.}
\label{fig:D4h-C4vA2}
\end{figure*}
\begin{figure*}
\includegraphics[width=.6\textwidth]{D4h-C4vB1-3orders.pdf}
\includegraphics[width=.7\textwidth]{D4h-C4vB1-3orders-table.png}
\caption{Convergence data for elongated, moving frame $d=(0,0,1)$, group $C_{4v}$, and irrep $B_1$.}
\label{fig:D4h-C4vB1}
\end{figure*}
\begin{figure*}
\includegraphics[width=.6\textwidth]{D4h-C4vB2-3orders.pdf}
\includegraphics[width=.7\textwidth]{D4h-C4vB2-3orders-table.png}
\caption{Convergence data for elongated, moving frame $d=(0,0,1)$, group $C_{4v}$, and irrep $B_2$.}
\label{fig:D4h-C4vB2}
\end{figure*}
\begin{figure*}
\includegraphics[width=.6\textwidth]{D4h-C4vE-4orders.pdf}
\includegraphics[width=.7\textwidth]{D4h-C4vE-4orders-table.png}
\caption{Convergence data for elongated, moving frame $d=(0,0,1)$, group $C_{4v}$, and irrep $E$.}
\label{fig:D4h-C4vE}
\end{figure*}
%
\begin{figure*}
\includegraphics[width=.5\textwidth]{D4h-C2vA1-5orders.pdf}
\includegraphics[width=.8\textwidth]{D4h-C2vA1-5orders-table.png}
\caption{Convergence data for elongated, moving frame $d=(1,1,0)$, group $C_{2v}$, and irrep $A_1$.}
\label{fig:D4h-C2vA1}
\end{figure*}
\begin{figure*}
\includegraphics[width=.6\textwidth]{D4h-C2vA2-3orders.pdf}
\includegraphics[width=.7\textwidth]{D4h-C2vA2-3orders-table.png}
\caption{Convergence data for elongated, moving frame $d=(1,1,0)$, group $C_{2v}$, and irrep $A_2$.}
\label{fig:D4h-C2vA2}
\end{figure*}
\begin{figure*}
\includegraphics[width=.5\textwidth]{D4h-C2vB1-4orders.pdf}
\includegraphics[width=.8\textwidth]{D4h-C2vB1-4orders-table.png}
\caption{Convergence data for elongated, moving frame $d=(1,1,0)$, group $C_{2v}$, and irrep $B_1$.}
\label{fig:D4h-C2vB1}
\end{figure*}
\begin{figure*}
\includegraphics[width=.5\textwidth]{D4h-C2vB2-4orders.pdf}
\includegraphics[width=.8\textwidth]{D4h-C2vB2-4orders-table.png}
\caption{Convergence data for elongated, moving frame $d=(1,1,0)$, group $C_{2v}$, and irrep $B_2$.}
\label{fig:D4h-C2vB2}
\end{figure*}
%
%
\begin{figure*}
\includegraphics[width=.5\textwidth]{D4h-C1v111A1-5orders.pdf}
\includegraphics[width=.8\textwidth]{D4h-C1v111A1-5orders-table.png}
\caption{Convergence data for elonagted, moving frame $d=(1,1,1)$, group $C_{1v}$, and irrep $A_1$.}
\label{fig:D4h-C1v111A1}
\end{figure*}
\begin{figure*}
\includegraphics[width=0.55\textwidth]{D4h-C1v111A2-4orders.pdf}
\includegraphics[width=0.75\textwidth]{D4h-C1v111A2-4orders-table.png}
\caption{Convergence data for elonagted, moving frame $d=(1,1,1)$, group $C_{1v}$, and irrep $A_2$.}
\label{fig:D4h-C1v111A2}
\end{figure*}
%
%

\begin{figure*}
\includegraphics[width=0.5\textwidth]{D4h-C1v012A1-5orders.pdf}
\includegraphics[width=0.8\textwidth]{D4h-C1v012A1-5orders-table.png}
\caption{Convergence data for elongated, moving frame $d=(0,1,2)$, group $C_{1v}$, and irrep $A_1$.}
\label{fig:D4h-C1v012A1}
\end{figure*}
\begin{figure*}
\includegraphics[width=0.5\textwidth]{D4h-C1v012A2-4orders.pdf}
\includegraphics[width=0.8\textwidth]{D4h-C1v012A2-4orders-table.png}
\caption{Convergence data for elongated, moving frame $d=(0,1,2)$, group $C_{1v}$, and irrep $A_2$.}
\label{fig:D4h-C1v012A2}
\end{figure*}
\end{widetext}

\end{supplement}
\end{document}